\documentclass[fleqn,10pt]{article}
\usepackage{latexsym, graphicx, epsfig, amsmath, amssymb,amsfonts}
\usepackage{natbib,amsthm,version}
\usepackage{amsbsy,bm,multirow,enumerate}
\usepackage[titletoc,page]{appendix}
\usepackage[mathscr]{eucal}
\usepackage{mathtools}
\usepackage{color}
\usepackage[utf8]{inputenc}
\usepackage[english]{babel}
\usepackage{amsthm}
\usepackage{enumerate}
\usepackage[hidelinks]{hyperref}
\usepackage{url}
\usepackage{subfigure}
\usepackage[lined,boxed]{algorithm2e}

\newcommand{\mbs}[1]{\mathbf{#1}}

\newtheorem{remark}{Remark}[section]

\theoremstyle{definition}

\title{
  On a Simple and Effective Thermal Open Boundary
  Condition for Convective Heat Transfer Problems
} 
\author{
  Xiaoyu Liu$^{1,2}$, \ Zhi Xie$^1$,
  Suchuan Dong$^2$\thanks{Author of correspondence.
    Email: sdong@purdue.edu}
  \\
  $^1$Collge of Infomation Science and Engineering \\
  Northeastern University, China \\
  $^2$Center for Computational and Applied Mathematics \\
  Department of Mathematics \\
  Purdue University, USA 
 } 

\date{(October 21, 2019)}
\begin{document}
\maketitle



\begin{abstract}

  We present an effective thermal open boundary condition for convective
  heat transfer problems on domains involving outflow/open boundaries.
  This boundary condition is energy-stable, and it ensures that the contribution
  of the open boundary will not cause an ``energy-like'' temperature functional
  to increase over time, irrespective of the state of flow on the
  open boundary.
  It is effective in coping with thermal open boundaries even in
  flow regimes where strong vortices or backflows are prevalent
  on such boundaries, and it is straightforward to implement.
  Extensive numerical simulations are presented to demonstrate the stability
  and effectiveness of our method for heat transfer
  problems with strong vortices and backflows occurring on the open
  boundaries. Simulation results are compared with previous works
  to demonstrate the accuracy of the presented method.

\end{abstract}


\vspace{0.05cm}

\section{Introduction}
\label{sec:intro}

%
%
%
%

In this work we focus on the numerical simulation of convective heat transfer
in fluid flows on domains involving outflow/open boundaries.
The domain boundary is open in the sense that the fluid and the heat can freely leave
or enter the domain through such a boundary.
This type of problems are typically encountered in flows with physically unbounded
domains, such as wakes, jets, shear layers, and cardiovascular or
respiratory networks~\cite{DongKC2014}.
Buoyancy-driven flows such as the natural convection in open-ended
cavities or open channels are other
examples~\cite{ChanT1985a,ChanT1986,Desrayaudetal2013,Zhangetal2017},
which have widespread applications in solar energy receivers,
cooling of electronics, and control of smoke or fire.
To numerically simulate such problems, it is necessary to
truncate the domain to a finite size, and some outflow/open boundary condition (OBC)
will be needed for the artificial boundary.
How to properly deal with the open boundary oftentimes holds the
key to successful simulations of these problems.
This turns out to be a very challenging problem~\cite{Gresho1991,SaniG1994}
as the Reynolds number increases to moderate and high values.
A well-known issue, at least for the flow simulations, is the backflow
issue and the so-called backflow instability~\cite{Dong2015clesobc,NiYD2019}.
This refers to the difficulty encountered in flow simulations
when strong vortices or backflows occur at the outflow/open boundary
at moderate and high Reynolds numbers.
Many open boundary conditions that work well
at low Reynolds numbers, such as the traction-free
condition~\cite{TaylorRM1985,Gartling1990,EngelmanJ1990,
Leone1990,BehrLST1991,SaniG1994,GuermondMS2005}
and the convective condition~\cite{Sommerfeld1949,Orlanski1976,Gresho1991,
KeskarL1999,OlshanskiiS2000,ForestierPPS2000,
RuithCM2004}, cease to work and become
unstable when strong vortices and backflows are present
at the outflow/open boundary.
It is observed that
an otherwise stable computation can instantly
blow up when a strong vortex passes through the outflow/open
boundary~\cite{DongK2005,DongKER2006,VargheseFF2007,DongKC2014,Dong2015clesobc}.
The backflow instability issue has attracted 
a number of  efforts in the past years.
A class of methods, the so-called energy-stable open
boundary conditions~\cite{DongKC2014,Dong2014obc,DongS2015,Dong2015clesobc,DongW2016,YangD2018,NiYD2019},
turn out to be particularly effective for overcoming the backflow
instability; see also related works
in~\cite{BruneauF1994,BruneauF1996,BazilevsGHMZ2009,LanzendorferS2011,Moghadametal2011,PorporaZVP2012,GravemeierCYIW2012,IsmailGCW2014,BertoglioC2014,FeistauerN2013,Fouchet2014,BraackM2014}, among others.
These energy-stable open boundary conditions,
by design, guarantee that the contributions from the open boundary
will not cause the total system energy to increase over time,
irrespective of the flow situations occurring at the open
boundary (e.g.~presence of backflows or strong vortices).
Therefore, stable results can be obtained with these methods
even when strong vortices or backflows occur on the outflow/open
boundary at high Reynolds numbers.
More importantly from the practical standpoint,
these energy-stable OBCs can be implemented in a straightforward way
with the commonly-used semi-implicit splitting type (or fractional-step) schemes
for the incompressible Navier-Stokes equations~\cite{Dong2015clesobc,DongS2015}.


For open-boundary  convective heat transfer problems,
a survey of literature indicates that how to deal with the thermal
open boundary, especially
for moderate and high Reynolds numbers where strong vortices or
backflows are prevalent at the open boundary, seems to be much less developed
when compared with that
for the fluid flows as outlined above.
The Sommerfeld radiation (or convective) condition (see e.g.~\cite{Orlanski1976,CominiN1998})
and the Neumann type zero-flux condition (see e.g.~\cite{DavalathB1987,YoungV1998,AbbassiTN2001})
are traditional boundary conditions for the temperature
applied to the open/outflow boundary.
In \cite{ChanT1985a,ChanT1985b} the natural convection in an open
cavity has been studied numerically with an extended large computational
domain and with a smaller domain consisting of the cavity only.
On the extended domain the Neumann type zero-flux condition is imposed
for both the velocity and the temperature on the outflow boundary~\cite{ChanT1985a}.
For the smaller domain,
on the open boundary of the cavity
the authors distinguish the sections where the
flow enters the cavity (inflow) and the sections
where the flow leaves the cavity (outflow),
and impose a temperature Dirichlet condition on
the inflow portion and the Neumann zero-flux condition for the temperature
on the outflow portion~\cite{ChanT1985b}.
Additionally, 
the authors therein employ a Neumann zero-flux condition
for the tangential velocity and the divergence-free condition for
the normal velocity component on the open boundary~\cite{ChanT1985b}.
While the use of extended computational domains pushes the open
boundary farther away and can
partially alleviate the issue that the true boundary condition
is unknown~\cite{Kettleborough1972}, this can be computationally costly because of
the increased domain size~\cite{Desrayaudetal2013};
see also e.g.~\cite{VafaiE1990,Gan2010} for investigations
of the domain size effect on the computed physical quantities.
The boundary condition of~\cite{ChanT1985b}
and its many variants have been widely adopted in 
studies of natural convection in subsequent years
and have been one of  the pre-dominant methods
for handling thermal open boundaries
where backflows may be present; 
see e.g.~\cite{LiT1994,KhanaferV2002,BilgenO2005,LalR2009,ChungV2010,FontanaSM2011,Desrayaudetal2013,ShirvanMMEV2017,Zhangetal2017}, among others.


In the current paper we present a new thermal open
boundary condition that is energy-stable and effective for
simulating convective heat transfer problems involving outflow/open
boundaries, even at high (or moderate) Reynolds numbers
when strong vortices or backflows occur on the open boundary. 
This boundary condition is formulated  such that
the contribution of the open/outflow boundary
will not cause an ``energy-like'' temperature function 
to increase over time, regardless of the state of flow
at the open boundary.
The form of this thermal open boundary condition has much
been inspired by the open boundary conditions from~\cite{Dong2015clesobc}
for the incompressible Navier-Stokes equations.
In particular, it contains an inertial term (time derivative of temperature)
and an extra nonlinear term combining the velocity and the temperature,
apart from the temperature directional derivative at the
boundary. The nonlinear term in the thermal open boundary
condition can be analogized to a term in those open boundary conditions
for the incompressible
Navier-Stokes equations~\cite{Dong2015clesobc,DongS2015,NiYD2019,BruneauF1994},
and it also bears a similarity to the conditions
considered in~\cite{PerezTBC2008a,Neustupa2017,CeretaniR2019}.
%
%

The presented thermal open boundary condition can be implemented
in a straightforward fashion. In the current paper we discretize
this open boundary condition and the heat transfer equation
based on a semi-implicit scheme. Upon discretization, this open boundary condition
becomes a Robin-type condition for the temperature, and
is implemented using a high-order spectral element
technique~\cite{KarniadakisS2005,SherwinK1995,ZhengD2011}.
The discretized system of algebraic
equations involves a coefficient matrix that is constant and time-independent
and can be pre-computed.
The current scheme for the thermal open boundary condition,
with no change, also applies to finite element-type
techniques.
%
We note that the current thermal open boundary condition is
much simpler to implement than the commonly-used boundary condition
from~\cite{ChanT1985b}. The condition of~\cite{ChanT1985b}
imposes a temperature Dirichlet condition on the backflow region
of the boundary. Since such a region is dynamic and changes over time,
this in general will require the re-computation and re-factorization (at least partially)
of the temperature coefficient matrix every time step in the implementation.
For finite element type methods, the dynamic nature of the backflow
region on the open boundary can make the implementation of the temperature
Dirichlet condition especially difficult.

We combine the
presented thermal open boundary condition, together with
the open boundary condition from~\cite{Dong2015clesobc} for
the incompressible Navier-Stokes equations,
to simulate convective heat transfer on domains involving
outflow/open boundaries.
Only one-way coupling between the velocity and the temperature
(velocity influencing temperature, but not the other way)
is considered in the current work.
We have performed
extensive numerical experiments to test the
presented method, especially in regimes of high or
fairly high Reynolds numbers, when strong vortices and backflows
become prevalent at the open boundary and
the backflow instability becomes a severe issue to conventional
methods. We compare our simulations with previous works
to demonstrate the accuracy of the current method.
The long-term stability of this method has been demonstrated
in the presence of strong vortices and backflows at the outflow/open
boundaries. We show that in such situations the current
thermal open boundary condition leads to reasonable simulation results,
while the Neumann-type zero-flux condition produces
unphysical temperature distributions.

The contributions of this paper lie in the thermal open
boundary condition developed herein and the numerical scheme
for treating the presented open boundary condition.
Particularly noteworthy are
the effectiveness of the method in coping with thermal open boundaries
where strong vortices or backflows may be present,
and its ease in implementation.

The rest of this paper is organized as follows.
In Section \ref{sec:method} we present the thermal open
boundary condition, look into its energy stability,
and develop a semi-implicit scheme for implementing
this boundary condition together with the heat transfer equation.
In Section \ref{sec:tests} we demonstrate the convergence
rates, and test the current method using two-dimensional
simulations of the heat transfer in
two canonical flows: the flow past a circular cylinder
and a jet impinging on a wall. Simulations are compared with previous
works to show the accuracy of our method.
We also demonstrate the method's long-term stability in regimes
where strong vortices and backflows are prevalent at the outflow/open boundaries.
Section \ref{sec:summary} concludes the presentation
with some closing remarks.
Appendix A provides a summary of the open boundary condition
and the numerical scheme from~\cite{Dong2015clesobc} for
the incompressible Navier-Stokes equations, which are employed
in the current work.


\section{Energy-Stable Thermal Open Boundary Condition}
\label{sec:method}

\subsection{Heat Transfer Equation and Energy-Stable Open Boundary Condition}

Consider a domain $\Omega$ in two or three dimensions, and
an incompressible flow contained within. We focus on the heat transfer
in this system. The problem is described by the following system
of equations (in non-dimensional form):
\begin{subequations}
  \begin{align}
    &
    \frac{\partial \mbs u}{\partial t} + \mbs u\cdot\nabla\mbs u
    + \nabla p -\nu\nabla^2\mbs u = \mbs f(\mbs x,t),
    \label{equ:nse} \\
    &
    \nabla\cdot\mbs u = 0,
    \label{equ:div} \\
    &
    \frac{\partial T}{\partial t} + \mbs u\cdot\nabla T = \alpha\nabla^2 T
    + g(\mbs x,t),
    \label{equ:tem}
  \end{align}
\end{subequations}
where $\mbs u(\mbs x,t)$ is the velocity, $p(\mbs x,t)$ is the pressure,
$T(\mbs x,t)$ is the temperature,
$\mbs f$ is an external body force, $g(\mbs x,t)$ is
an external volumetric heat source term,
and $\mbs x$ denotes the spatial coordinate and $t$ is time.
$\nu$ is the inverse of the Reynolds number ($Re$) or non-dimensional viscosity,
\begin{equation}\label{equ:def_nu}
  \nu = \frac{1}{Re} = \frac{\nu_f}{U_0L}
\end{equation}
where $\nu_f$ is the kinematic viscosity of the fluid, $U_0$ is
the velocity scale and $L$ is the length scale.
$\alpha$ is the inverse of the Peclet number
or the non-dimensional thermal diffusivity,
\begin{equation}\label{equ:peclet}
  \alpha = \frac{1}{P_e} = \frac{\alpha_f}{U_0L},
\end{equation}
where $\alpha_f$ is the thermal diffusivity of the fluid.
We assume that both $\nu$ and $\alpha$
are constants.
In the current work we will consider only the one-way
coupling between the flow and temperature. In other words,  the flow
influences the temperature distribution, while the effect
of the temperature on the flow will not be accounted for.
In addition, some other effects such as the heat production
due to the viscous dissipation will also be ignored.
Note that equations \eqref{equ:nse} and \eqref{equ:div}
are the incompressible Navier-Stokes equations describing
the motion of the fluid.


Let $\partial\Omega$ denote the boundary of the domain $\Omega$.
We assume that $\partial\Omega$ consists of two types
(non-overlapping with each other),
$\partial\Omega = \partial\Omega_d \cup\partial\Omega_o$,
with the following properties:
\begin{itemize}

\item
  $\partial\Omega_d$ is the inflow 
  or solid-wall boundary.
  On $\partial\Omega_d$ the velocity $\mbs u$ is known.
  In terms of the temperature, we assume that $\partial\Omega_d$
  further consists of two sub-types (non-overlapping),
  $\partial\Omega_d = \partial\Omega_{dd}\cup\partial\Omega_{dn}$.
  On $\partial\Omega_{dd}$ the temperature is known, and
  on $\partial\Omega_{dn}$ the heat flux is known.

\item
  $\partial\Omega_o$ is the outflow/open boundary.
  On $\partial\Omega_o$ none of the field variables (velocity, pressure,
  temperature) is known. 

\end{itemize}
How to deal with
the thermal open/outflow boundary $\partial\Omega_o$ is the subject of the current study.
Open boundary conditions for the incompressible Navier-Stokes equations
have been studied extensively in a number of previous works
(see e.g.~\cite{Gresho1991,SaniG1994,BruneauF1996,DongKC2014,DongS2015,Dong2015clesobc,NiYD2019},
among others).
In this paper, for the Navier-Stokes equations, we will employ
the open boundary condition developed in \cite{Dong2015clesobc}.
This boundary condition, together with a corresponding numerical
algorithm, is summarized  in the Appendix A
for the sake of completeness.

We now concentrate on how to deal with
the open/outflow boundary for the heat transfer equation~\eqref{equ:tem}.
Multiplying equation \eqref{equ:tem} by $T$ and integrating over
the domain $\Omega$, we obtain the following balance equation,
\begin{equation}\label{equ:tem_eng}
  \begin{split}
  \frac{\partial}{\partial t}\int_{\Omega} \frac12 \left|T\right|^2d\Omega
  =& -\alpha\int_{\Omega}\left|\nabla T\right|^2 d\Omega
  + \int_{\Omega} g(\mbs x,t) T d\Omega
  + \int_{\partial\Omega_{dd}\cup\partial\Omega_{dn}}\left[
    \alpha\mbs n\cdot\nabla T - \frac12(\mbs n\cdot\mbs u)T
    \right]T dA \\
  &
  + \int_{\partial\Omega_o}\underbrace{\left[
    \alpha\mbs n\cdot\nabla T - \frac12(\mbs n\cdot\mbs u)T
    \right]T}_{\text{outflow boundary term (OBT)}} dA, 
  \end{split}
\end{equation}
where $\mbs n$ is the outward-pointing unit vector normal to the
boundary, and we have used integration by part, equation \eqref{equ:div}
and the divergence theorem.
The quantity $\frac12|T|^2$ can be considered as an effective
``energy'' for the heat transfer equation.
The last surface integral on the right hand side (RHS) 
represents the contribution of the open/outflow boundary
to this balance equation for the effective energy.
This term (OBT) is indefinite, and can be
positive or negative depending on the imposed boundary condition
 and the flow state  on $\partial\Omega_o$.
In particular, with the Neumann-type zero-flux
condition (see e.g.~\cite{DavalathB1987,YoungV1998,AbbassiTN2001}),
\begin{equation}\label{equ:zero_obc}
  \mbs n\cdot\nabla T = 0, \quad \text{on}\ \partial\Omega_o,
\end{equation}
this open-boundary term would become positive locally when backflow occurs
(i.e.~$\mbs n\cdot\mbs u<0$) on the outflow/open boundary,
e.g.~when strong vortices pass through $\partial\Omega_o$ at
moderate or high Reynolds numbers.
This can cause un-controlled growth in the effective energy, leading
to poor simulation results
or numerical instabilities.

We are interested in seeking open boundary conditions for
the temperature such that the open-boundary term in
the balance equation \eqref{equ:tem_eng}
is always non-positive, regardless of the state of flow 
on the open boundary $\partial\Omega_o$. 
As such, 
the contribution from the outflow/open boundary
will not cause the effective energy $\frac12|T|^2$ to grow over time,
in the absence of the external heat source and with appropriate
boundary conditions for the other types of boundaries.
This will be conducive to the stability of computations.
We refer to such conditions as
energy-stable thermal open boundary conditions.

In the current work, we consider the following
 open boundary condition
for the temperature, 
  \begin{align}
    &
    \alpha D_0\frac{\partial T}{\partial t}
    +\alpha\mbs n\cdot\nabla T -
    \left[(\mbs n\cdot\mbs u)T\right]\Theta_0(\mbs n,\mbs u)
    = 0,
    \quad \text{on} \ \partial\Omega_o.
    \label{equ:obc_A}
  \end{align}
In this equation $D_0\geqslant 0$ is a chosen constant, and
  $U_c = \frac{1}{D_0}$ plays the role of a convection velocity scale
  on the outflow/open boundary $\partial\Omega_o$. In practice, one
  can first estimate the convection velocity scale $U_c$
  on $\partial\Omega_o$ and then set $D_0=\frac{1}{U_c}$ in
  the boundary condition \eqref{equ:obc_A}. 
$\Theta_0(\mbs n,\mbs u)$ is a smoothed step function given by
(see~\cite{DongS2015,Dong2015clesobc}),
\begin{equation}\label{equ:def_Theta0}
  \Theta_0(\mbs n,\mbs u) = \frac12\left(
  1 - \tanh\frac{\mbs n\cdot\mbs u}{U_0\delta}
  \right);
  \qquad
  \lim_{\delta \rightarrow 0}\Theta_0(\mbs n,\mbs u) = \Theta_{s0}(\mbs n,\mbs u)
  = \left\{
  \begin{array}{ll}
    1, & \text{if} \ \mbs n\cdot\mbs u < 0, \\
    0, & \text{if} \ \mbs n\cdot\mbs u > 0,
  \end{array}
  \right.
\end{equation}
where $U_0$ is the characteristic velocity scale, and $\delta>0$ is
a small constant that controls the sharpness of the
smoothed step function. The function is sharper with a smaller $\delta$.
As $\delta \rightarrow 0$, $\Theta_0(\mbs n,\mbs u)$ approaches
the step function $\Theta_{s0}(\mbs n,\mbs u)$, taking the unit value
if $\mbs n\cdot\mbs u<0$ and vanishing otherwise.
Therefore the term involving $\Theta_0$ in the boundary condition
\eqref{equ:obc_A} 
only takes effect
in the regions of backflow on the outflow/open boundary $\partial\Omega_o$.


The form of this thermal open boundary condition~\eqref{equ:obc_A} has much
been inspired by the boundary condition for the incompressible
Navier-Stokes equations from~\cite{Dong2015clesobc}.
The boundary condition \eqref{equ:obc_A}, 
with $\delta$ sufficiently small,
is an energy-stable open boundary condition for the heat transfer equation.
With this boundary condition on $\partial\Omega_o$
the balance equation \eqref{equ:tem_eng} is reduced
  to, under the assumption that $g(\mbs x,t)=0$,
  $\mbs u=0$ on $\partial\Omega_d=\partial\Omega_{dd}\cup\partial\Omega_{dn}$,
  $T=0$ on $\partial\Omega_{dd}$
  and $\mbs n\cdot\nabla T=0$ on $\partial\Omega_{dn}$,
  \begin{equation}\label{equ:eng_A}
    \begin{split}
    &\frac{\partial}{\partial t}\left(
    \int_{\Omega} \frac12\left|T \right|^2 d\Omega
    + \alpha D_0 \int_{\partial\Omega_o} \frac12\left|T \right|^2 dA
    \right) \\
    &= -\alpha\int_{\Omega} \left|\nabla T  \right|^2 d\Omega
    + \int_{\partial\Omega_o}\frac12(\mbs n\cdot\mbs u)T^2 \left[
      2 \Theta_{s0}(\mbs n,\mbs u) - 1
      \right] dA \\
    &=
    -\alpha\int_{\Omega} \left|\nabla T  \right|^2 d\Omega
    -\int_{\partial\Omega_o}\frac12\left|\mbs n\cdot\mbs u \right|T^2 dA,
    \quad \text{as} \ \delta \rightarrow 0.
    \end{split}
  \end{equation}


\begin{remark}\label{rem:rem_2}

  One can also consider the following more general form of
  open boundary condition for the temperature,
  \begin{equation}\label{equ:gobc}
    \alpha D_0\frac{\partial T}{\partial t} + \alpha\mbs n\cdot\nabla T
    - \left[\frac{\theta}{2}(\mbs n\cdot\mbs u)T\right]\Theta_0(\mbs n,\mbs u)
    = 0,
    \quad \text{on} \ \partial\Omega_o,
  \end{equation}
  where the parameter $\theta$ is a chosen constant satisfying
  $\theta \geqslant 1$.
  The boundary condition \eqref{equ:obc_A} 
  corresponds to \eqref{equ:gobc} with $\theta=2$. 
  Analogous to equation \eqref{equ:eng_A},
  we can show that equation \eqref{equ:gobc},
  with $\theta\geqslant 1$ and $\delta$ sufficiently small,
  represents a family of energy-stable thermal open boundary conditions,
  because in this case equation \eqref{equ:tem_eng} is reduced to:
  \begin{equation}\label{equ:eng_gobc}
    \begin{split}
    &\frac{\partial}{\partial t}\left(
    \int_{\Omega} \frac12\left|T \right|^2 d\Omega
    + \alpha D_0 \int_{\partial\Omega_o} \frac12\left|T \right|^2 dA
    \right) \\
    &= -\alpha\int_{\Omega} \left|\nabla T  \right|^2 d\Omega
    + \int_{\partial\Omega_o}\frac12(\mbs n\cdot\mbs u)T^2 \left[
      \theta \Theta_{s0}(\mbs n,\mbs u) - 1
      \right] dA \\
    &\leqslant
    -\alpha\int_{\Omega} \left|\nabla T  \right|^2 d\Omega,
    \quad \text{as} \ \delta \rightarrow 0.
    \end{split}
  \end{equation}  
    
\end{remark}

%


Apart from the outflow/open boundary, we impose
the following Dirichlet condition for the temperature
on $\partial\Omega_{dd}$,
\begin{equation}\label{equ:dbc}
  T = T_d(\mbs x,t), \quad \text{on} \ \partial\Omega_{dd},
\end{equation}
where $T_d(\mbs x,t)$ denotes the boundary temperature distribution,
and the following Neumann type condition on $\partial\Omega_{dn}$,
\begin{equation}\label{equ:nbc}
  \mbs n\cdot\mbs \nabla T = g_c(\mbs x, t), \quad \text{on} \ \partial\Omega_{dn},
\end{equation}
where $g_c$ is a prescribed term associated with the
boundary heat flux.
In addition, we assume the following initial condition for
the temperature,
\begin{equation}\label{equ:ic}
  T(\mbs x,0) = T_{in}(\mbs x)
\end{equation}
where $T_{in}$ denotes the initial temperature distribution.


Besides the temperature, the incompressible Navier-Stokes
equations \eqref{equ:nse}--\eqref{equ:div} also require appropriate
boundary  conditions  and 
initial conditions.
The boundary and initial conditions for the Navier-Stokes
equations employed in the current work are summarized in the
Appendix A.

\subsection{Numerical Algorithm and Implementation}

Let us now consider how to numerically solve
the heat transfer equation \eqref{equ:tem}, together with
the open boundary condition \eqref{equ:obc_A} (or \eqref{equ:gobc})
for $\partial\Omega_o$, and the boundary conditions
\eqref{equ:dbc} for $\partial\Omega_{dd}$ and
\eqref{equ:nbc} for $\partial\Omega_{dn}$.
We assume that the velocity $\mbs u$ has already been computed
by solving the incompressible Navier-Stokes equations
\eqref{equ:nse}--\eqref{equ:div}, together with appropriate
boundary conditions for $\partial\Omega_d$ and $\partial\Omega_o$.
The numerical algorithm employed in the current work for
solving the incompressible Navier-Stokes equations stems from
our previous work~\cite{Dong2015clesobc}, which has been 
summarized in Appendix A as mentioned before, 

We next focus on the solution of the temperature field.
We re-write the boundary conditions \eqref{equ:obc_A} and \eqref{equ:gobc} into
a unified form,
\begin{equation}\label{equ:obc}
  \alpha D_0\frac{\partial T}{\partial t} +
  \alpha\mbs n\cdot\nabla T - H(\mbs n,\mbs u, T) = g_b(\mbs x,t),
  \quad \text{on} \ \partial\Omega_o,
\end{equation}
where
\begin{equation}\label{equ:def_H}
  H(\mbs n,\mbs u, T) = \left\{
  \begin{array}{ll}
    \left[(\mbs n\cdot\mbs u)T  \right]\Theta_0(\mbs n,\mbs u),
    & \text{for boundary condition \eqref{equ:obc_A}},
    \\
    \left[\frac{\theta}{2}(\mbs n\cdot\mbs u)T  \right]\Theta_0(\mbs n,\mbs u),
    & \text{for general form \eqref{equ:gobc}}, \\
  \end{array}
  \right.
\end{equation}
and $g_b$ is a prescribed source term for the purpose of numerical
testing only, which will be set to $g_b=0$ in actual simulations.

Let $n\geqslant 0$ denote the time step index, and
$(\cdot)^n$ denote the variable $(\cdot)$ at time step $n$.
Let $J$ ($J=1$ or $2$) denote the temporal order of accuracy.
Given $T^n$ and $\mbs u^{n+1}$ (computed using the algorithm
from Appendix A), we compute $T^{n+1}$ based on the following
scheme:
\begin{subequations}
  \begin{align}
    &
    \frac{\gamma_0T^{n+1}-\hat T}{\Delta t}
    + \mbs u^{n+1}\cdot\nabla T^{*,n+1}
    = \alpha\nabla^2 T^{n+1} + g^{n+1};
    \label{equ:tem_1} \\
    &
    \alpha D_0\frac{\gamma_0T^{n+1}-\hat{T}}{\Delta t} +
    \alpha \mbs n\cdot\nabla T^{n+1} - H(\mbs n,\mbs u^{n+1}, T^{*,n+1}) = g_b^{n+1},
    \quad \text{on} \ \partial\Omega_o;
    \label{equ:tem_2} \\
    &
    T^{n+1} = T_d^{n+1}, \quad \text{on} \ \partial\Omega_{dd};
    \label{equ:tem_3} \\
    &
    \mbs n\cdot\nabla T^{n+1} = g_c^{n+1}, \quad \text{on} \ \partial\Omega_{dn}.
    \label{equ:tem_4}
  \end{align}
\end{subequations}
%
In the above equations
$\Delta t$ is the time step size.
$\frac{1}{\Delta t}(\gamma_0T^{n+1}-\hat T)$ is an approximation of
$\left.\frac{\partial T}{\partial t} \right|^{n+1}$ based on
the $J$-th order backward differentiation formula (BDF), in which
\begin{equation}\label{equ:def_hat}
  \gamma_0 = \left\{
  \begin{array}{ll}
    1, & J=1, \\
    3/2, & J=2;
  \end{array}
  \right.
  \quad
  \hat T = \left\{
  \begin{array}{ll}
    T^n, & J=1, \\
    2T^n - \frac12 T^{n-1}, & J=2.
  \end{array}
  \right.
\end{equation}
$T^{*,n+1}$ is a $J$-th order explicit approximation of $T^{n+1}$, specifically
given by
\begin{equation}\label{equ:def_star}
  T^{*,n+1} = \left\{
  \begin{array}{ll}
    T^n, & J=1, \\
    2T^n-T^{n-1}, & J=2.
  \end{array}
  \right.
\end{equation}
Note that $H(\mbs n,\mbs u^{n+1},T^{*,n+1})$ is given by equation
\eqref{equ:def_H}.

In the current work we employ $C^0$-continuous
high-order spectral elements~\cite{KarniadakisS2005,ZhengD2011} for spatial discretizations.
Let $\varphi(\mbs x)$ denote an arbitrary test function that vanishes on
$\partial\Omega_{dd}$, i.e.~$\left.\varphi \right|_{\partial\Omega_{dd}}=0$.
Multiplying $\varphi$ to equation \eqref{equ:tem_1} and integrating
over the domain $\Omega$, we obtain the weak form about $T^{n+1}$,
\begin{multline}\label{equ:weak_T}
  \int_{\Omega}\nabla T^{n+1}\cdot\nabla\varphi d\Omega
  + \frac{\gamma_0}{\alpha\Delta t}\int_{\Omega} T^{n+1}\varphi d\Omega
  + \frac{\gamma_0D_0}{\Delta t}\int_{\partial\Omega_o}T^{n+1}\varphi dA \\
  = \frac{1}{\alpha}\int_{\Omega}\left(
  g^{n+1} + \frac{\hat T}{\Delta t} - \mbs u^{n+1}\cdot\nabla T^{*,n+1}
  \right)\varphi d\Omega 
  + \int_{\partial\Omega_{dn}} g_c^{n+1}\varphi dA \\
  + \int_{\partial\Omega_o}\left[
    \frac{D_0}{\Delta t}\hat{T} +
    \frac{1}{\alpha}g_b^{n+1} + \frac{1}{\alpha}H(\mbs n,\mbs u^{n+1},T^{*,n+1})
    \right]\varphi dA,
  \qquad \forall \varphi \ \text{with}\ \varphi|_{\partial\Omega_{dd}}=0,
\end{multline}
where we have used integration by part, the divergence theorem,
and the
equations \eqref{equ:tem_2} and \eqref{equ:tem_4}.

The weak form \eqref{equ:weak_T}, 
together with the Dirichlet condition \eqref{equ:tem_3},
can be discretized using $C^0$ spectral elements
in the standard way~\cite{KarniadakisS2005}.
Within a time step we first compute the velocity
$\mbs u^{n+1}$ and pressure $p^{n+1}$
using the algorithm
from the Appendix A, and
then solve equation \eqref{equ:weak_T} 
together with \eqref{equ:tem_3} for
the temperature $T^{n+1}$.

\begin{remark}
  \label{rem:rem_3}
  When $D_0=0$, the boundary conditions \eqref{equ:obc_A}
  and \eqref{equ:gobc}
  are reduced to
  \begin{align}
    &
    \alpha\mbs n\cdot\nabla T -
    \left[(\mbs n\cdot\mbs u)T\right]\Theta_0(\mbs n,\mbs u)
    = 0,
    \quad \text{on} \ \partial\Omega_o;
    \label{equ:obc_C}
    \\
    &
    \alpha\mbs n\cdot\nabla T -
    \left[\frac{\theta}{2}(\mbs n\cdot\mbs u)T\right]\Theta_0(\mbs n,\mbs u)
    = 0,
    \quad \text{on} \ \partial\Omega_o.
    \label{equ:obc_D}
  \end{align}
  The scheme given in \eqref{equ:tem_1}--\eqref{equ:tem_4}
  equally applies to these two forms of boundary conditions,
  by simply setting $D_0=0$ and $g_b=0$ within.
  In this case, the weak form is still given by
  \eqref{equ:weak_T}, with $D_0=0$ and $g_b=0$.
  
\end{remark}


\section{Representative  Simulations}
\label{sec:tests}

We next test the performance of the method
 presented in the previous
section using several two-dimensional convective heat transfer
problems involving outflow/open boundaries.
In particular, we study flow regimes where
strong vortices or backflows are prevalent at the outflow/open
boundary. In such cases, how to handle the open
boundary is the key to successful simulations of these
problems.
We show that the current method produces stable and reasonable
results for the temperature field,
while the Neumann type zero-flux condition
leads to unphysical temperature distributions
on the outflow/open boundary when the vortices pass through.

\subsection{Convergence Rates}

\begin{figure}
  \centerline{
    \includegraphics[width=3.8in]{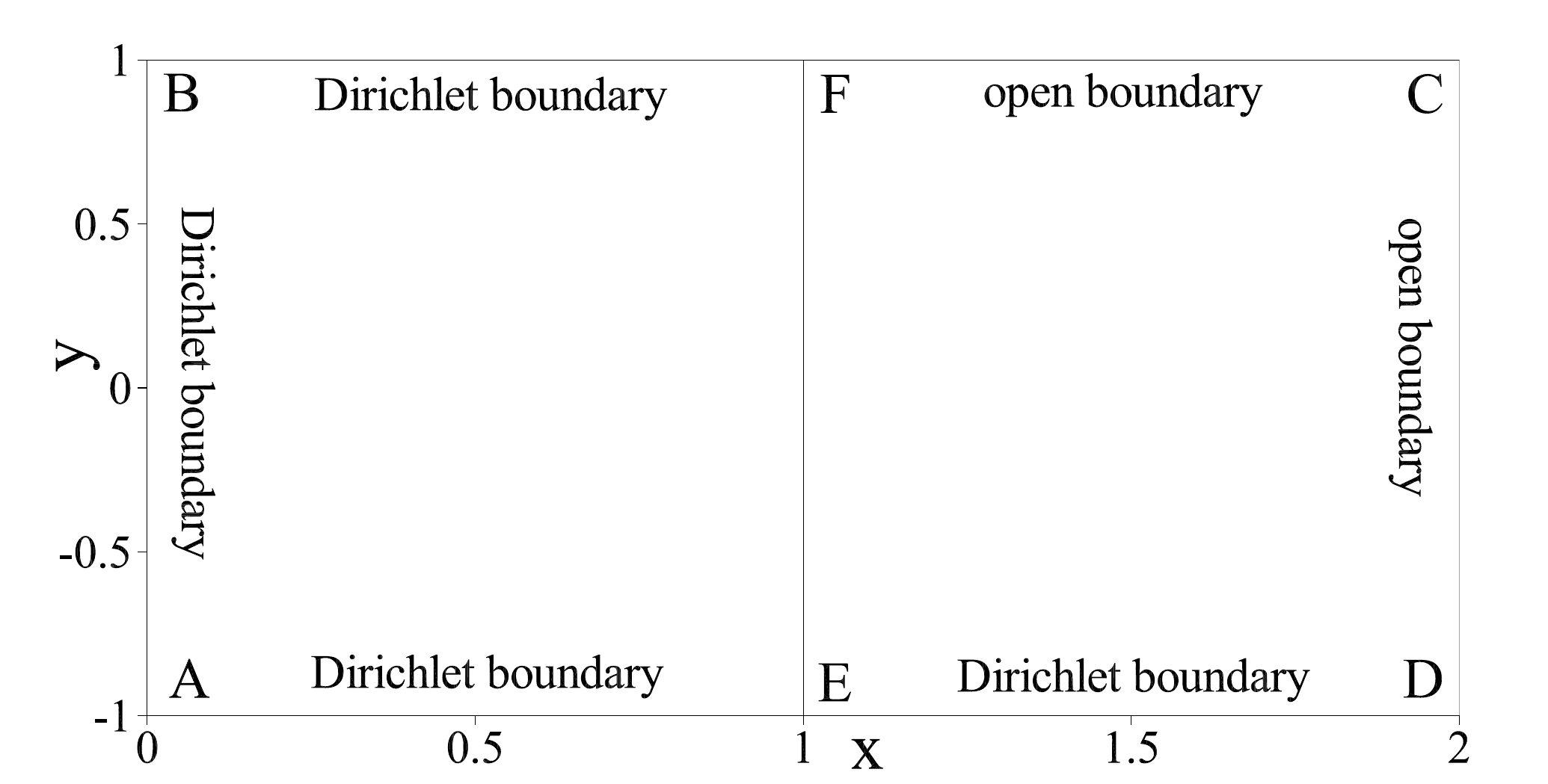}(a)
  }
  \centerline{
    \includegraphics[width=3.in]{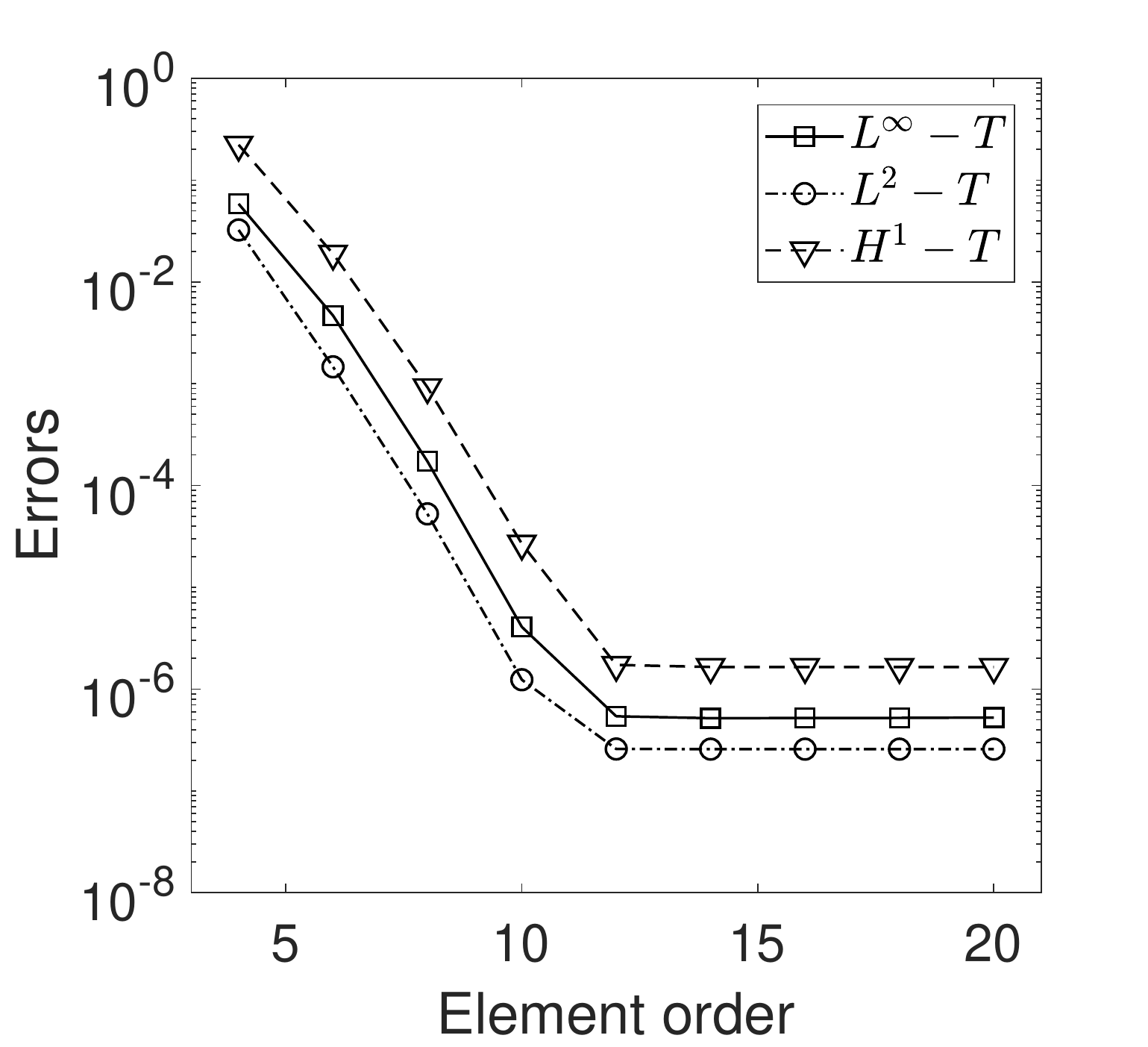}(b)
    \includegraphics[width=3.in]{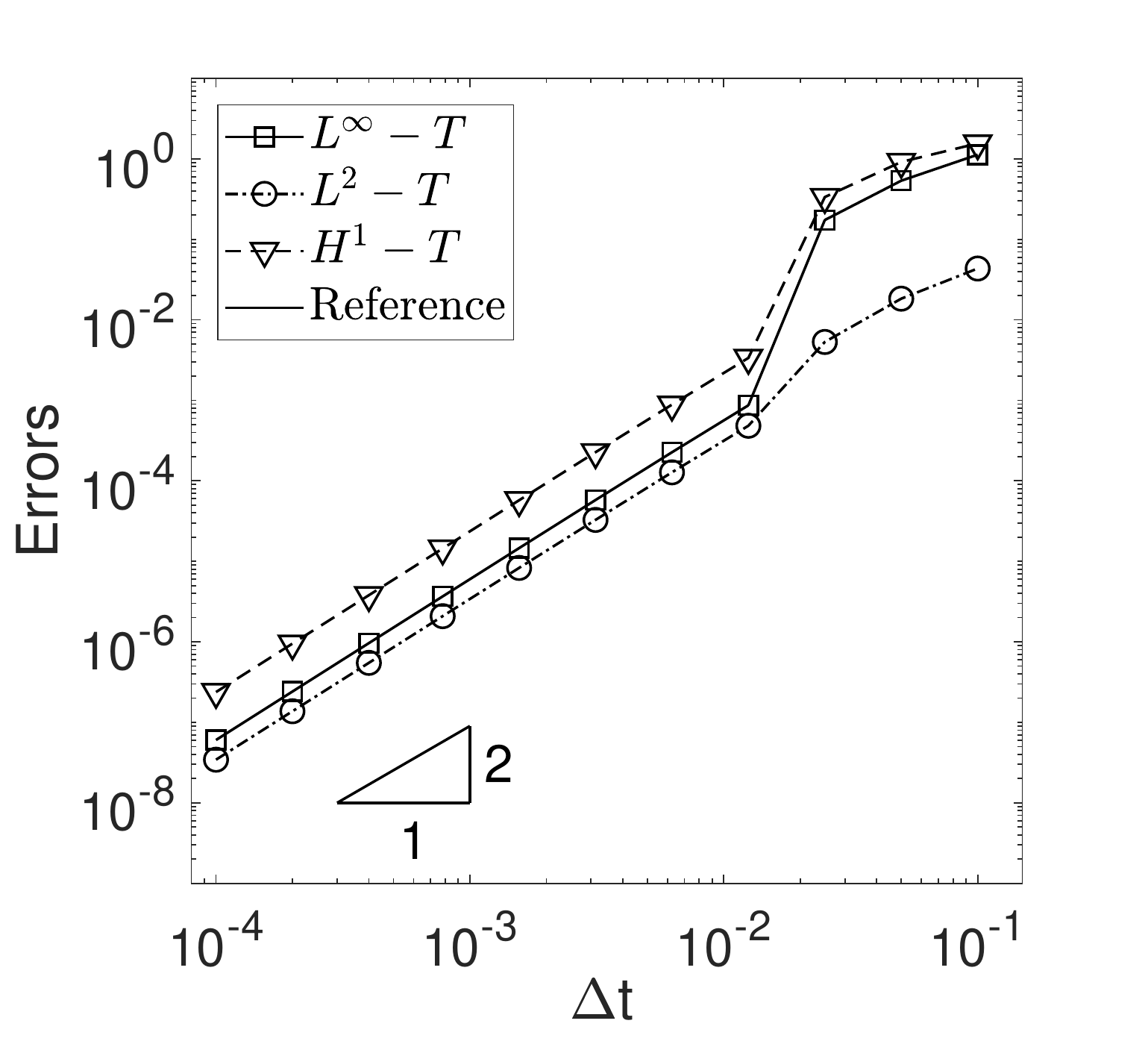}(c)
  }
  \caption{
    Convergence tests:
    (a) Flow domain and configuration.
    (b) Temperature errors ($L^{\infty}$, $L^2$ and $H^1$ norms)
    as a function of the element order (fixed $t_f=0.1$
    and $\Delta t=0.001$).
    (c) Temperature errors as a function of $\Delta t$ (fixed $t_f=0.5$
    and element order $16$).
  }
  \label{fig:conv}
\end{figure}

We first demonstrate the spatial and temporal convergence rates
of the method from Section \ref{sec:method} by using a manufactured
analytic solution to the heat transfer equation.
Consider the rectangular domain $\overline{ABCD}$ shown in Figure~\ref{fig:conv}(a),
$0\leqslant x\leqslant 2$ and $-1\leqslant y\leqslant 1$,
and the flow and heat transfer problem on
this domain. We employ the following expressions for
the flow/temperature fields of a manufactured solution
to the governing equations~\eqref{equ:nse}--\eqref{equ:tem},
\begin{equation}
  \left\{
  \begin{split}
    &
    u = 2\sin(\pi x)\cos(\pi y)\sin(2t), \\
    &
    v = -2\cos(\pi x)\sin(\pi y)\sin(2t), \\
    &
    p = 2\sin(\pi x)\sin(\pi y)\cos(2t), \\
    &
    T = 2\cos(\pi x)\sin(\pi y)\sin(2t),
  \end{split}
  \right.
  \label{equ:anal}
\end{equation}
where $\mbs u=(u,v)$. 
In equations \eqref{equ:nse} and \eqref{equ:tem} the external
force $\mbs f$ and the source term $g$ are chosen such that
the expressions from \eqref{equ:anal} satisfy these equations.
Note that the $(u,v)$ expressions in \eqref{equ:anal} also satisfy the
equation \eqref{equ:div}.

We discretize the domain $\overline{ABCD}$ using two
quadrilateral elements of the same size; see Figure \ref{fig:conv}(a).
On the sides $\overline{AD}$, $\overline{AB}$ and $\overline{BF}$
we impose the Dirichlet  condition~\eqref{equ:dbc}
for the temperature and the Dirichlet condition~\eqref{equ:dbc_v}
(in Appendix A) for the velocity, where the boundary
temperature and velocity are chosen according to the
analytical expressions from \eqref{equ:anal}.
The sides $\overline{FC}$ and $\overline{CD}$ are assumed to be open
boundaries, and we impose the boundary condition~\eqref{equ:obc}
for the temperature and the condition~\eqref{equ:obc_v}
for the velocity. In equation \eqref{equ:obc}, $H(\mbs n,\mbs u, T)$
is taken to be the first expression from equation \eqref{equ:def_H},
and $g_b$ is chosen such that the analytical expressions
from \eqref{equ:anal} satisfy the equation \eqref{equ:obc}
on the open boundary.
In equation \eqref{equ:obc_v}, $\mbs E(\mbs n,\mbs u)$
is given by \eqref{equ:def_E}, and $\mbs f_b$ is chosen
such that the analytic expressions of \eqref{equ:anal}
satisfy the equation \eqref{equ:obc_v} on the open boundary.
The initial conditions for the temperature and velocity
are given by~\eqref{equ:ic} and \eqref{equ:ic_v}, respectively,
in which $T_{in}$ and $\mbs u_{in}$ are chosen according to
the analytic expressions from \eqref{equ:anal}
by setting $t=0$.


The scheme from Section \ref{sec:method} is employed
to solve for the temperature field, and
the algorithm from the Appendix A is employed
to solve for the velocity field, in time from $t=0$
to $t_f$ (to be specified below).
Then the numerical solution of the temperature
at $t=t_f$ is compared with the analytic expression
from \eqref{equ:anal}, and the errors in
the $L^{\infty}$, $L^2$ and $H^1$ norms are computed and monitored.
In the numerical tests below we employ a fixed
non-dimensional viscosity $\nu=0.01$ and
thermal diffusivity $\alpha=0.01$.
Other parameter values include
$D_0=1.0$, $U_0=1.0$ and $\delta=0.05$ in
equations~\eqref{equ:obc}, \eqref{equ:obc_v} and \eqref{equ:def_Theta0}.
The element order and the time step size
are varied respectively in the spatial and temporal
convergence tests to study their effects on the
numerical errors.


Figure \ref{fig:conv}(b) illustrates the behavior of the method
in spatial convergence tests. In this group of tests
we employ a fixed $t_f=0.1$ and $\Delta t = 0.001$, and vary
the element order systematically between $4$ and $20$.
The figure shows the $L^{\infty}$, $L^2$ and $H^1$
errors of the temperature at $t=t_f$ as a function of
the element order. For element orders below $12$ the errors decrease
exponentially with increasing element order. For
element orders above $12$ the numerical errors are observed
to remain at a constant level, due to the saturation of
the temporal truncation error.


Figure \ref{fig:conv}(c) illustrates the temporal convergence behavior
of the method. In this group of tests a fixed $t_f=0.5$ and an element order
$16$ are employed, and the time step size $\Delta t$ is varied
systematically between $\Delta t=0.1$ and $\Delta t=1e-4$.
The figure shows the numerical errors of the temperature as
a function of $\Delta t$ from these tests.
It is evident that the method exhibits a
temporal second-order convergence rate for the temperature
as $\Delta t$ becomes small.

\subsection{Flow Past a Warm Circular Cylinder}
\label{sec:cyl}


\begin{figure}[tb]
  \centerline{
    \includegraphics[height=2in]{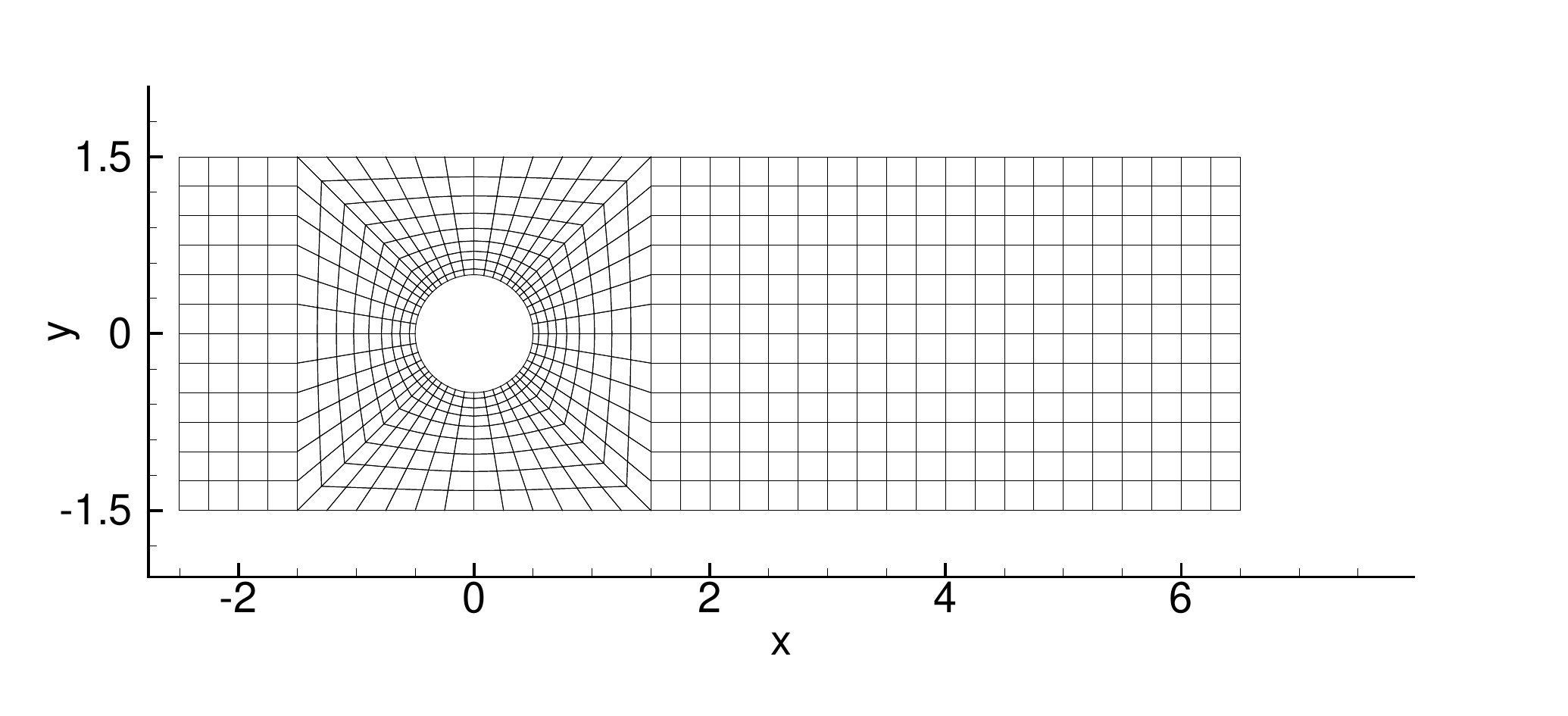}
  }
  \caption{
    Cylinder flow: Flow domain and the mesh of $720$ quadrilateral elements.
  }
  \label{fig:cyl_config}
\end{figure}

In this section
we use a canonical problem, the heat transfer in the flow past a circular
cylinder, to test the performance of the thermal open boundary condition 
and the numerical scheme herein.

Consider the flow domain  shown in Figure~\ref{fig:cyl_config},
occupying the region $-2.5d\leqslant x\leqslant 6.5d$
and $-1.5d\leqslant y\leqslant 1.5d$,
where $d=1$ is the cylinder diameter.
The cylinder center coincides with the origin of the coordinate
system. A cooler fluid with temperature $T_0=20$ degrees Celsius
enters the domain from the left
with a uniform velocity $U_0=1$ along the $x$ direction.
The flow exits the domain from the right side.
The surface of the cylinder is maintained at a higher
temperature $T_h=80$ degrees Celsius.
The top and bottom sides of the domain ($y=\pm 1.5d$)
are assumed to be periodic. 
This configuration mimics the flow past an infinite array of
circular cylinders in the $y$ direction.
We would like to study the heat transfer in this flow.
We are particularly interested
in the regimes of moderate to fairly high Reynolds numbers,
when the vortices generated at the cylinder can persist
in the entire wake region and exit the domain on the right.
How the current thermal open boundary condition
perform in such situations will be studied. 


We choose the inflow velocity $U_0$ as the velocity scale,
the cylinder diameter $d$ as the length scale, and
the unit temperature $T_d=1$ degree Celsius as
the temperature scale. Then all the physical variables and
parameters are normalized accordingly.
So the Reynolds number and the Peclet number are defined
in terms of the cylinder diameter in this problem.


We discretize the domain using a mesh of $720$ quadrilateral elements;
see Figure \ref{fig:cyl_config}.
On the left boundary ($x=-2.5d$),
Dirichlet boundary conditions~\eqref{equ:dbc} and \eqref{equ:dbc_v}
are imposed for the temperature and the velocity, respectively,
in which the boundary temperature ($T_d$) and velocity ($\mbs w$) are set
to the inflow temperature and inflow velocity as given above.
On the cylinder surface, no-slip condition is imposed
for the velocity, and the Dirichlet condition~\eqref{equ:dbc}
with $T_d=T_h=80$ is imposed for the temperature. 
Periodic conditions are imposed on the top/bottom boundaries
for all field variables.
%
On the right boundary, the open boundary condition~\eqref{equ:obc_A}
is imposed for the temperature, in which we set $D_0=\frac{1}{U_0}=1$
and $\delta =0.05$ for this problem.
For the Navier-Stokes equations
we impose the open boundary condition~\eqref{equ:obc_v},
with $\mbs f_b=0$ and $\mbs E(\mbs n,\mbs u)$ given by \eqref{equ:def_E}.


We employ the algorithm from Section \ref{sec:method} to solve
the temperature equation \eqref{equ:tem} with $g=0$  and the algorithm from
the Appendix A to solve the Navier-Stokes
equations~\eqref{equ:nse}--\eqref{equ:div} with $\mbs f=0$.
We have conducted simulations at three Reynolds numbers
($Re=300$, $2000$ and $5000$)
and two Peclet numbers (corresponding to $\alpha=0.01$ and $0.005$).
The element order, the time step size,
and other simulation parameters are varied in the simulations
to study their effects on the results.
For any given set of parameter values we have performed
long-time simulations  so that the flow has
reached a statistically stationary state.
Therefore, the initial conditions have no effect on the reported
results.


\begin{figure}[tb]
  \centerline{
    \includegraphics[width=3.2in]{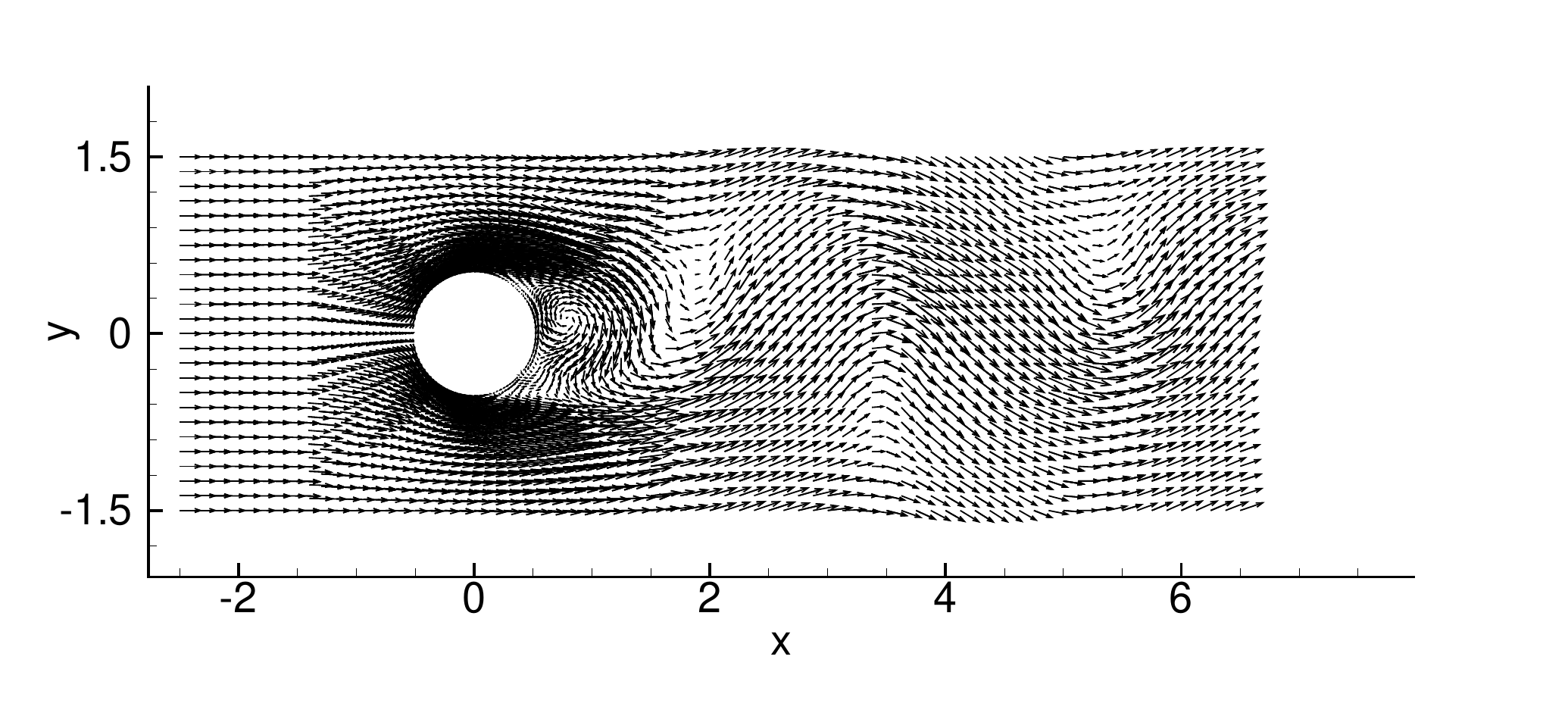}(a)
    \includegraphics[width=3.2in]{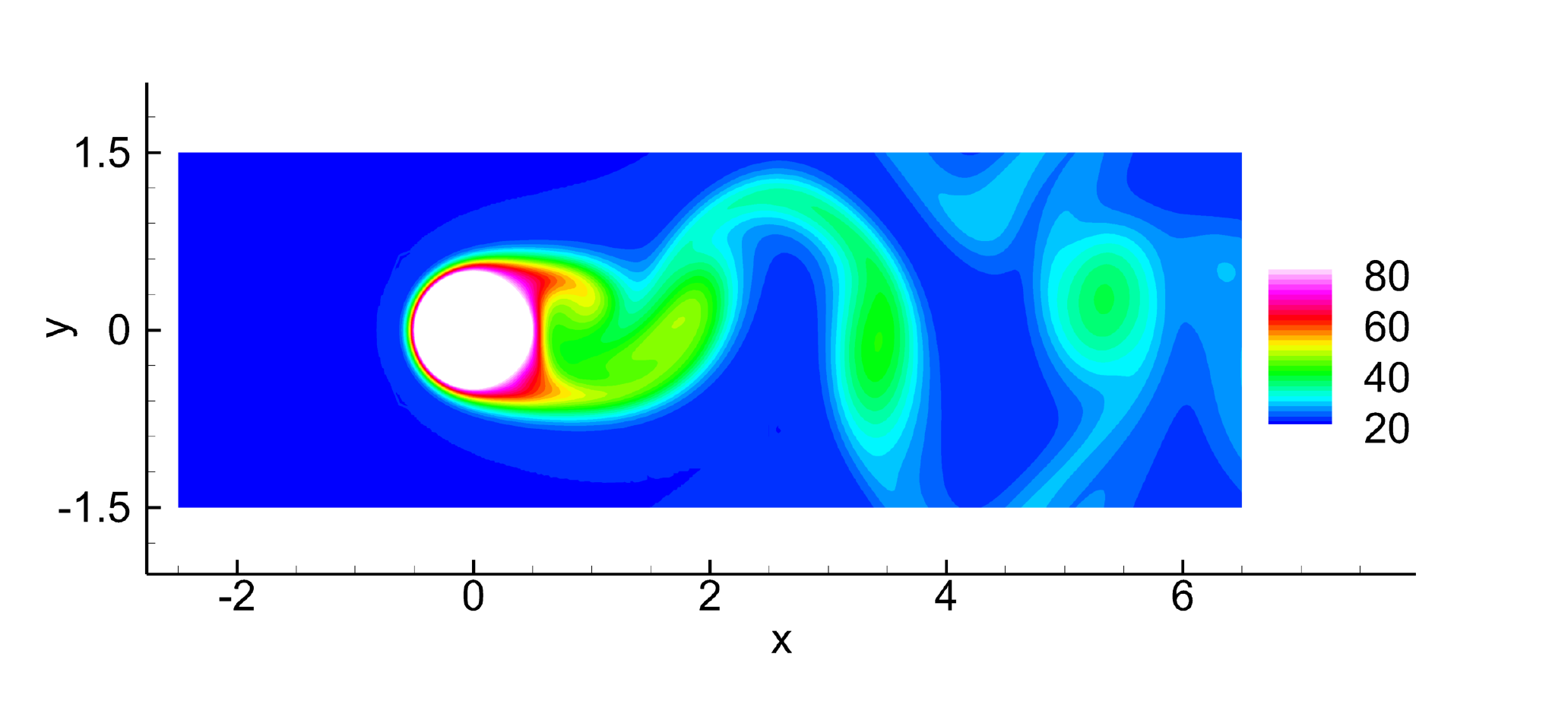}(b)
  }
  \centerline{
    \includegraphics[width=3.2in]{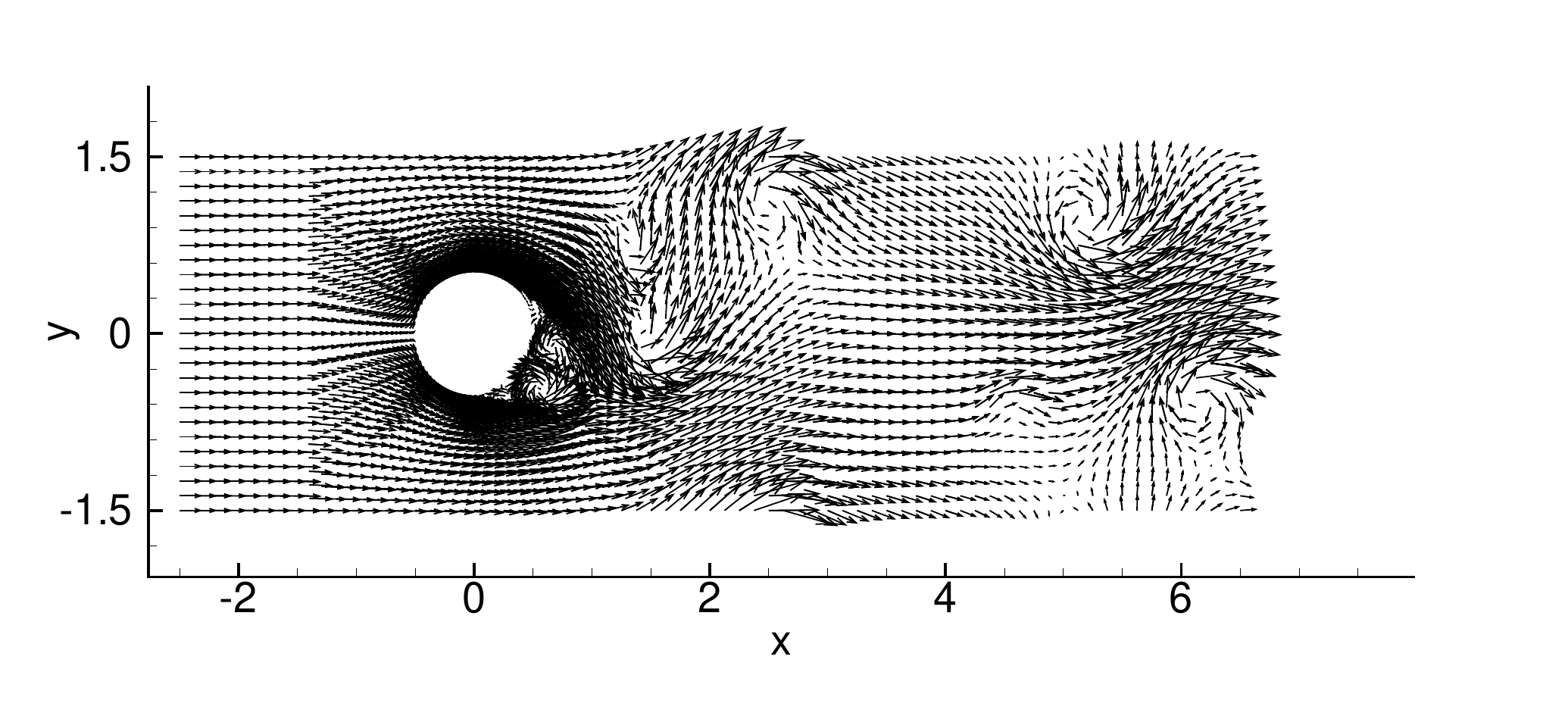}(c)
    \includegraphics[width=3.2in]{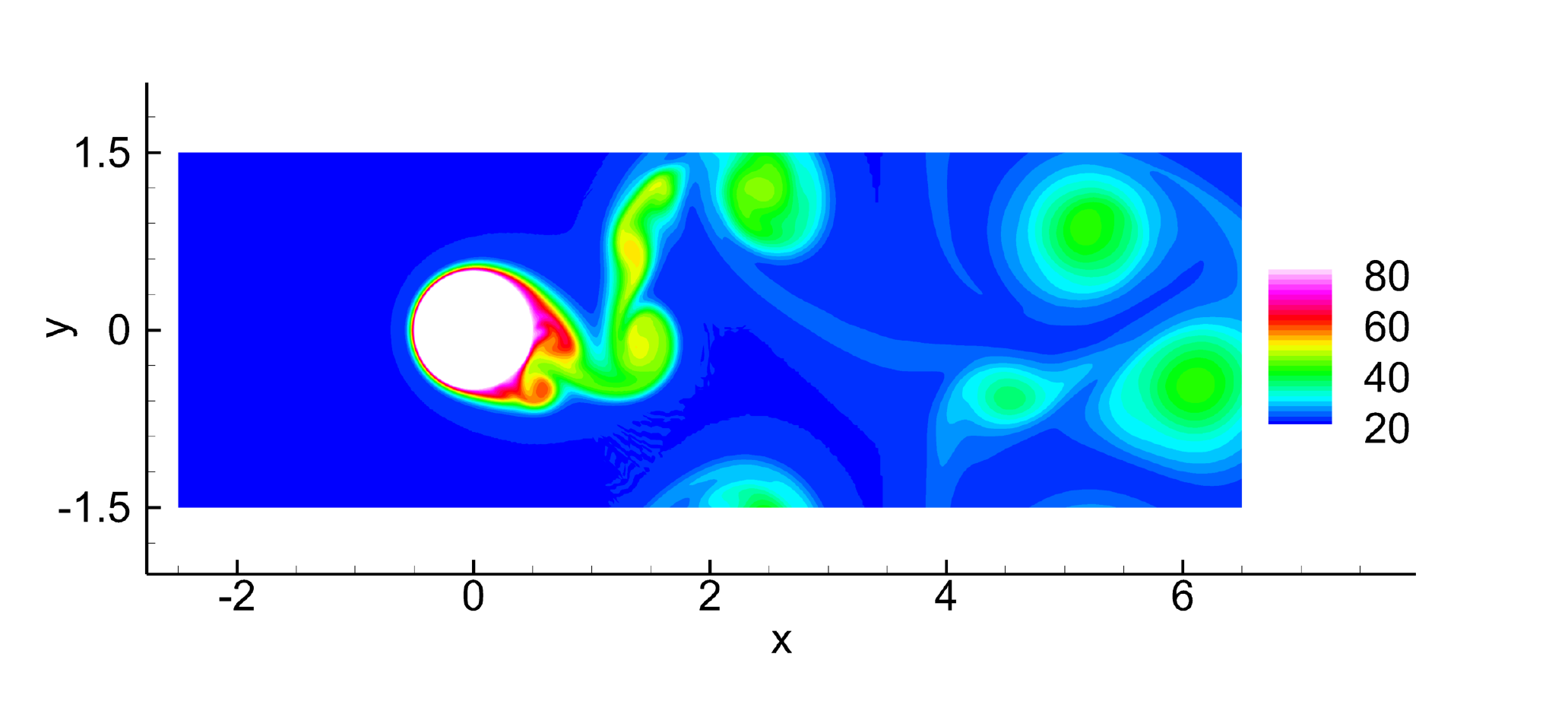}(d)
  }
  \caption{ (color online)
    Cylinder flow: Instantaneous velocity distributions (plots (a) and (c))
    and temperature distributions (plots (b) and (d)).
    Plots (a) and (b) are for $Re=300$ and $\alpha=0.01$,
    and plots (c) and (d) are for $Re=5000$ and $\alpha=0.005$.
    Velocity vectors are shown on a sparser set of
    mesh points for clarity. 
  }
  \label{fig:cyl_char}
\end{figure}


Figure \ref{fig:cyl_char} provides an overview of the
characteristics of the flow and temperature fields for this problem.
It shows the distributions of
the instantaneous velocity (left column) and temperature
(right column) at Reynolds numbers $Re=300$ (top row)
and $Re=5000$ (bottom row).
For $Re=300$ (Figures \ref{fig:cyl_char}(a)-(b)),
the non-dimensional thermal diffusivity is
$\alpha=0.01$,
and the simulations are performed using
an element order $6$ and a time step size $\Delta t=0.001$.
For $Re=5000$ (Figures \ref{fig:cyl_char}(c)-(d)),
the non-dimensional thermal diffusivity is $\alpha=0.005$,
and the results are computed using an element order $8$
and a time step size $\Delta t=2.5e-4$.
The flow is unsteady at these Reynolds numbers
and is characterized by regular or irregular vortex shedding
in the cylinder wake. At the lower $Re=300$,
the vortices are quite weak, and no backflow is observed at
the outflow boundary (Figure \ref{fig:cyl_char}(a)).
At the higher Reynolds number $Re=5000$, the vortices
persist in the entire wake region, and strong backflows are observed
at the outflow boundary while these vortices are passing
through. Because the vortices are generated at the cylinder
and shed into the wake, the vortex cores contain warmer fluids,
as is evident from
the temperature distributions in Figures~\ref{fig:cyl_char}(b) and (d).


\begin{figure}
  \centering
  \includegraphics[width=4in]{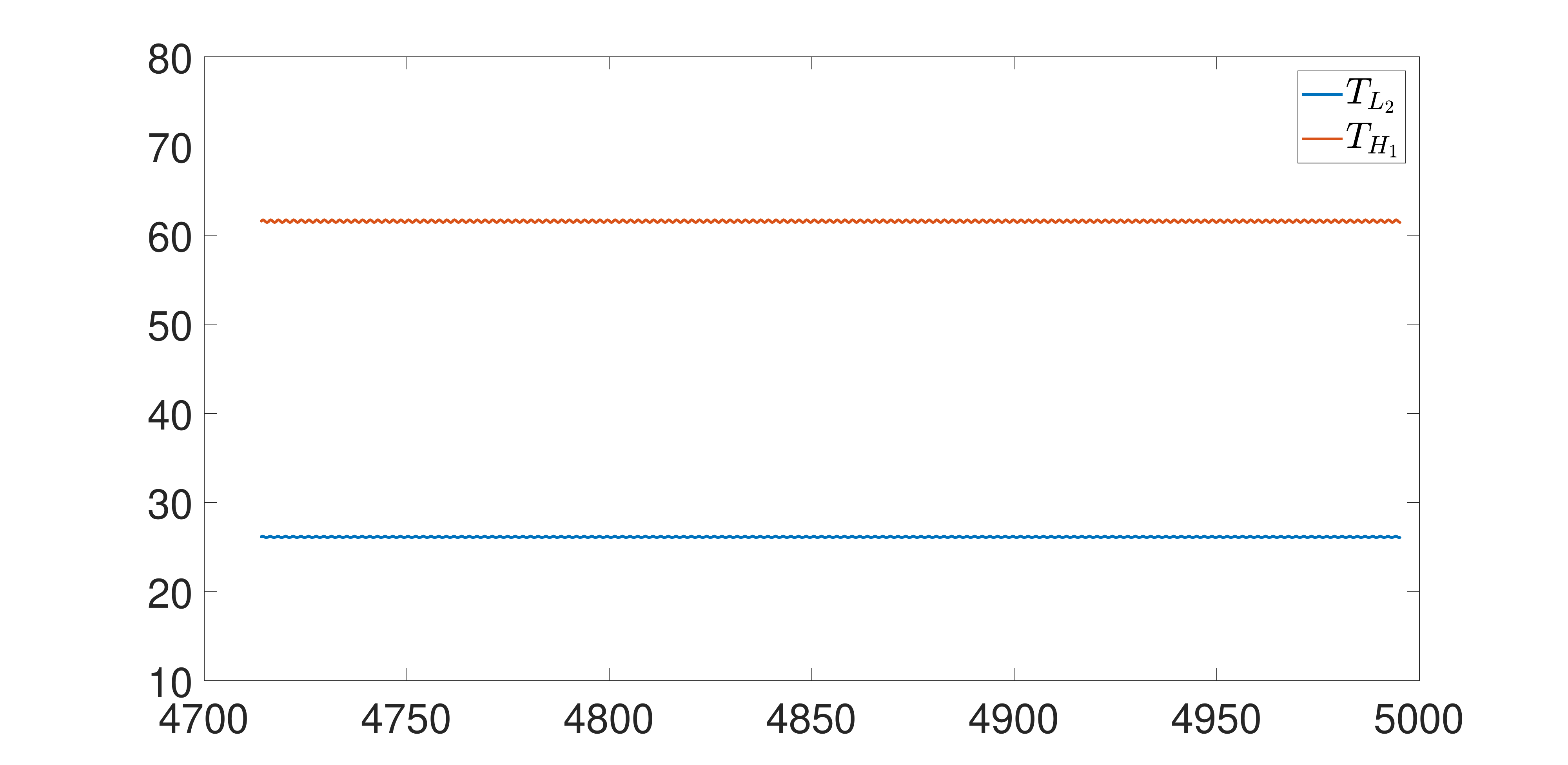}(a)
  \includegraphics[width=4in]{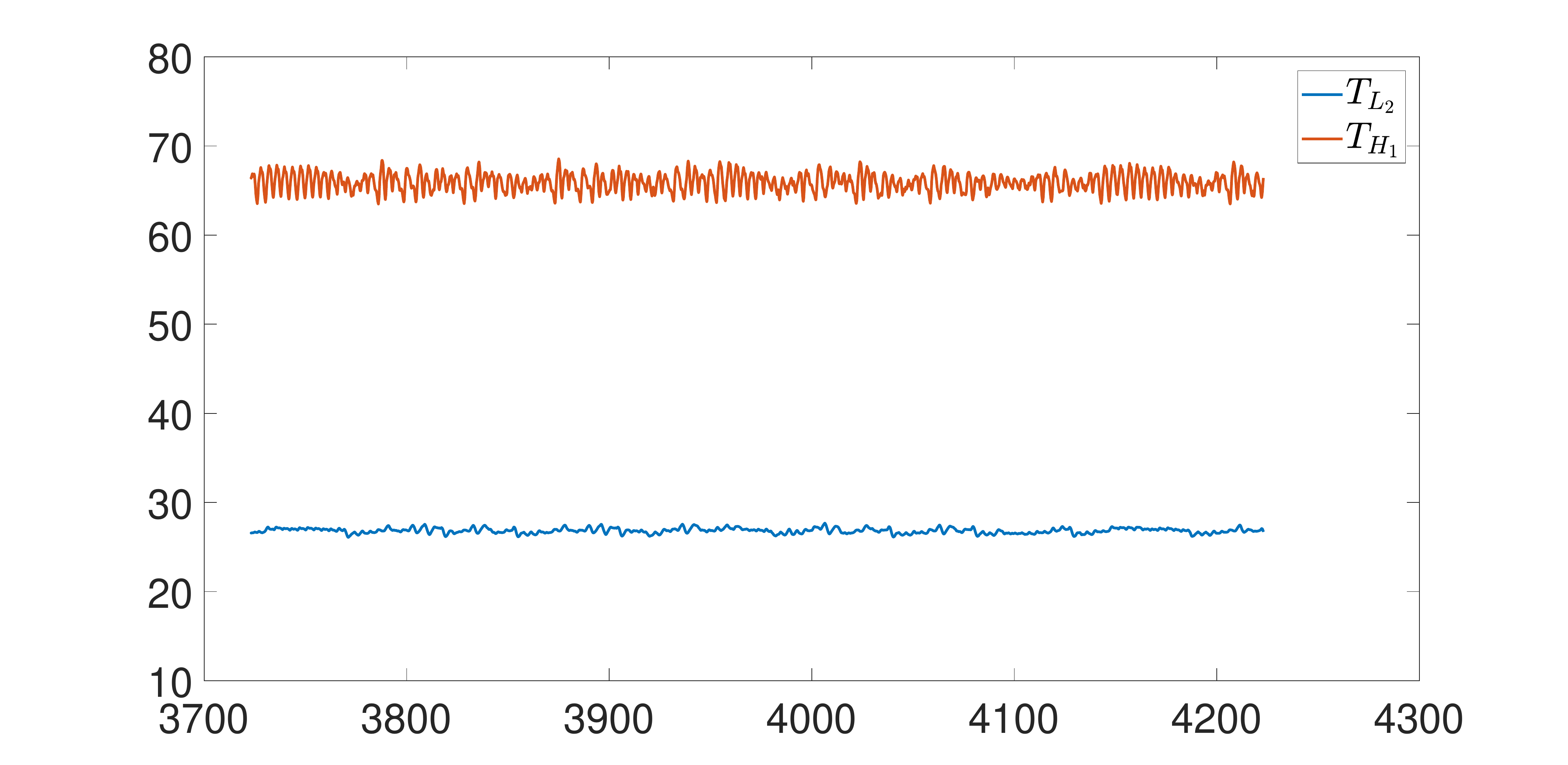}(b)
  \caption{
    Cylinder flow: Time histories of the $T_{L_{2}}(t)$
    and $T_{H_{1}}(t)$ at Reynolds numbers (a) $Re=300$ and (b) $Re=2000$.
    Thermal diffusivity is $\alpha=0.01$ for both cases.
  }
  \label{fig:cyl_hist}
\end{figure}

To characterize  the overall evolution of
the temperature field, we have computed and monitored
the following quantities:
\begin{equation}
  T_{L_{2}}(t) = \sqrt{
    \frac{1}{V_{\Omega}}\int_{\Omega} \left[ T(\mbs x,t)\right]^2 d\Omega
  },
  \qquad
  T_{H_{1}}(t) = \sqrt{
    \frac{1}{V_{\Omega}}\int_{\Omega} \left[
      \left(T(\mbs x,t) \right)^2
      + \left|\nabla T  \right|^2
      \right] d\Omega
  },
  \label{equ:T_norm}
\end{equation}
where
$V_{\Omega} = \int_{\Omega}d\Omega$ is the volume of the domain.
These are basically the $L^2$ and $H^1$
norms of the temperature field.
Figure \ref{fig:cyl_hist} shows  time histories of
$T_{L_{2}}(t)$ and $T_{H_{1}}(t)$ at two Reynolds numbers
$Re=300$ and $Re=2000$. The non-dimensional thermal diffusivity
is $\alpha=0.01$ for both cases.
The results for $Re=300$ (Figure \ref{fig:cyl_hist}(a)) are computed with
an element order $5$, and those for $Re=2000$ (Figure \ref{fig:cyl_hist}(b)) 
are obtained with an element order $7$.
At $Re=300$,
both temperature norms exhibit a regular fluctuation in time
with a small amplitude.
At the higher Reynolds number $Re=2000$, the fluctuations in
these temperature signals  are much
stronger and irregular, especially with $T_{H_1}(t)$.
The long histories in these plots signify the stability of
our simulations. The constant mean level and the invariant characteristics
of the fluctuations suggest that the temperature field
has reached a statistically stationary state.


\begin{table}[tb]
  \centering
  \begin{tabular}{llllll}
    \hline
    Reynolds number & Element order & $\overline{T}_{L_2}$ & $T'_{L_2}$
    & $\overline{T}_{H_1}$ & $T'_{H_1}$ \\ \hline
    300 & 3 & 26.124 & 5.27e-2 & 61.563 & 0.104 \\
    & 4 & 26.128 & 5.33e-2 & 61.563 & 0.103 \\
    & 5 & 26.130 & 5.37e-2 & 61.566 & 0.104 \\
    & 6 & 26.129 & 5.36e-2 & 61.564 & 0.104 \\ \hline
    2000 & 5 & 26.786 & 0.286 & 65.895 & 1.014 \\
    & 6 & 26.764 & 0.314 & 65.833 & 0.908 \\
    & 7 & 26.825 & 0.283 & 65.908 & 1.016 \\
    & 8 & 26.854 & 0.284 & 65.932 & 1.046 \\
    & 9 & 26.836 & 0.280 & 65.935 & 1.037 \\ 
    \hline
  \end{tabular}
  \caption{
    Cylinder flow: Time-averaged mean and root-mean-square (rms)
    of $T_{L_2}(t)$ and $T_{H_1}(t)$ computed using
    various element orders at two Reynolds numbers.
    Thermal diffusivity is $\alpha=0.01$.
  }
  \label{tab:cyl_order}
\end{table}


From the time histories of $T_{L_2}(t)$ and $T_{H_1}(t)$ we
can compute the time-averaged mean and root-mean-square (rms)
of these temperatures, which can be compared quantitatively
to assess the effect of the simulation parameters.
Table~\ref{tab:cyl_order} lists the mean ($\overline{T}_{L_2}$
and $\overline{T}_{H_1}$) and rms ($T'_{L_2}$ and $T'_{H_1}$) values
of $T_{L_2}(t)$ and $T_{H_1}(t)$ corresponding to
a range of element orders, for Reynolds numbers
$Re=300$ and $Re=2000$ with a thermal diffusivity $\alpha=0.01$.
We have employed a time step size
$\Delta t=0.001$ for $Re=300$ and $\Delta t=5e-4$ for
$Re=2000$ in this set of simulations.
We observe that, for both Reynolds numbers,
the change in the mean and rms temperatures is not significant
with increasing element order, indicating a convergence
of simulation results with respect to the mesh resolution.
In the results reported below, the majority of simulations
are performed with an element order $5$
for $Re=300$ and with an element order $7$ for higher Reynolds numbers.

\begin{figure}
  \centerline{
    \includegraphics[width=3.1in]{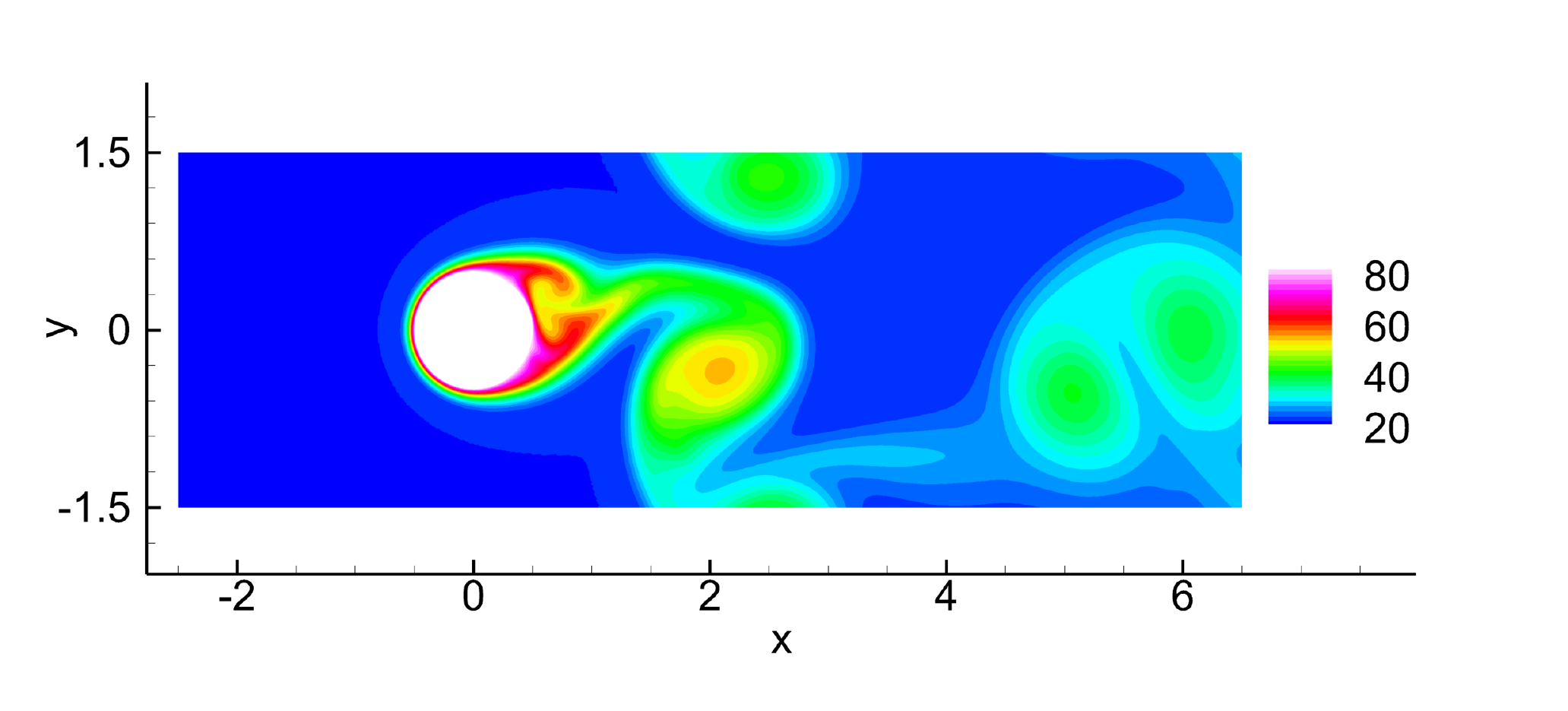}(a)
    \includegraphics[width=3.1in]{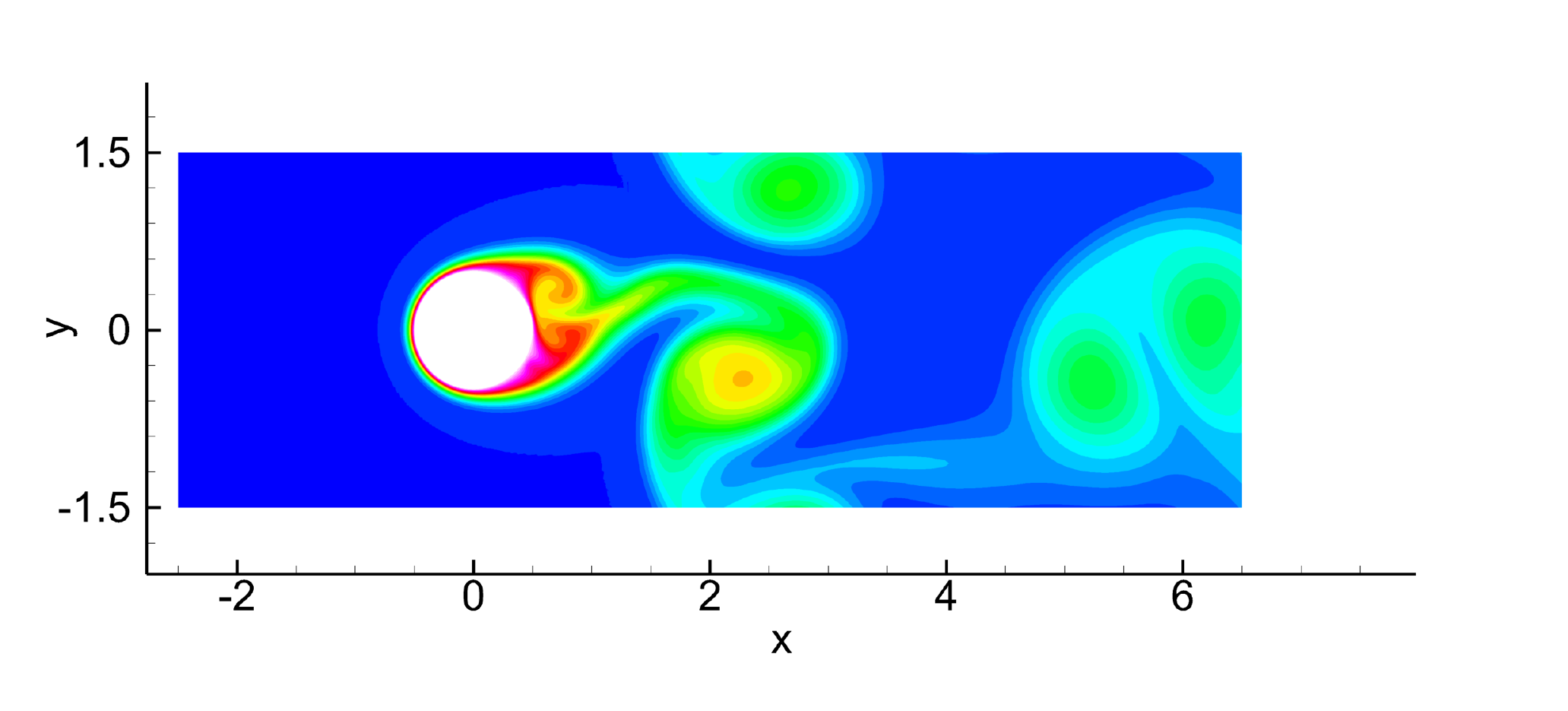}(b)
  }
  \centerline{
    \includegraphics[width=3.1in]{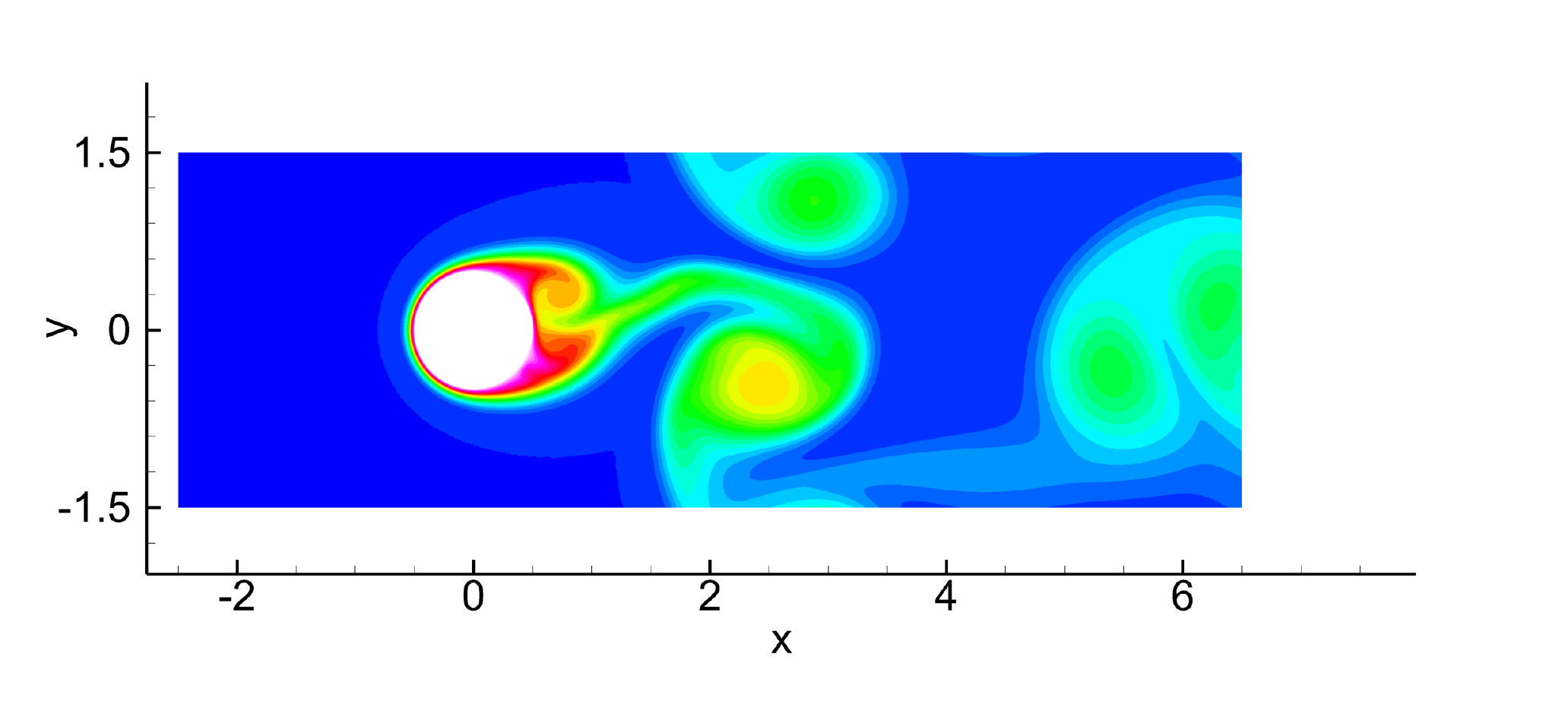}(c)
    \includegraphics[width=3.1in]{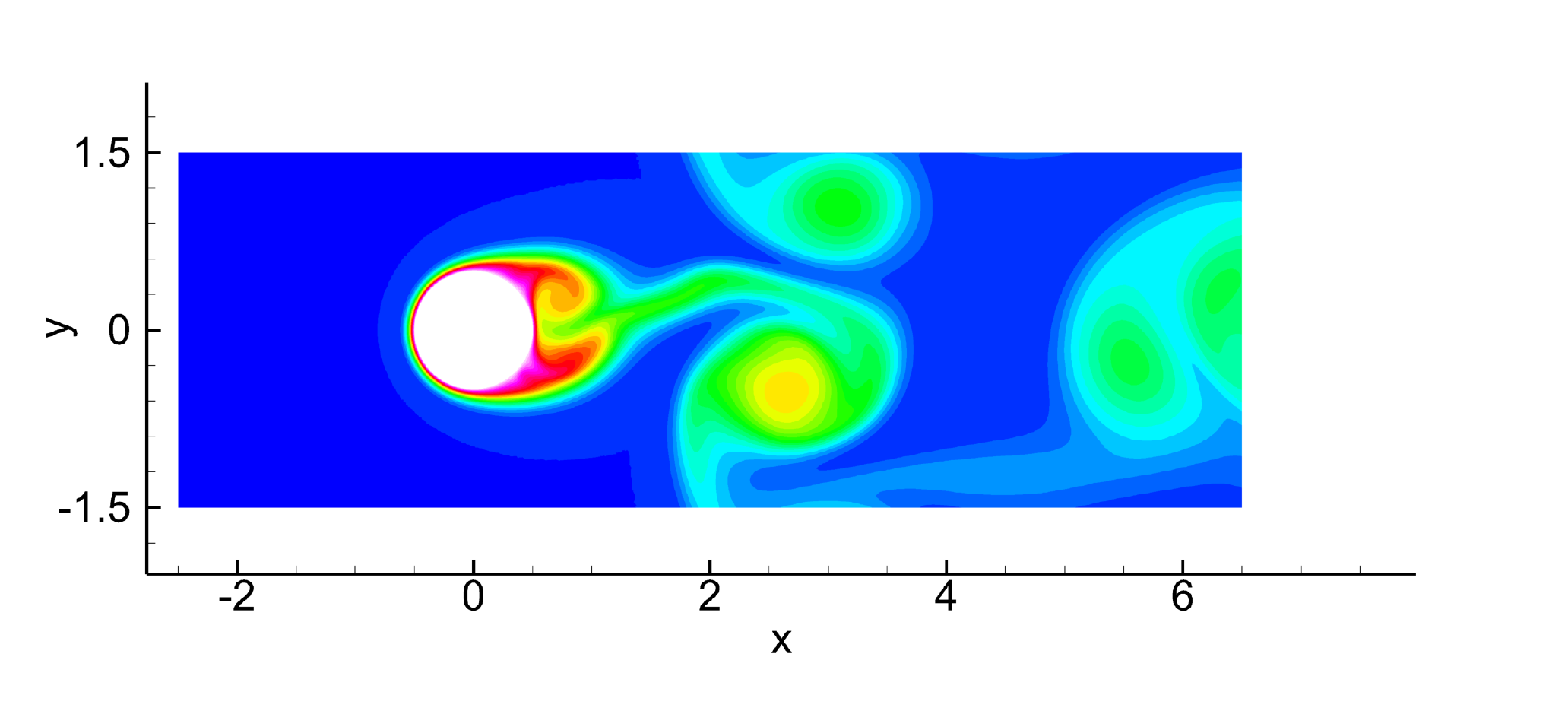}(d)
  }
  \centerline{
    \includegraphics[width=3.1in]{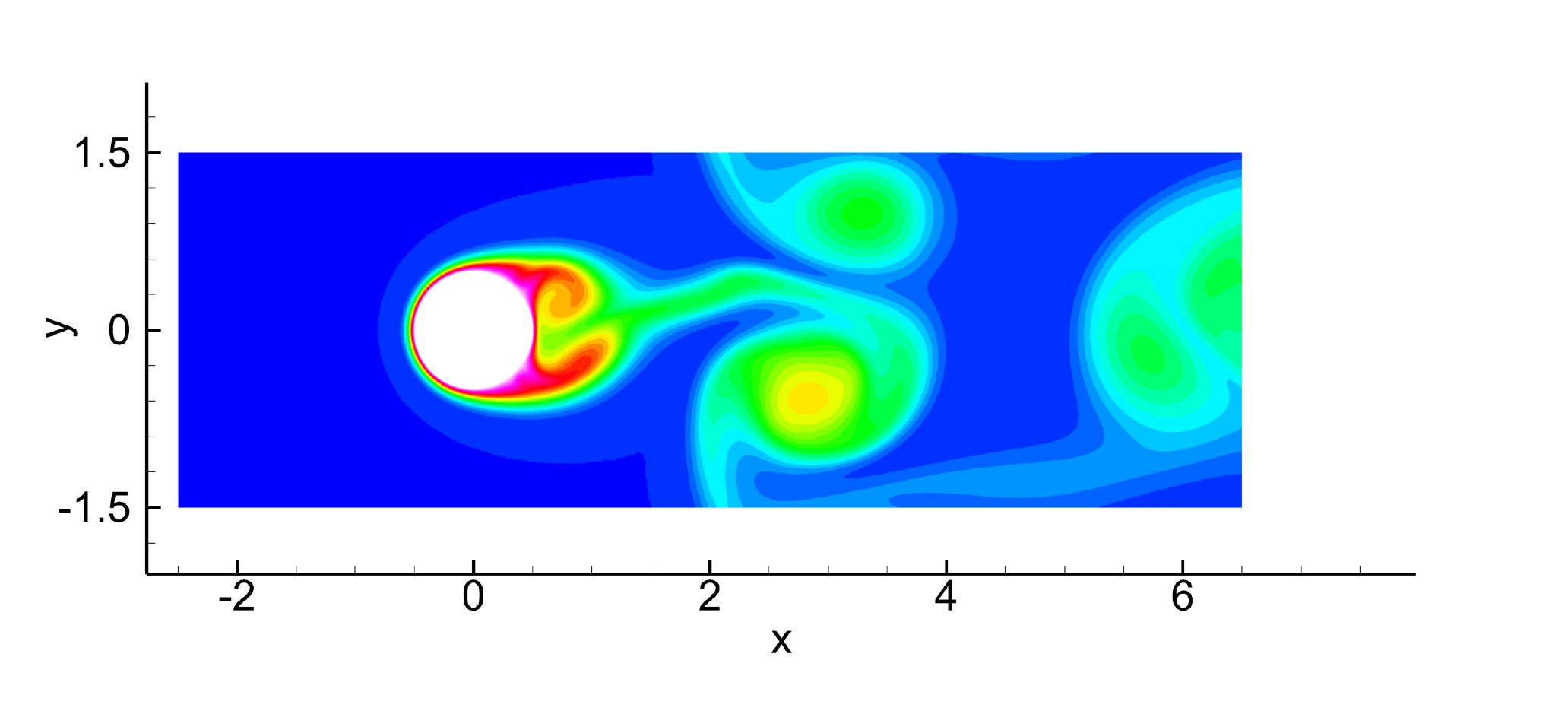}(e)
    \includegraphics[width=3.1in]{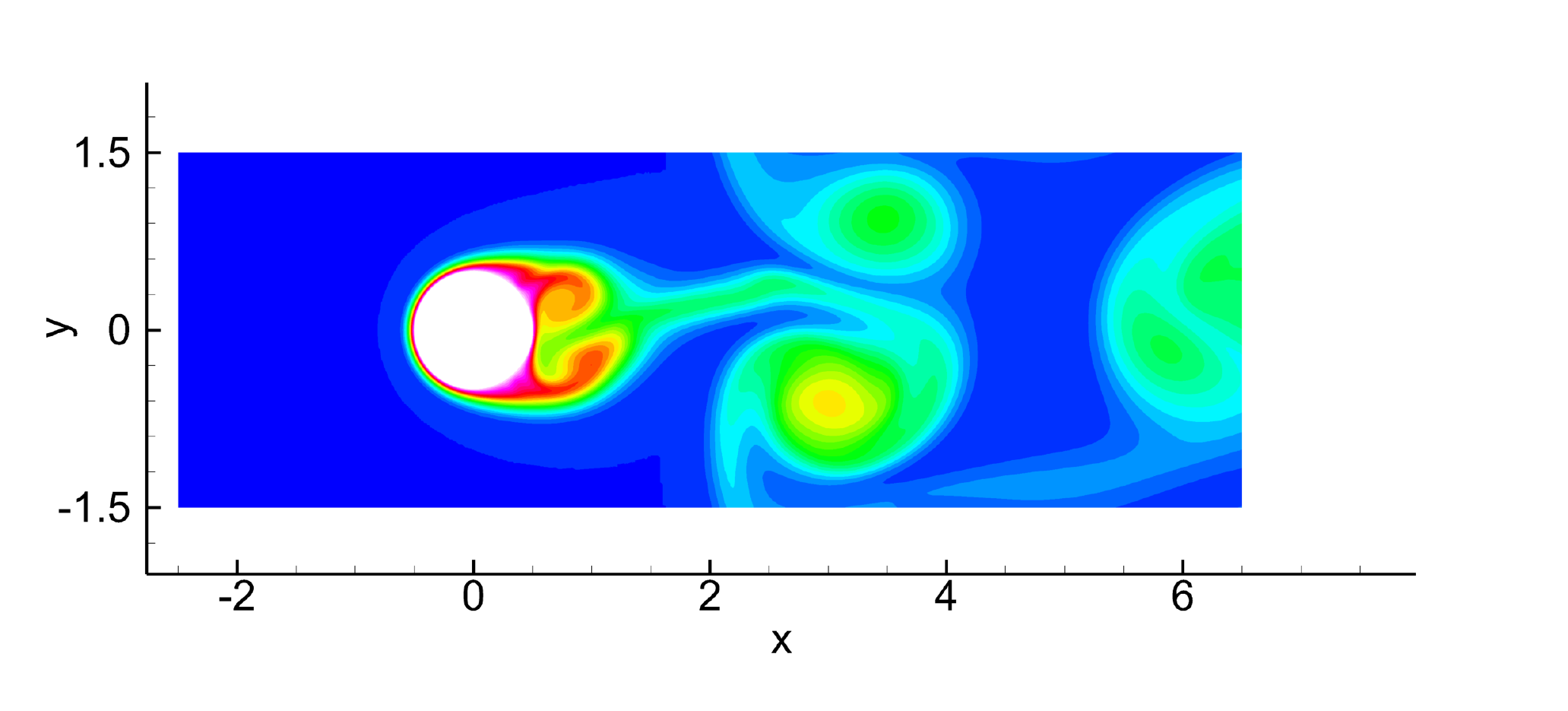}(f)
  }
  \centerline{
    \includegraphics[width=3.1in]{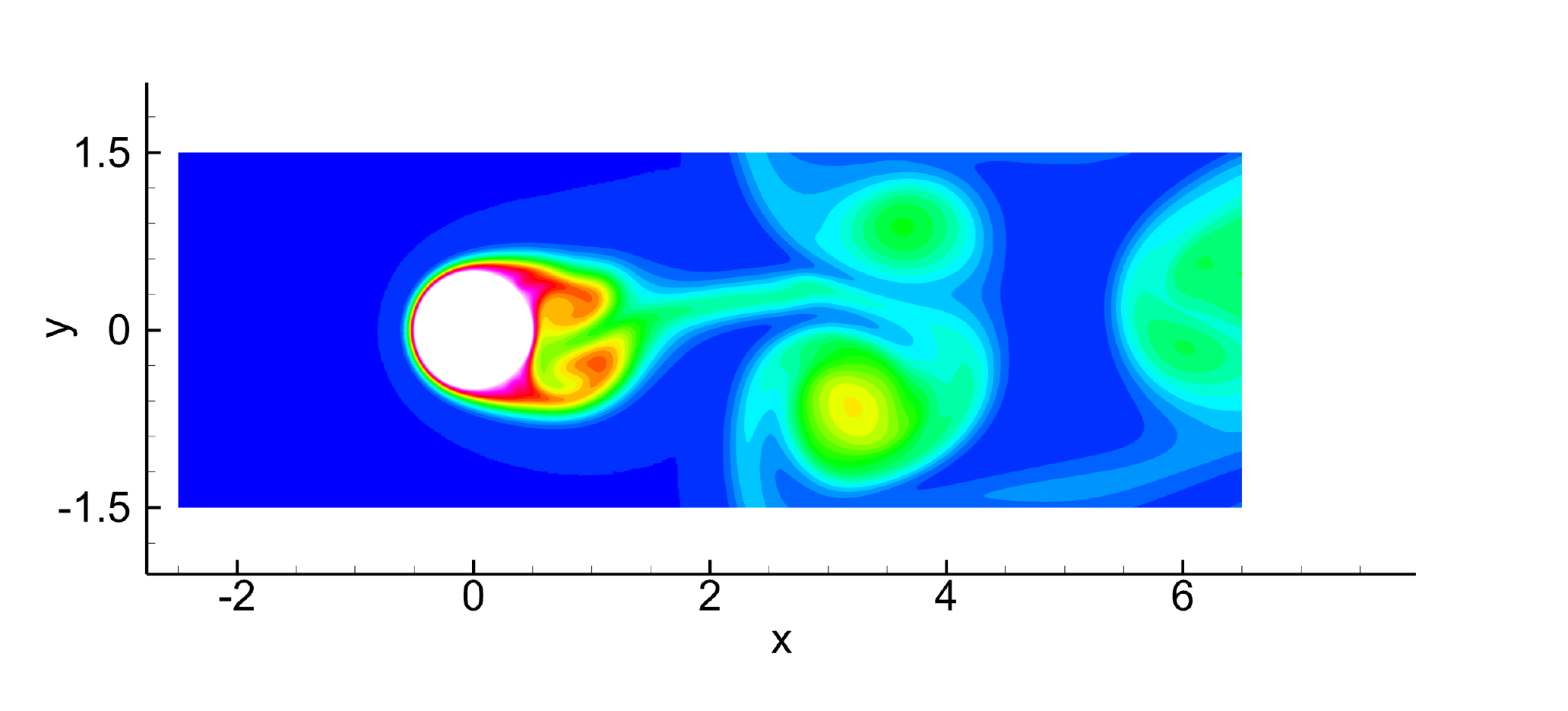}(g)
    \includegraphics[width=3.1in]{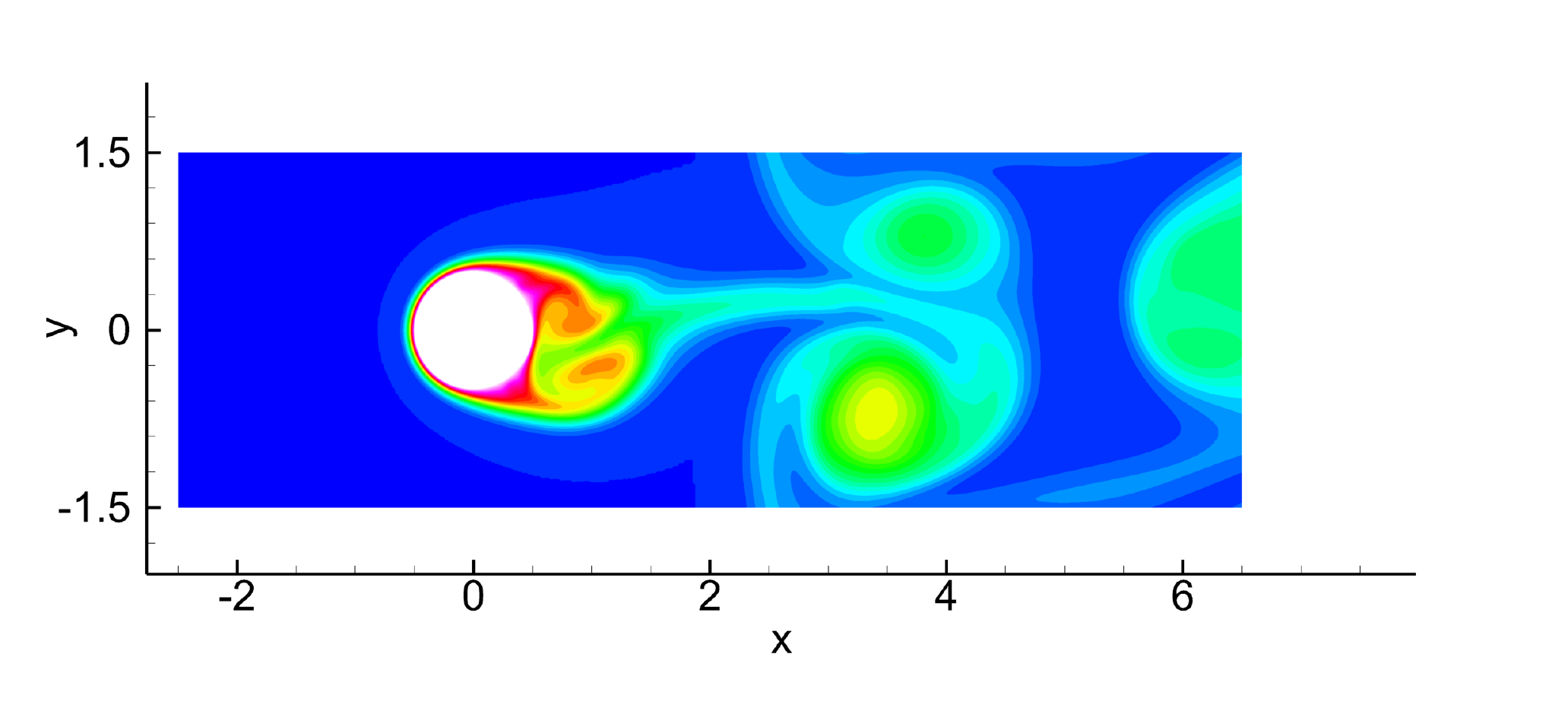}(h)
  }
  \centerline{
    \includegraphics[width=3.1in]{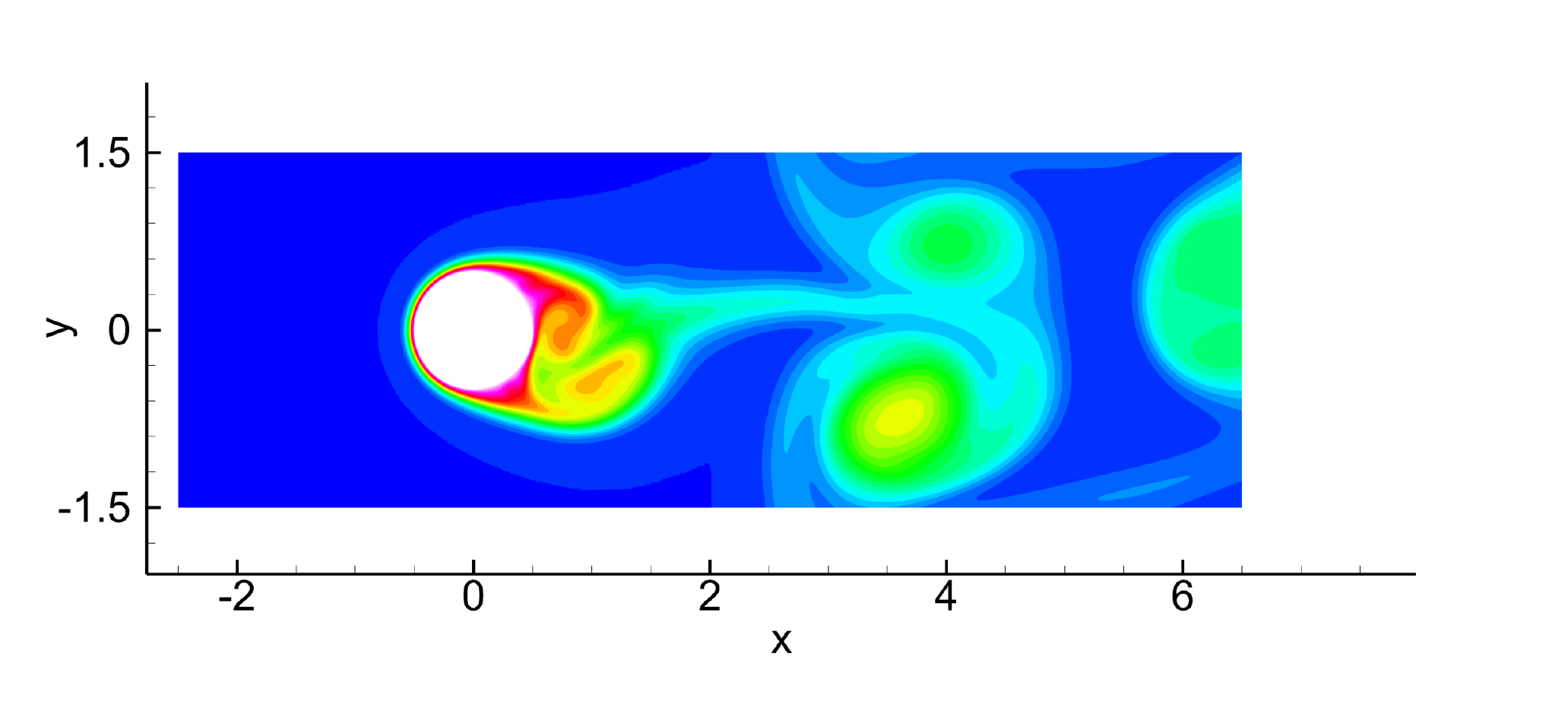}(i)
    \includegraphics[width=3.1in]{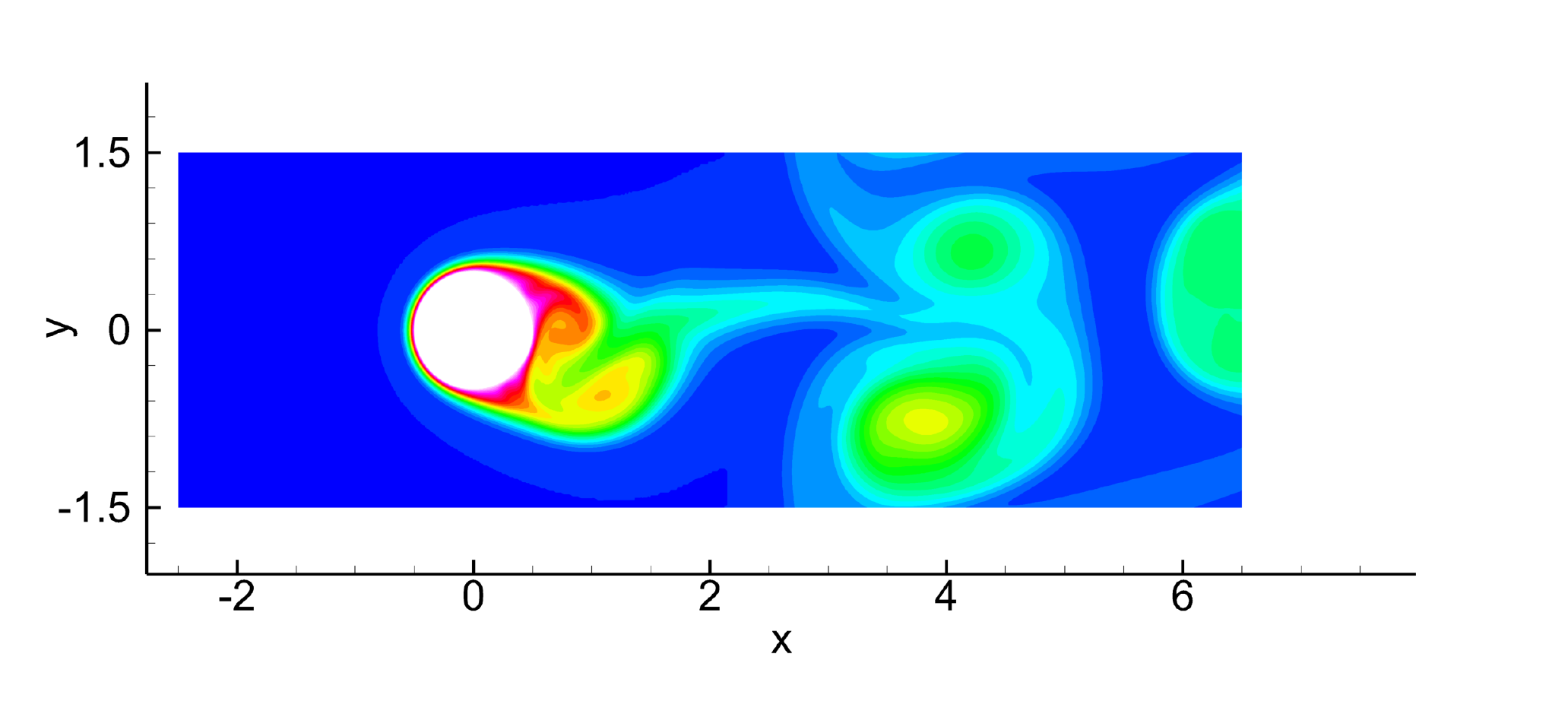}(j)
  }
  \caption{(color online)
    Cylinder flow (Re=2000): temporal sequence of snapshots of
    the temperature distribution computed using the current thermal OBC.
    (a) $t=t_0$, (b) $t=t_0+0.2$, (c) $t=t_0+0.4$, (d) $t=t_0+0.6$,
    (e) $t=t_0+0.8$, (f) $t=t_0+1.0$, (g) $t=t_0+1.2$, (h) $t=t_0+1.4$,
    (i) $t=t_0+1.6$, (j) $t=t_0+1.8$.
    Thermal diffusivity is $\alpha=0.01$.
  }
  \label{fig:cyl_Tobc}
\end{figure}

\begin{figure}
  \centerline{
    \includegraphics[width=3.1in]{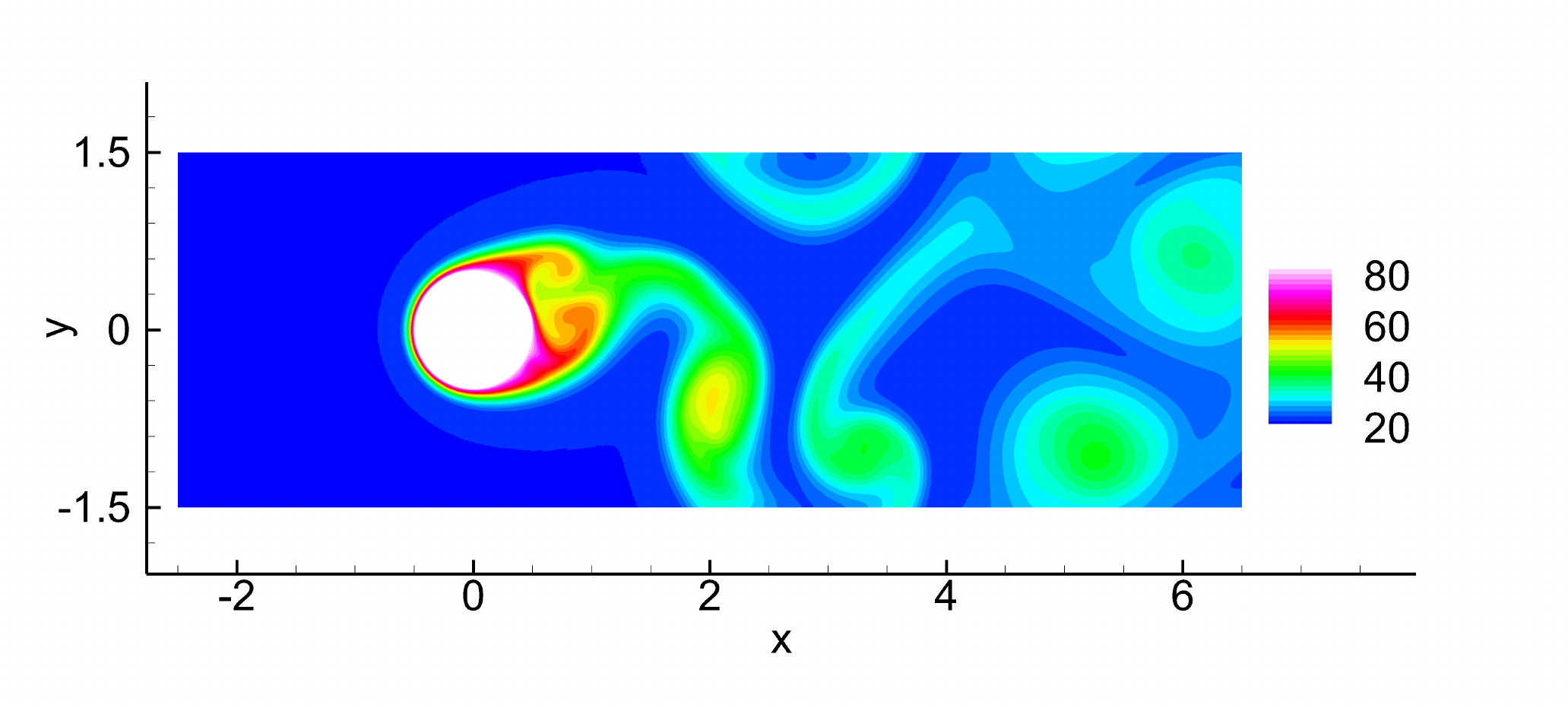}(a)
    \includegraphics[width=3.1in]{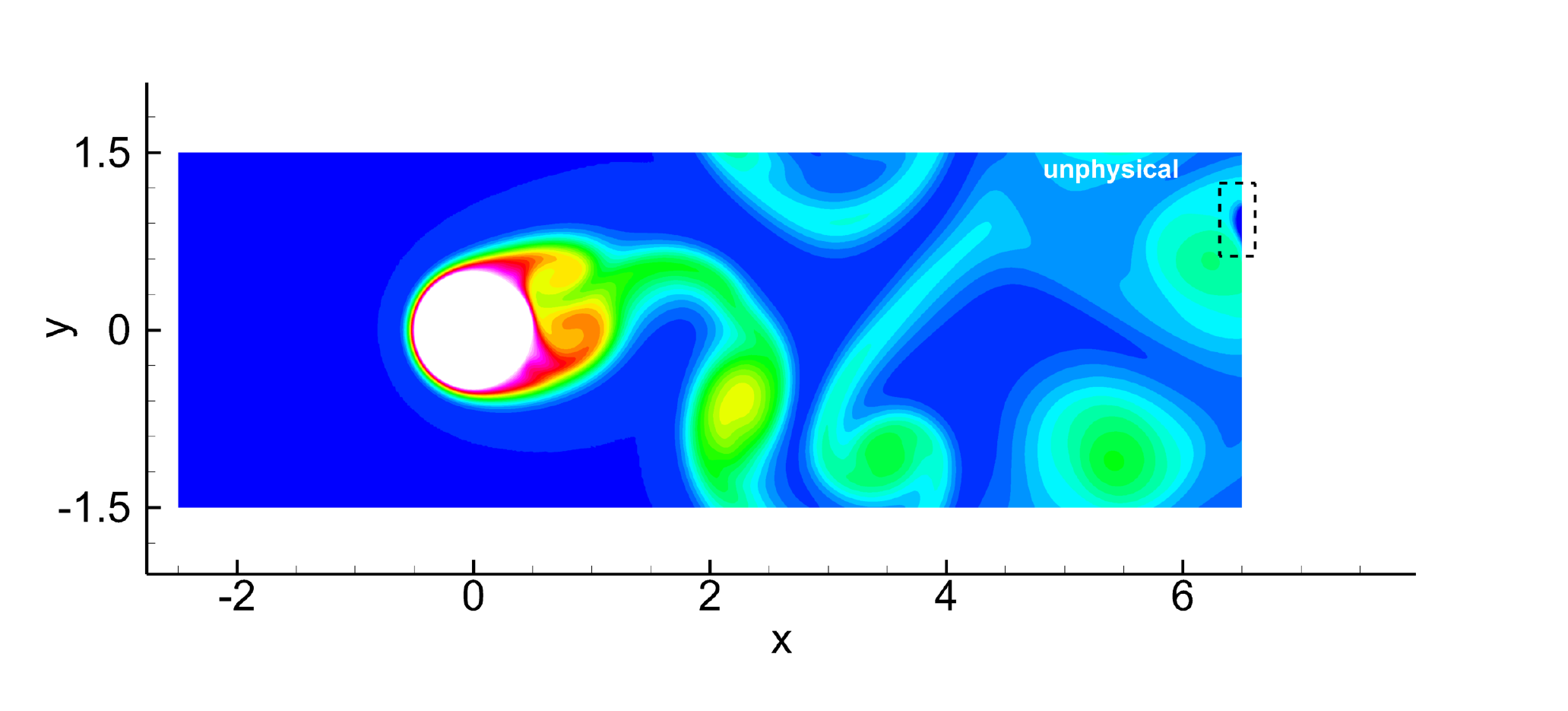}(b)
  }
  \centerline{
    \includegraphics[width=3.1in]{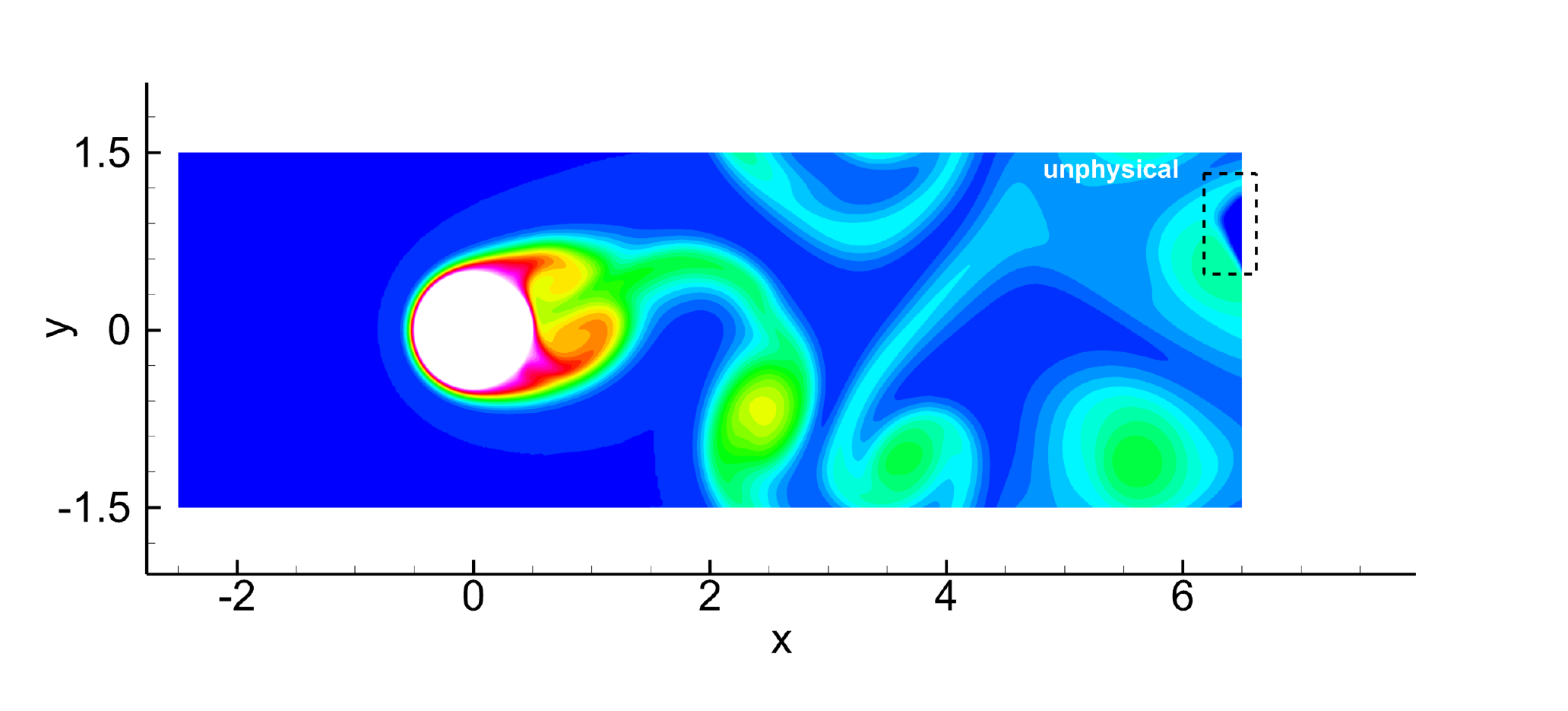}(c)
    \includegraphics[width=3.1in]{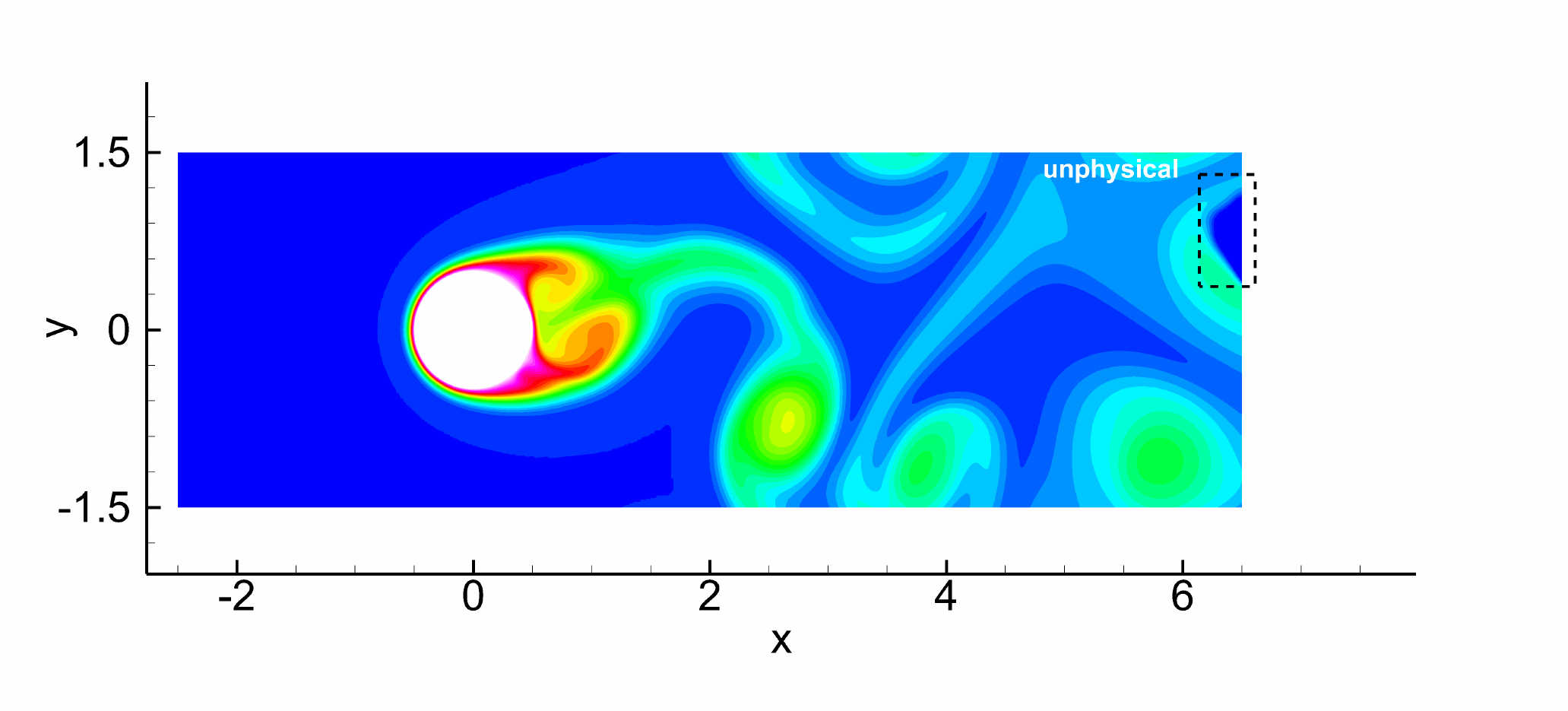}(d)
  }
  \centerline{
    \includegraphics[width=3.1in]{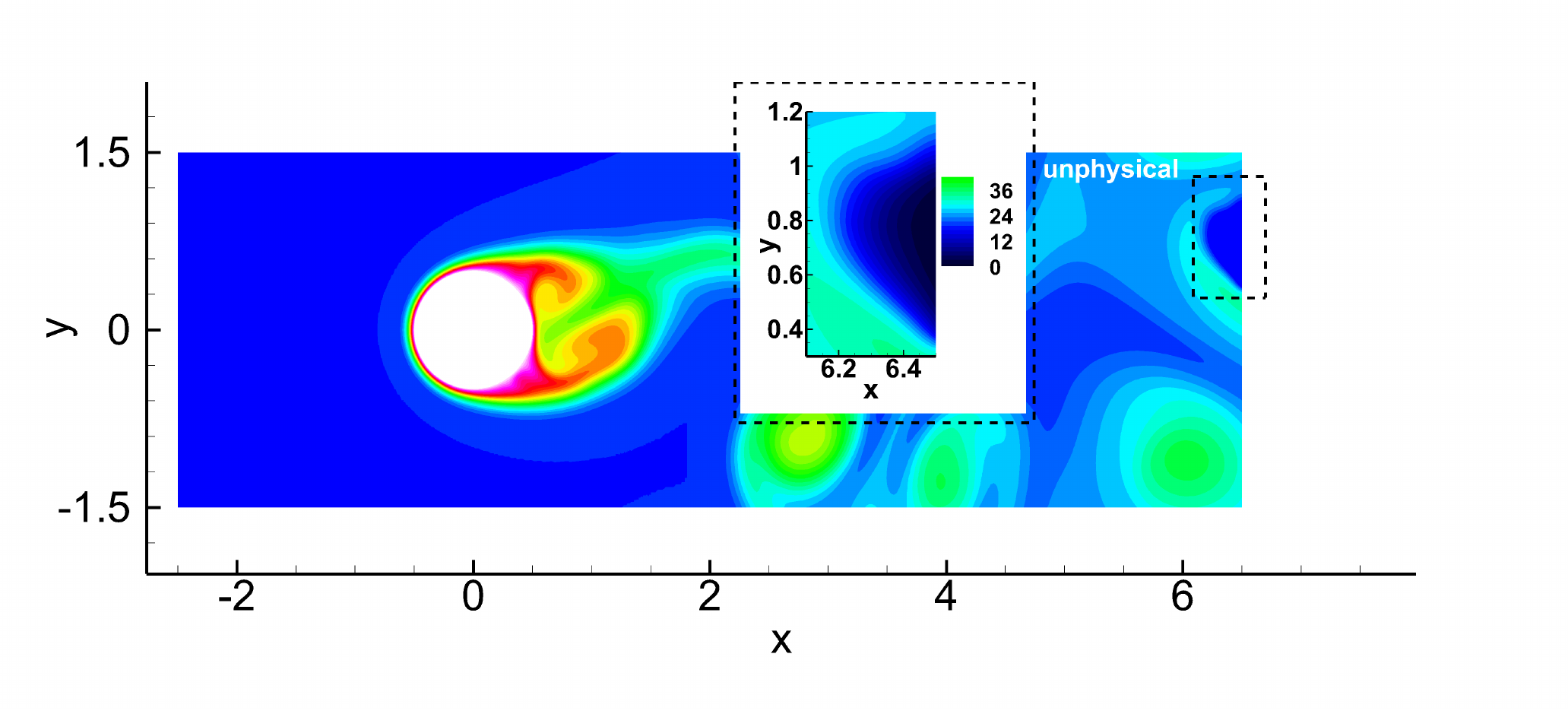}(e)
    \includegraphics[width=3.1in]{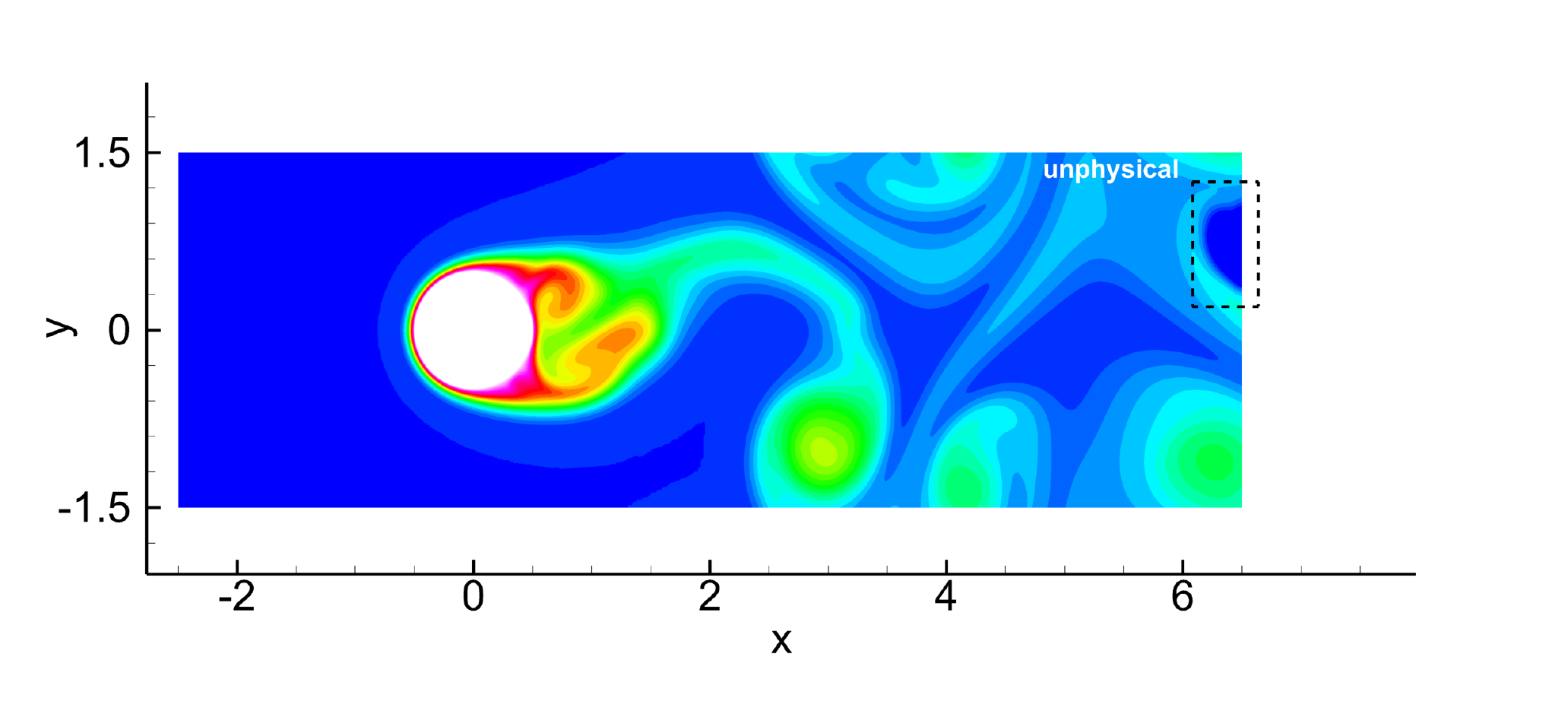}(f)
  }
  \centerline{
    \includegraphics[width=3.1in]{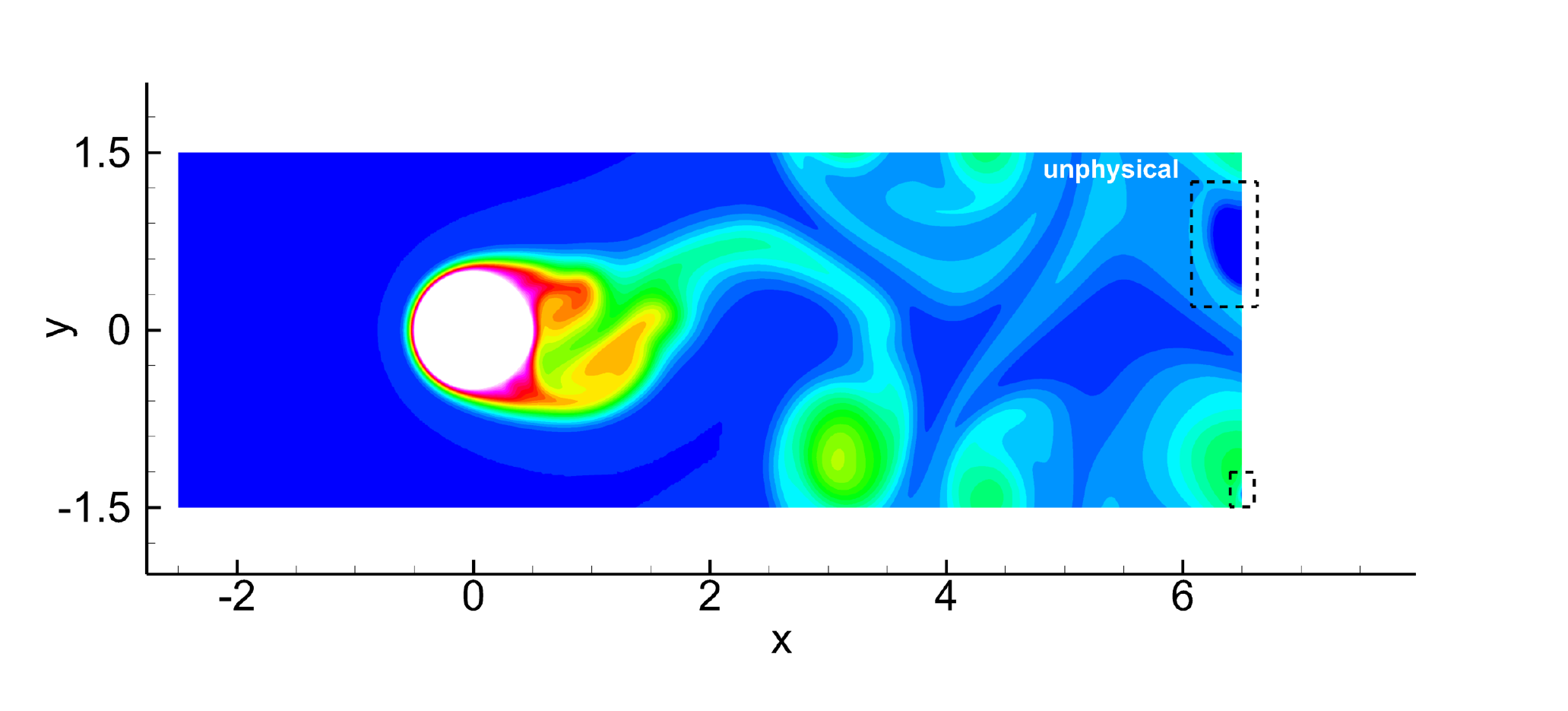}(g)
    \includegraphics[width=3.1in]{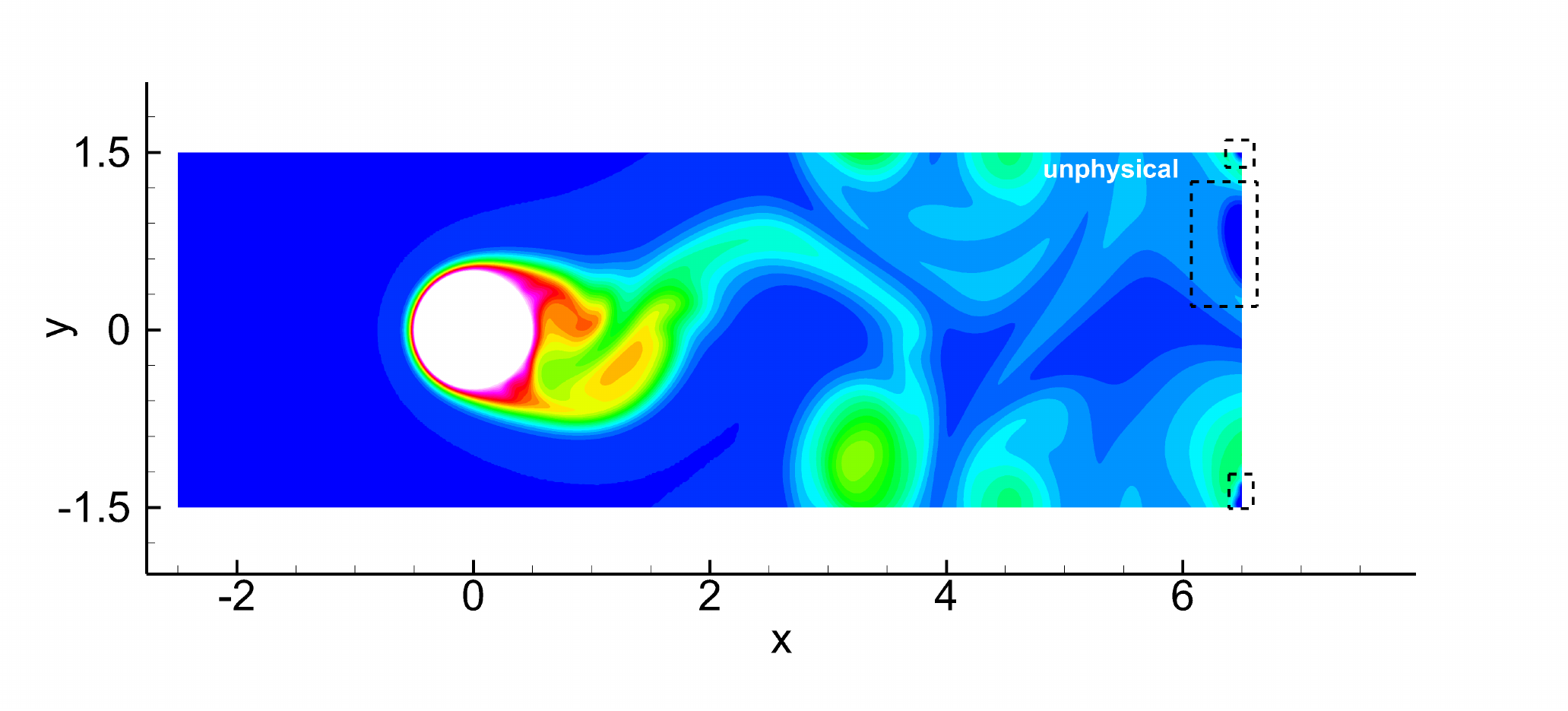}(h)
  }
  \centerline{
    \includegraphics[width=3.1in]{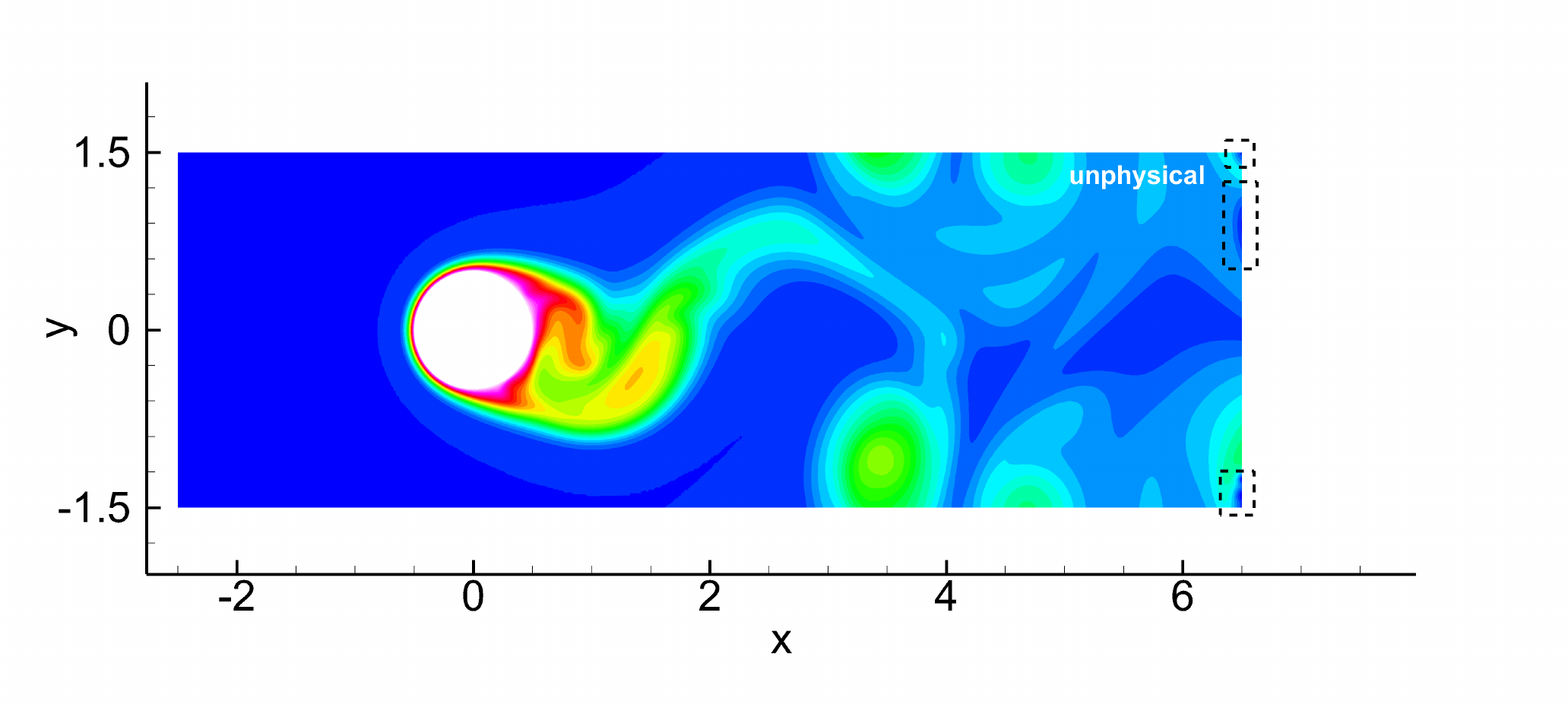}(i)
    \includegraphics[width=3.1in]{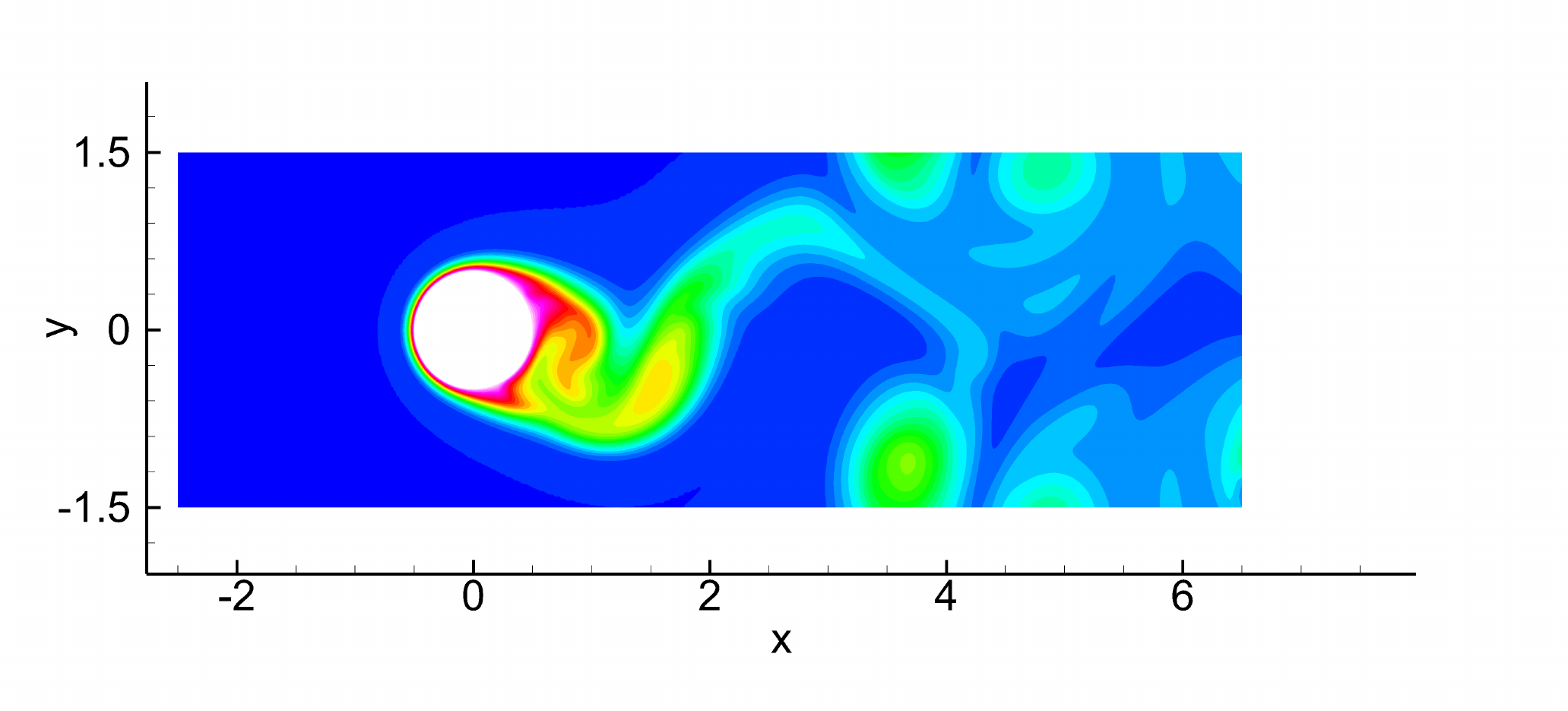}(j)
  }
  \caption{ (color online)
    Cylinder flow (Re=2000): temporal sequence of snapshots of
    the temperature distribution computed using the Neumann-type
    zero-flux OBC (equation~\eqref{equ:zero_obc}).
    (a) $t=t_1$, (b) $t=t_1+0.2$, (c) $t=t_1+0.4$, (d) $t=t_1+0.6$,
    (e) $t=t_1+0.8$, (f) $t=t_1+1.0$, (g) $t=t_1+1.2$, (h) $t=t_1+1.4$,
    (i) $t=t_1+1.6$, (j) $t=t_1+1.8$.
    The inset of plot (e) shows a magnified view near the outflow boundary.
    Dashed boxes mark the regions with unphysical temperature distribution
    (below $20$ degrees Celsius).
    Thermal diffusivity is $\alpha=0.01$.
  }
  \label{fig:cyl_Tfree}
\end{figure}


Let us next look into the effect of the open boundary condition
on the simulated flow and temperature fields.
We observe that for this problem
the boundary condition imposed on the outflow boundary,
especially at fairly high 
Reynolds numbers where strong vortices or backflows
occur on such boundary, is critical to the stability
of computations and to the physical soundness
of the simulated temperature field. 
When the Reynolds number is low (e.g.~$Re=300$),
no vortex or backflow occurs at
the outflow boundary, and usual
open boundary conditions (e.g.~the Neumann type zero-flux
condition~\eqref{equ:zero_obc}) can work
well and  produce reasonable results. 
When the Reynolds number becomes fairly high, e.g.~with~$Re\approx 2000$ or
larger for the circular cylinder flow,
strong vortices or backflows can occur on the open boundary.
In such cases the adoption of the energy-stable  open boundary condition
developed herein for the temperature field and those
from e.g.~\cite{Dong2015clesobc} (see also Appendix A)
for the Navier-Stokes equations
are critical to the successful simulations of this problem.

These points are demonstrated by two temporal sequences of the
velocity and temperature distributions shown in
Figures~\ref{fig:cyl_Tobc} to \ref{fig:cyl_Tfree}.
These results correspond to the
Reynolds number $Re=2000$
and a non-dimensional thermal diffusivity $\alpha=0.01$, and
they are obtained respectively using the current thermal open
boundary condition~\eqref{equ:obc_A} and
the Neumann-type zero-flux condition~\eqref{equ:zero_obc}
for the outflow boundary.
Note that in both cases we have employed
the open boundary condition~\eqref{equ:obc_v} with $\mbs E(\mbs n,\mbs u)$
given in~\eqref{equ:def_E}
for the Navier-Stokes equations.
Figure \ref{fig:cyl_Tobc} shows the temporal
sequence of snapshots of the temperature fields
computed
using the current thermal open boundary condition~\eqref{equ:obc_A}.
Figure \ref{fig:cyl_Tfree} shows the temporal sequence
of snapshots of the temperature fields
 computed using the Neumann-type zero-flux condition~\eqref{equ:zero_obc}
for the temperature on the outflow boundary.
In these simulations we have employed an element order 8
and a time step size $\Delta t=5e-4$.
At this Reynolds number, strong vortices and backflows
can be clearly observed on the outflow boundary.
The use of the energy-stable OBC
from~\cite{Dong2015clesobc} (see Appendix A) for the Navier-Stokes
equations employed here is crucial
for the stable flow simulation.
What is striking concerns the temperature distribution
near the outflow boundary obtained by these two methods.
A comparison between Figures \ref{fig:cyl_Tobc}
and \ref{fig:cyl_Tfree} shows that the current thermal OBC~\eqref{equ:obc_A} and the
Neumann-type OBC~\eqref{equ:zero_obc} lead to quite different results
near the outflow boundary when the vortices
are passing through, and that 
in such cases the Neumann-type thermal OBC
produces apparently unphysical temperature distributions.
With the Neumann-type OBC~\eqref{equ:zero_obc},
we observe that when a vortex crosses the outflow boundary
the computed temperature  at the vortex core
can attain unphysical values.
In Figure \ref{fig:cyl_Tfree}
those regions 
with unphysical temperature distributions are
marked by the dashed boxes.
The computed temperature in those regions
can become nearly zero degrees Celsius
(see the dark blue region in the inset of
Figure \ref{fig:cyl_Tfree}(e)), which is
clearly unphysical given the inflow and
the cylinder temperature ($20$ and $80$ degrees Celsius, respectively).
We observe that
the unphysical temperature arises only when the vortices are passing through
the outflow boundary, where backflows occur.
After the vortices
exit the domain, the temperature field near the outflow
boundary computed by the Neumann-type OBC~\eqref{equ:zero_obc}
is restored to a reasonable distribution.
In contrast,
with the current thermal OBC~\eqref{equ:obc_A}, 
we observe that the computed temperature field at/near
the outflow boundary exhibits a reasonable distribution throughout the time,
particularly when strong vortices pass through the outflow
boundary and when backflows occur there.
This is evident from Figure \ref{fig:cyl_Tobc}.

The above observations are not limited to the Reynolds number $Re=2000$,
or thermal diffusivity $\alpha=0.01$ (we have also tested $\alpha=0.005$).
At higher Reynolds numbers (we have tested $Re=5000$,
$\alpha=0.01$ and $0.005$),
the same characteristics in the computed temperature distributions 
have been observed with regard to the current thermal OBC
and the Neumann-type OBC.
These results demonstrate a clear advantage of the current thermal
OBC for dealing with  thermal open boundaries
at high (and moderate) Reynolds numbers, when strong vortices or backflows
might occur there.
At low Reynolds numbers (we have tested~$Re=300$), when no vortices or backflows
occur at the outflow boundary, we observe that
both the current thermal OBC and
the Neumann-type zero-flux OBC produce reasonable temperature
distributions for this problem.


\subsection{Warm Jet Impinging on a Cool Wall}
\label{sec:jet}


\begin{figure}[tb]
  \centerline{
    \includegraphics[width=4in]{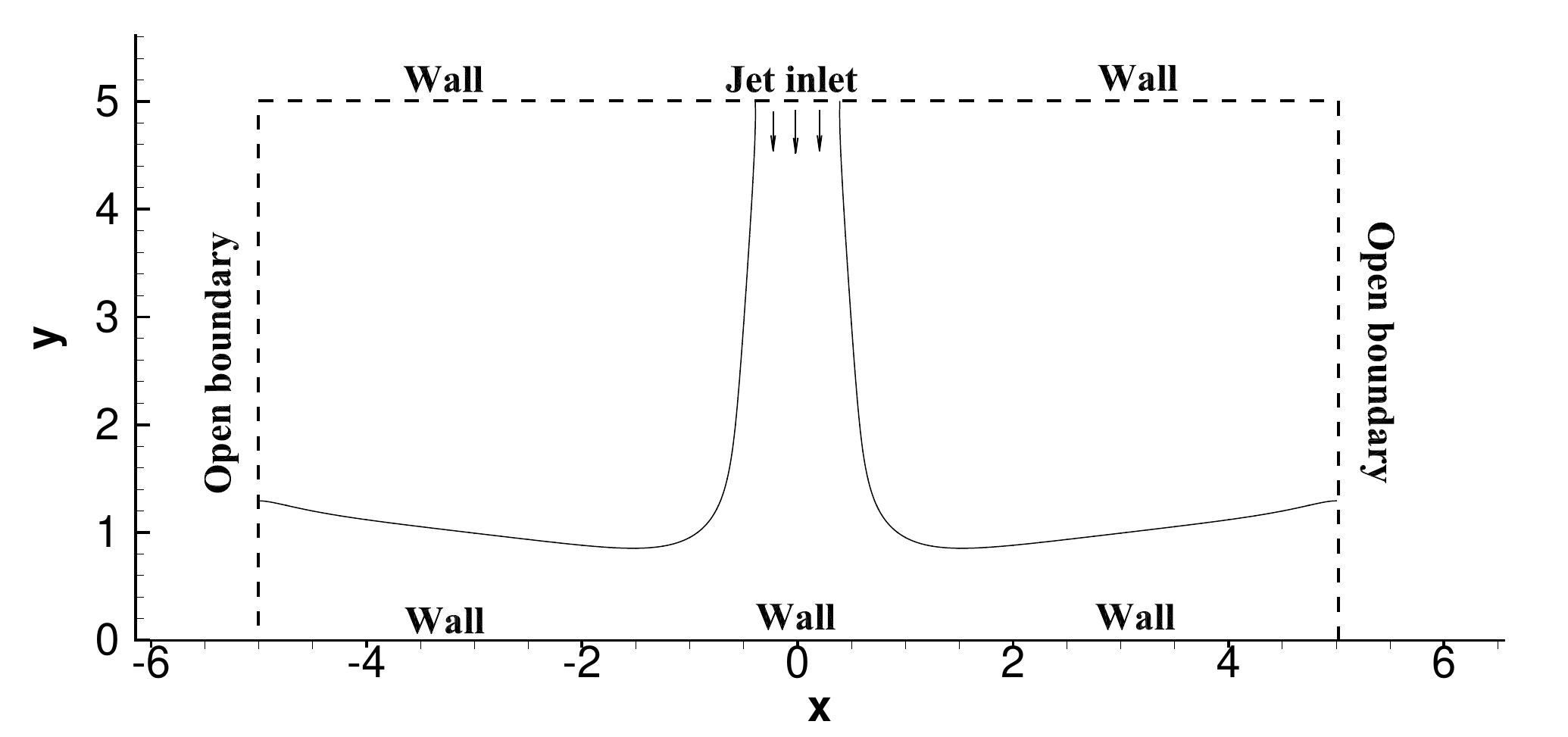}
  }
  \caption{
    Impinging jet: Flow configuration and boundary conditions.
  }
  \label{fig:jet_conf}
\end{figure}

In this subsection we test the presented method by simulating
the heat transfer in another canonical flow, a jet impinging
on a solid wall. At moderate and fairly high Reynolds numbers,
the physical instability of the jet and the vortices formed
along the jet profile, together with the open domain boundary,
pose a significant challenge to the flow and temperature
simulations.


We consider the domain depicted in Figure \ref{fig:jet_conf},
$-5d\leqslant x\leqslant 5d$ and $0\leqslant y\leqslant 5d$,
where $d=1$ is the jet inlet diameter.
The top and bottom of the domain are solid walls,
which are maintained at a constant temperature $T_w=20$ degrees
Celsius.
The left and right sides of the domain are open,
where the fluid can freely leave or enter the domain.
The initial fluid temperature in the domain is assumed to be $T_{in}=20$
degrees Celsius, and the fluid is initially assumed to be at rest.
In the middle of the top wall there is an orifice with a diameter $d$,
through which a jet of fluid is issued into the domain.
The jet velocity and temperature at the
inlet is assumed to have the following
distribution:
\begin{equation}
  \left\{
  \begin{split}
    &
    u = 0, \\
    &
    v = -U_0\left[
      \left(\mathcal{H}(x,0) - \mathcal{H}(x,R_0)  \right)\tanh\frac{1-x/R_0}{\sqrt{2}\epsilon}
      + \left(\mathcal{H}(x,-R_0) - \mathcal{H}(x,0) \right)\tanh\frac{1+x/R_0}{\sqrt{2}\epsilon}
      \right]\\
    &
    T = T_w + (T_h-T_w)\left(1-\frac{x^2}{R_0^2} \right),
  \end{split}
  \right.
  \label{equ:jet_VTbc}
\end{equation}
where $R_0=d/2$ is the jet radius, $U_0=1$ is the velocity scale, $\epsilon=\frac{1}{40}$,
and $T_h=80$ degrees Celsius is the centerline
temperature.
$\mathcal{H}(x,a)$ is the Heaviside step function, taking a unit value if $x\geqslant a$
and vanishing otherwise.
With the above expressions, the inlet velocity has a ``top-hat'' profile,
essentially $U_0$ except in a thin layer (thickness
controlled by the parameter $\epsilon$) near the wall.
With this configuration, the jet enters the domain
through the inlet, impinges on the bottom wall, and
then exits the domain through the open boundaries on
the left and right sides.
We would like to study the heat transfer in this flow using
the method from Section \ref{sec:method}.

\begin{figure}[tb]
  \centerline{
    \includegraphics[width=3.1in]{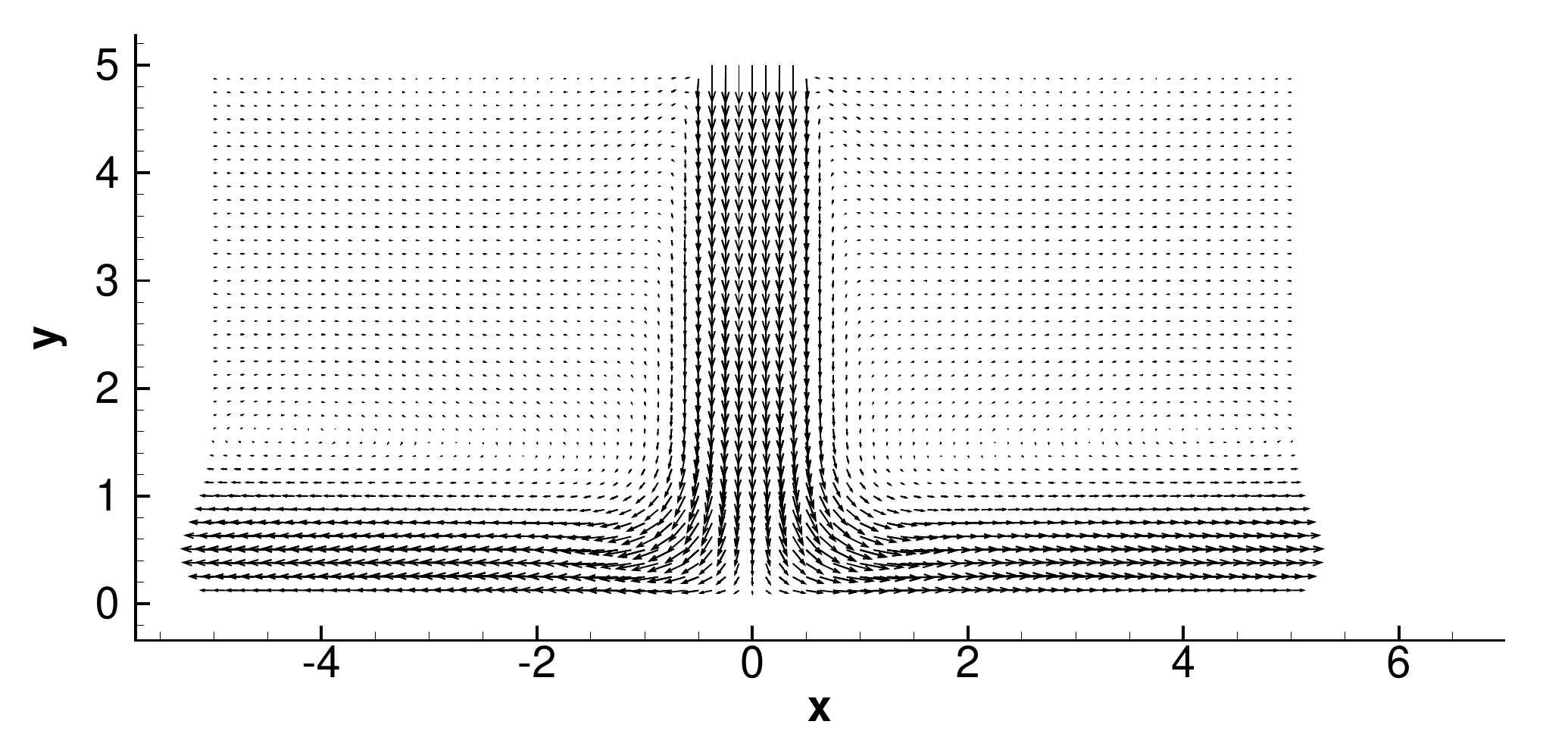}(a)
    \includegraphics[width=3.1in]{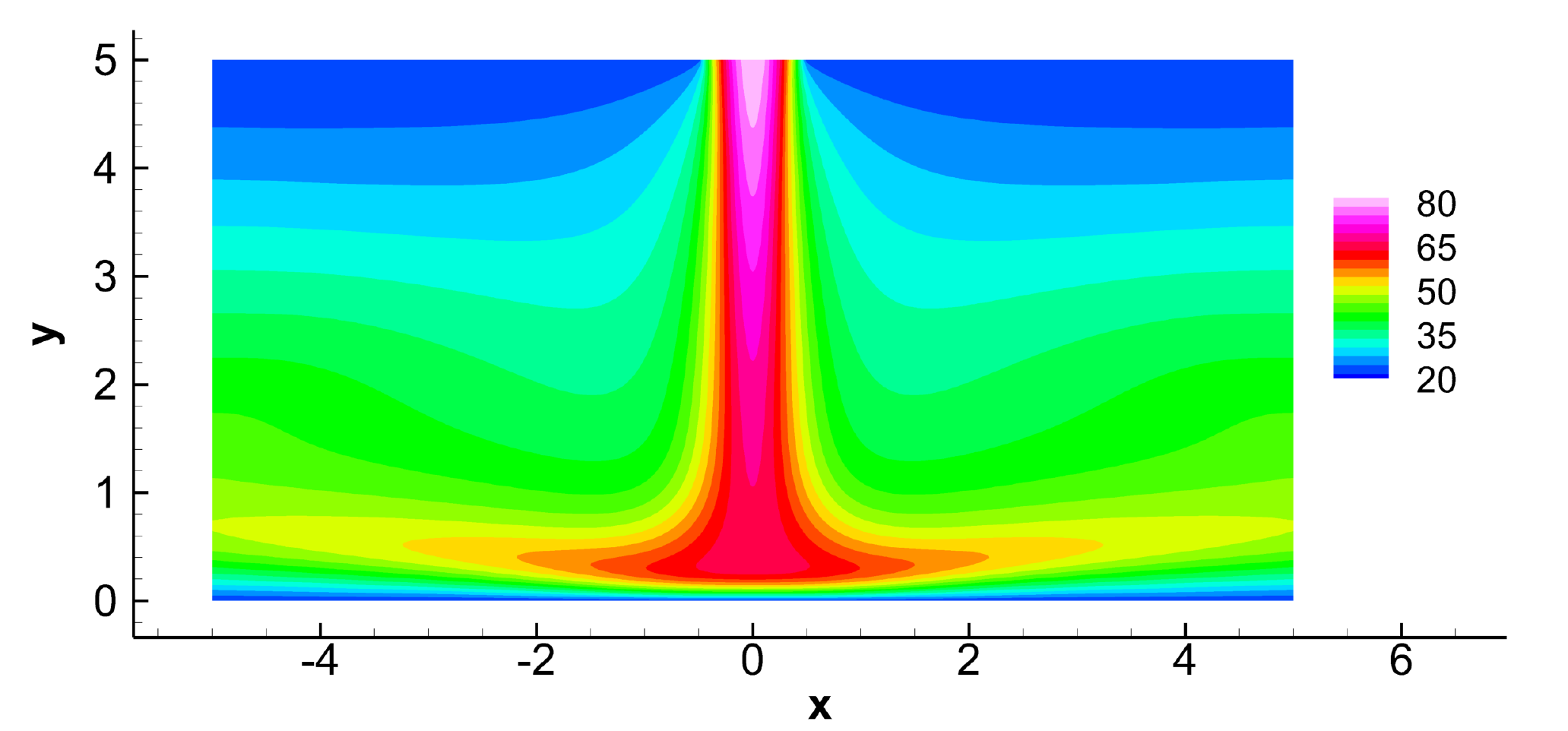}(b)
  }
  \centerline{
    \includegraphics[width=3.1in]{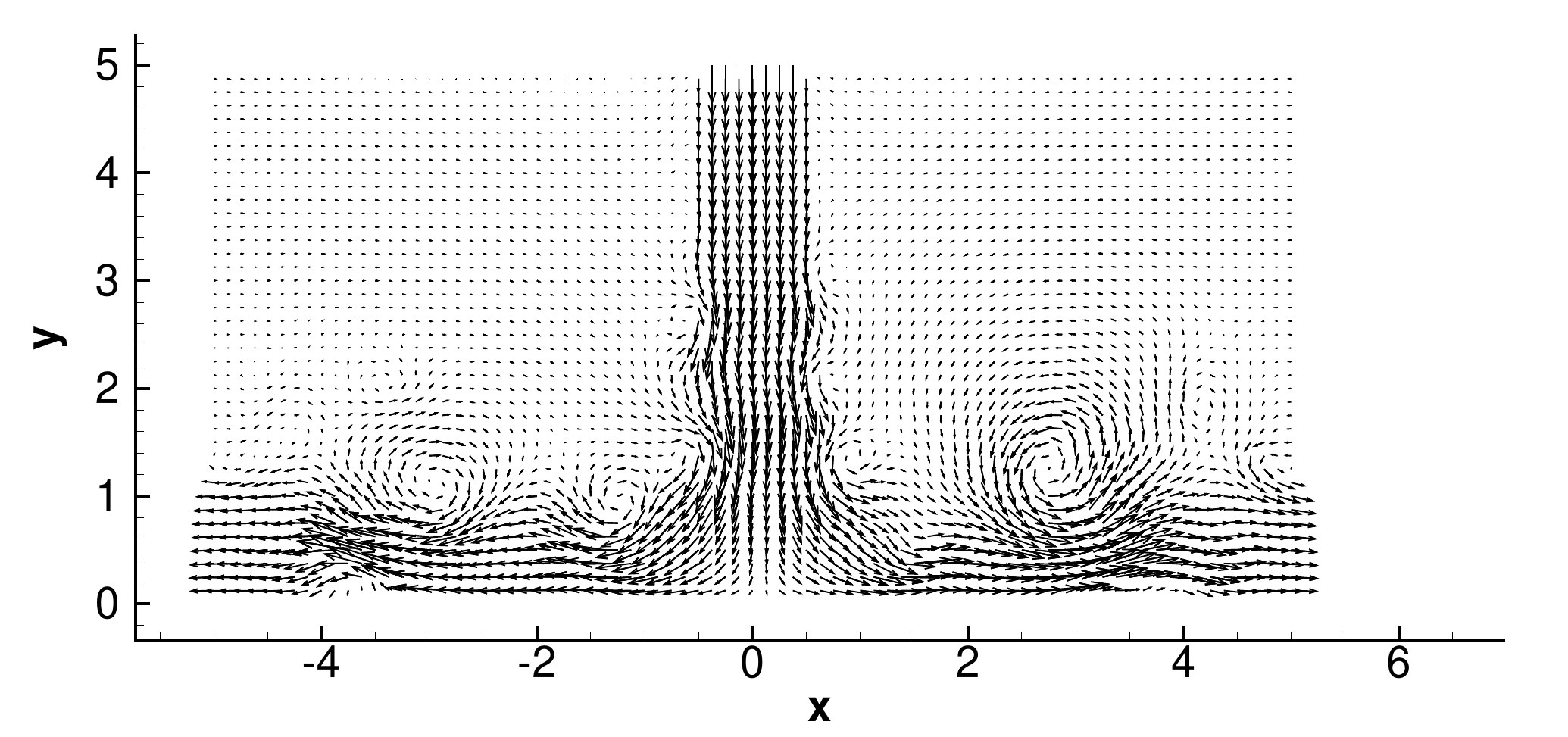}(c)
    \includegraphics[width=3.1in]{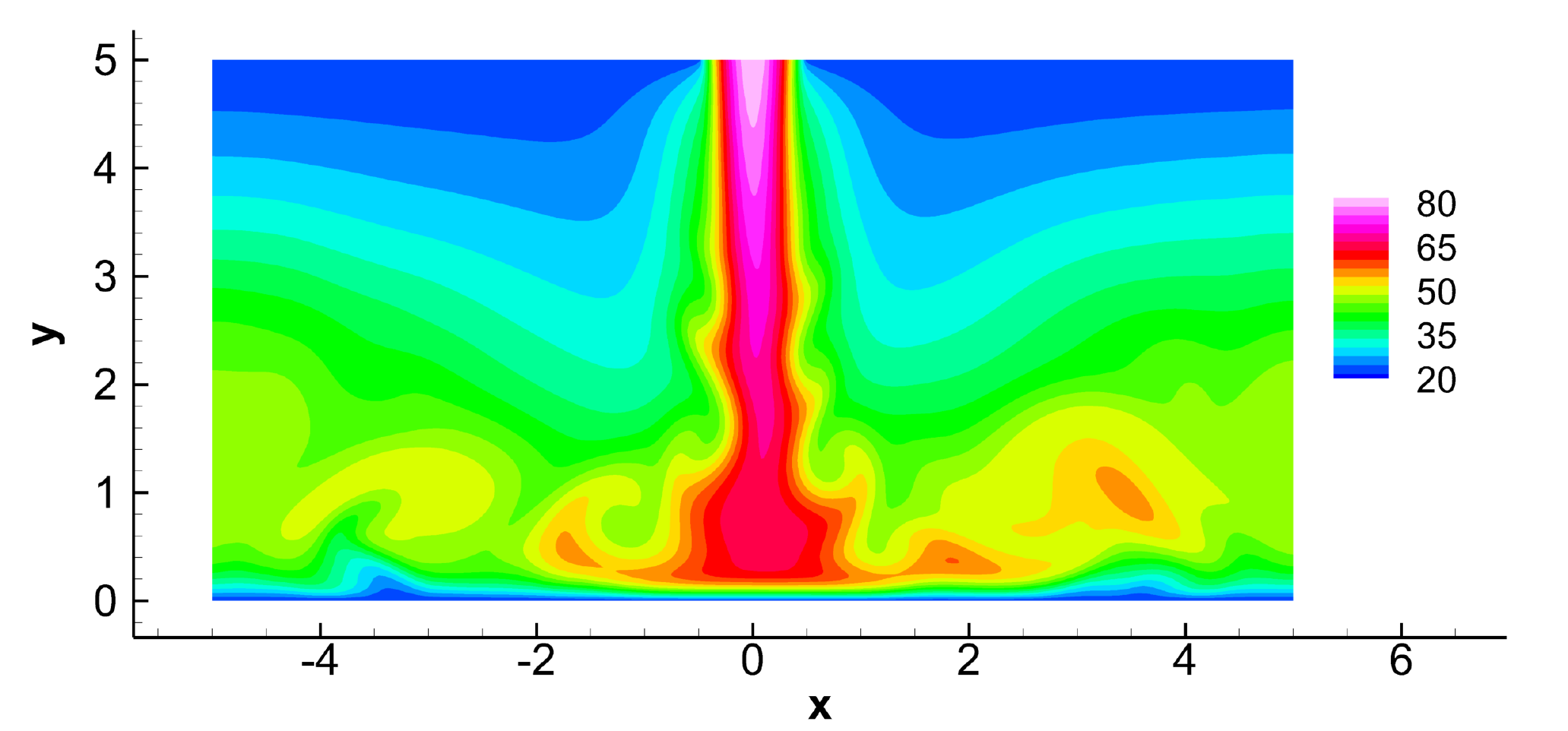}(d)
  }
  \caption{ (color online)
    Impinging jet: instantaneous  velocity
    field (plots (a) and (c)) and temperature field (plots (b) and (d)).
    Plots (a) and (b) are for Reynolds number $Re=300$.
    Plots (c) and (d) are for Reynolds number $Re=2000$.
    Thermal diffusivity is $\alpha=0.01$ with both Reynolds numbers.
    Velocity vectors are plotted on a sparser set of mesh points for clarity.
  }
  \label{fig:jet_char}
\end{figure}


We use $U_0$ as the velocity scale, $d$ as the length scale,
and $T_d=1$ degree Celsius as the temperature scale.
All the physical variables and parameters are then normalized
accordingly. So the Reynolds number is defined based on the jet inlet diameter
for this problem.


We employ the method presented in Section \ref{sec:method}
to solve the heat transfer equation \eqref{equ:tem} with $g=0$,
and the algorithm from the Appendix A to solve the Navier-Stokes equations
\eqref{equ:nse}--\eqref{equ:div} with $\mbs f=0$.
The domain is discretized using a mesh of $800$ uniform quadrilateral elements,
with $40$ elements along the $x$ direction and $20$ elements along the
$y$ direction. On the top and bottom walls,
no-slip condition (zero velocity) has been imposed for the velocity,
and the Dirichlet condition~\eqref{equ:dbc} with $T_d=T_w=20$
has been imposed for the temperature.
At the jet inlet, we impose the Dirichlet conditions~\eqref{equ:dbc_v}
for the velocity and~\eqref{equ:dbc} for the temperature,
with the boundary velocity and temperature chosen according to
the expressions given in \eqref{equ:jet_VTbc}.
On the left/right boundaries of the domain, we
impose the open boundary condition~\eqref{equ:obc_A} for the
temperature, with $D_0=\frac{1}{U_0}$ and $\delta=\frac{1}{20}$,
and the condition~\eqref{equ:obc_v} for the velocity,
with $\mbs f_b=0$ and $\mbs E(\mbs n,\mbs u)$ given by \eqref{equ:def_E}
(for $Re=300$ and $2000$)
or \eqref{equ:def_Egobc} with $(\beta_0,\beta_1,\beta_2)=(0,1,0)$
(for $Re=5000$).
%
The initial temperature is set to $T_{in}=20$ and the initial velocity is set to zero.
We have performed long-time simulations, and the flow and
temperature have reached a statistically stationary
state. So these initial conditions have no effect on the results reported
below.
The element order and the time step size are varied systematically
in the simulations to study their effects on the simulation results.
The problem corresponding to
three Reynolds numbers (Re=300, 2000 and 5000) and two thermal diffusivities
($\alpha=0.01$ and $0.005$) has been simulated.


Figure \ref{fig:jet_char} provides an overview of the distribution characteristics
of the velocity and temperature fields obtained using the current method
at two Reynolds numbers $Re=300$
and $Re=2000$ and a non-dimensional thermal diffusivity $\alpha=0.01$.
At low Reynolds numbers (e.g.~$Re=300$) this is a steady flow.
The vertical jet splits into two horizontal streams after impinging on
the bottom wall, which exit the domain through the open boundaries
on the left/right sides.
Strong flows are largely confined to regions of the vertical
jet and the near-wall horizontal streams (Figure \ref{fig:jet_char}(a)), and
the velocity field is quite weak
outside these regions.
Correspondingly, warm fluids are confined to the regions of the vertical
jet and the horizontal streams, and the temperature decays along
the profile of the jet streams (Figure \ref{fig:jet_char}(b)). 
As the Reynolds number increases (e.g.~$Re=2000$),
the vertical jet becomes unstable downstream
of the inlet due to the Kelvin-Helmholtz instability, and vortices can be observed
to form on both sides of the jet profile (Figure \ref{fig:jet_char}(c)).
These vortices are advected along the vertical and horizontal jet streams
and exit the domain through the left/right open boundaries.
The temperature distributions at these Reynolds numbers
are unsteady and exhibit more complicated features. The vortices 
are observed to carry warm fluids with them, forming hot spots
along the jet profile (Figure \ref{fig:jet_char}(d)).


\begin{figure}
  \centerline{
    \includegraphics[width=4in]{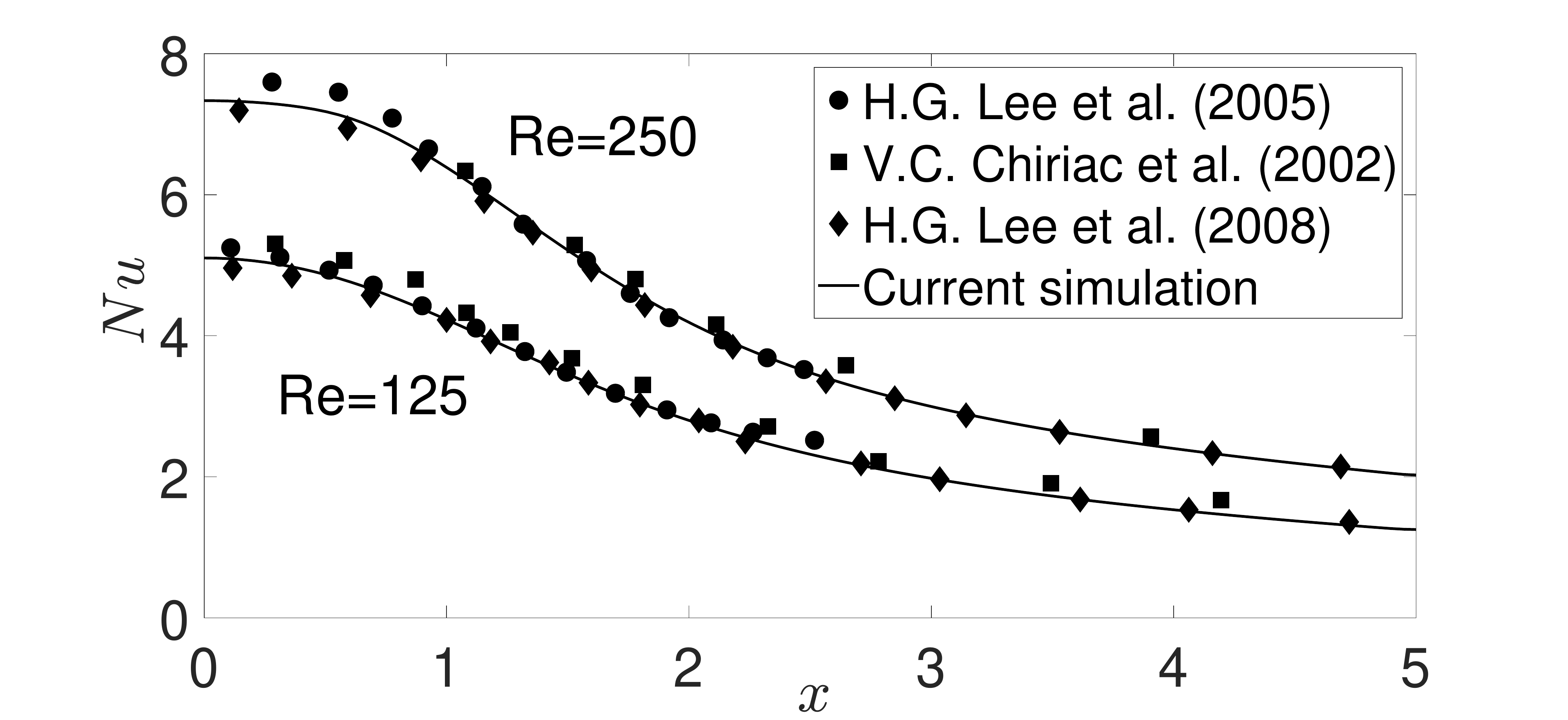}
  }
  \caption{
    Impinging jet: Comparison of the local Nusselt number
    at the lower wall between the current simulation and 
    the previous works~\cite{ChiriacO2002,LeeHY2005,LeeYH2008}
    for Reynolds numbers $Re=125$ and $Re=250$.
    Inlet height to inlet diameter ratio is $5$.
  }
  \label{fig:jet_Nu}
\end{figure}

Before proceeding further,
we first compare our simulations with previous works
for an assessment and verification of
the accuracy of our method.
Figure \ref{fig:jet_Nu} shows the profiles
of the local Nusselt number ($Nu$) at the bottom wall for Reynolds numbers
$Re=125$ and $250$
from the current simulation and from the previous
works of ~\cite{ChiriacO2002,LeeHY2005,LeeYH2008}.
For comparison, we have employed here the same flow setting and simulation conditions
as those from~\cite{LeeYH2008} for this set of results.
Note that these conditions are slightly different
from those given above for the rest of current simulations.
Specifically, here the upper wall is
adiabatic with the boundary condition $\mbs n\cdot\nabla T$=0,
and the lower wall is maintained at a constant temperature $T_w$.
The velocity and temperature at the jet inlet
are both uniform ($u=0$, $v=-U_0$, $T=T_h$), and
the Prandtl number is fixed
at $Pr=\nu/\alpha=0.7$~\cite{LeeYH2008}.
On the left and right open boundaries, we impose
the open boundary condition~\eqref{equ:obc_A}
with $D_0=\frac{1}{U_0}$ and $\delta =0.05$
for the temperature, and the boundary
condition~\eqref{equ:obc_v} with
$\mbs f_b=0$ and $\mbs E(\mbs n,\mbs u)$ given by
\eqref{equ:def_E}.
The local Nusselt number at the bottom wall is defined as
$Nu=\left.\frac{1}{T_h-T_w}\frac{\partial T}{\partial y}\right|_{y=0}$.
It is evident from Figure \ref{fig:jet_Nu}
that our simulation results are in good agreement
with those of ~\cite{LeeYH2008,ChiriacO2002,LeeHY2005}.


\begin{figure}[tb]
  \centerline{
    \includegraphics[width=4in]{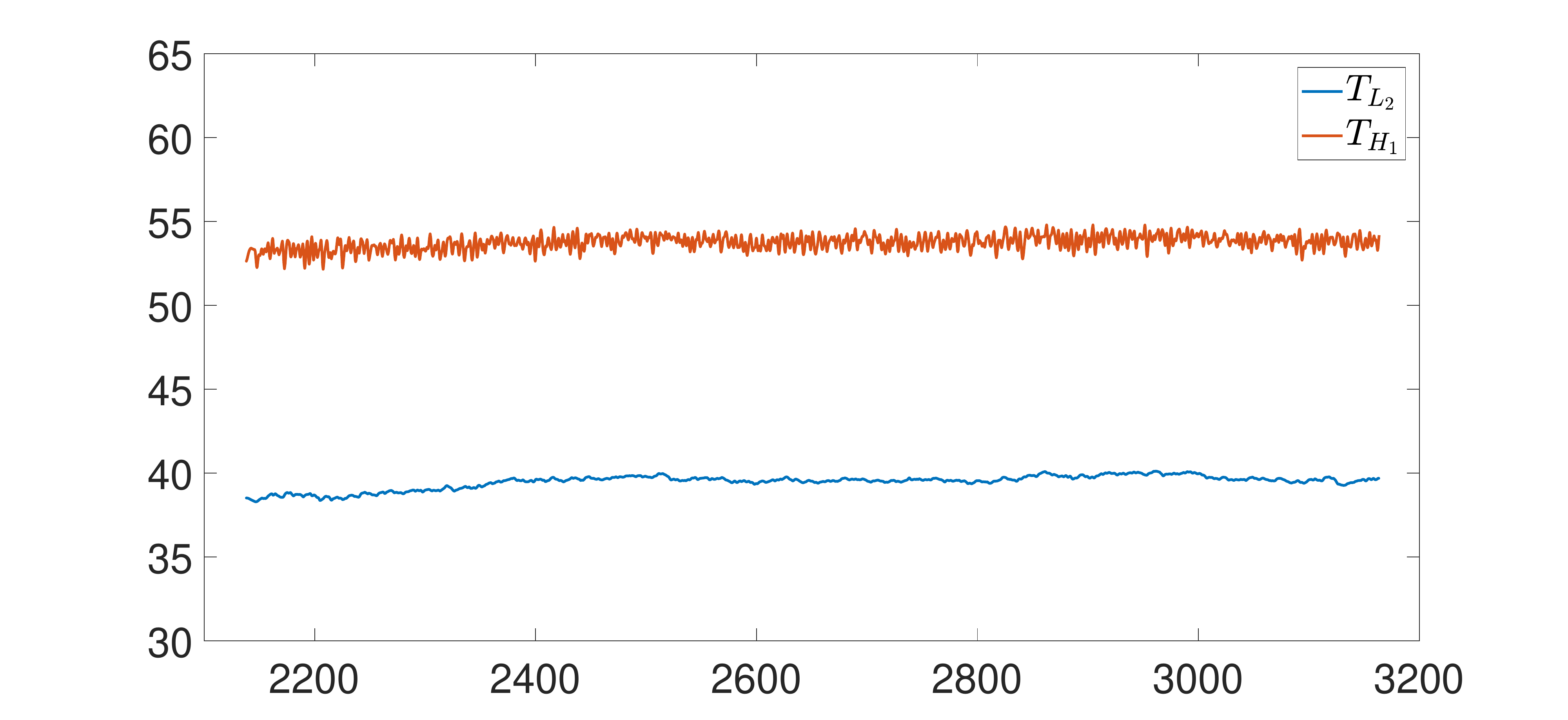}
  }
  \caption{
    Impinging jet: Time histories of $T_{L_2}(t)$ and $T_{H_1}(t)$
    (see equation~\eqref{equ:T_norm})
    at Reynolds number $Re=2000$ and thermal diffusivity $\alpha=0.01$.
  }
  \label{fig:jet_hist}
\end{figure}

\begin{table}[tb]
  \centering
  \begin{tabular}{llllll}
    \hline
    Reynolds number & Element order & $\overline{T}_{L_2}$ & $T'_{L_2}$
    & $\overline{T}_{H_1}$ & $T'_{H_1}$ \\ \hline
    300 & 3 & 39.599 & 0 & 53.341 & 0 \\
    & 4 & 38.513 & 0 & 52.611 & 0 \\
    & 5 & 38.502 & 0 & 52.607 & 0 \\
    & 6 & 38.498 & 0 & 52.606 & 0 \\
    & 7 & 38.508 & 0 & 52.613 & 0 \\
    & 8 & 38.559 & 0 & 52.645 & 0 \\
    \hline
    2000 & 5 & 39.934 & 0.228 & 54.036 & 0.355  \\
    & 6 & 39.745 & 0.139 & 53.915 & 0.318 \\
    & 7 & 39.663 & 0.175 & 53.846 & 0.367 \\
    & 8 & 39.703 & 0.135 & 53.887 & 0.356 \\
    & 9 & 39.743 & 0.171 & 53.902 & 0.342 \\
    \hline
  \end{tabular}
  \caption{
    Impinging jet: Time-averaged mean ($\overline{T}_{L_2}$ and $\overline{T}_{H_1}$)
    and root-mean-square (rms, $T'_{L_2}$ and $T'_{H_1}$)
    temperatures of $T_{L_2}(t)$ and $T_{H_1}(t)$ computed using
    various element orders.
    Thermal diffusivity is $\alpha=0.01$.
  }
  \label{tab:jet_stat}
\end{table}

Let us now focus on the study of the
temperature and flow features of the impinging jet.
For each set of physical and simulation parameters (Reynolds number, thermal
diffusivity, element order, time step size), we have performed
a long-time simulation, 
and the flow and temperature fields of the jet have reached a statistically stationary
state corresponding to that set of parameter values.
Figure \ref{fig:jet_hist} shows the time histories of the
temperature norms $T_{L_2}(t)$ and $T_{H_1}(t)$ defined in
equation \eqref{equ:T_norm} for the impinging jet problem.
This corresponds to the Reynolds number $Re=2000$ and
a thermal diffusivity $\alpha=0.01$, and is computed
with an element order $8$ and a time step size $\Delta t=2.5e-4$.
These temperature histories are fluctuational in time. But their values
all stay around some constant mean levels, and the overall characteristics
of these signals remain the same over time.
These results signify that our method is long-term stable and
that the temperature distribution has indeed reached a statistically
stationary state.



\begin{figure}
  \centerline{
    \includegraphics[height=3in]{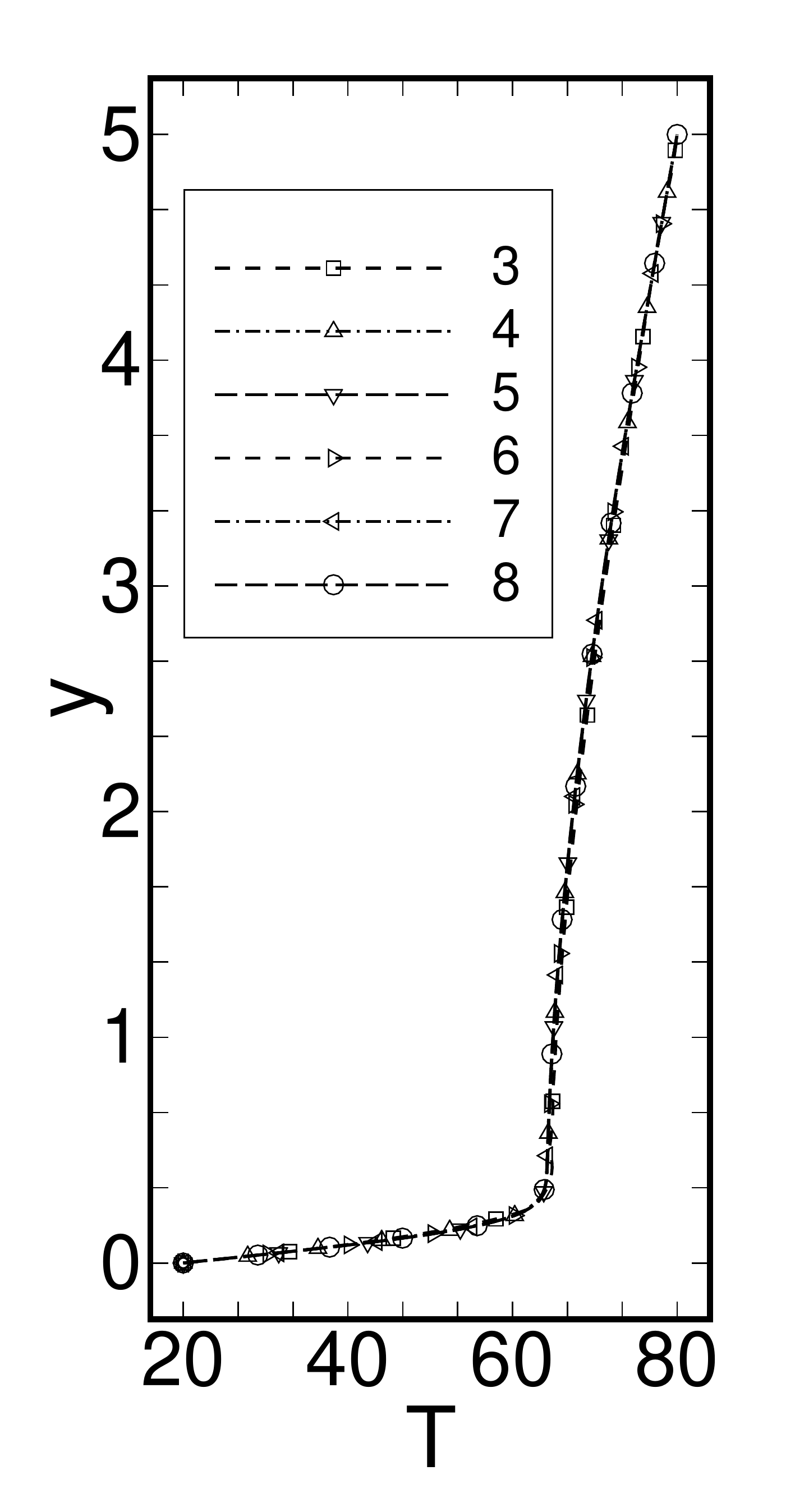}(a)
    \includegraphics[height=3in]{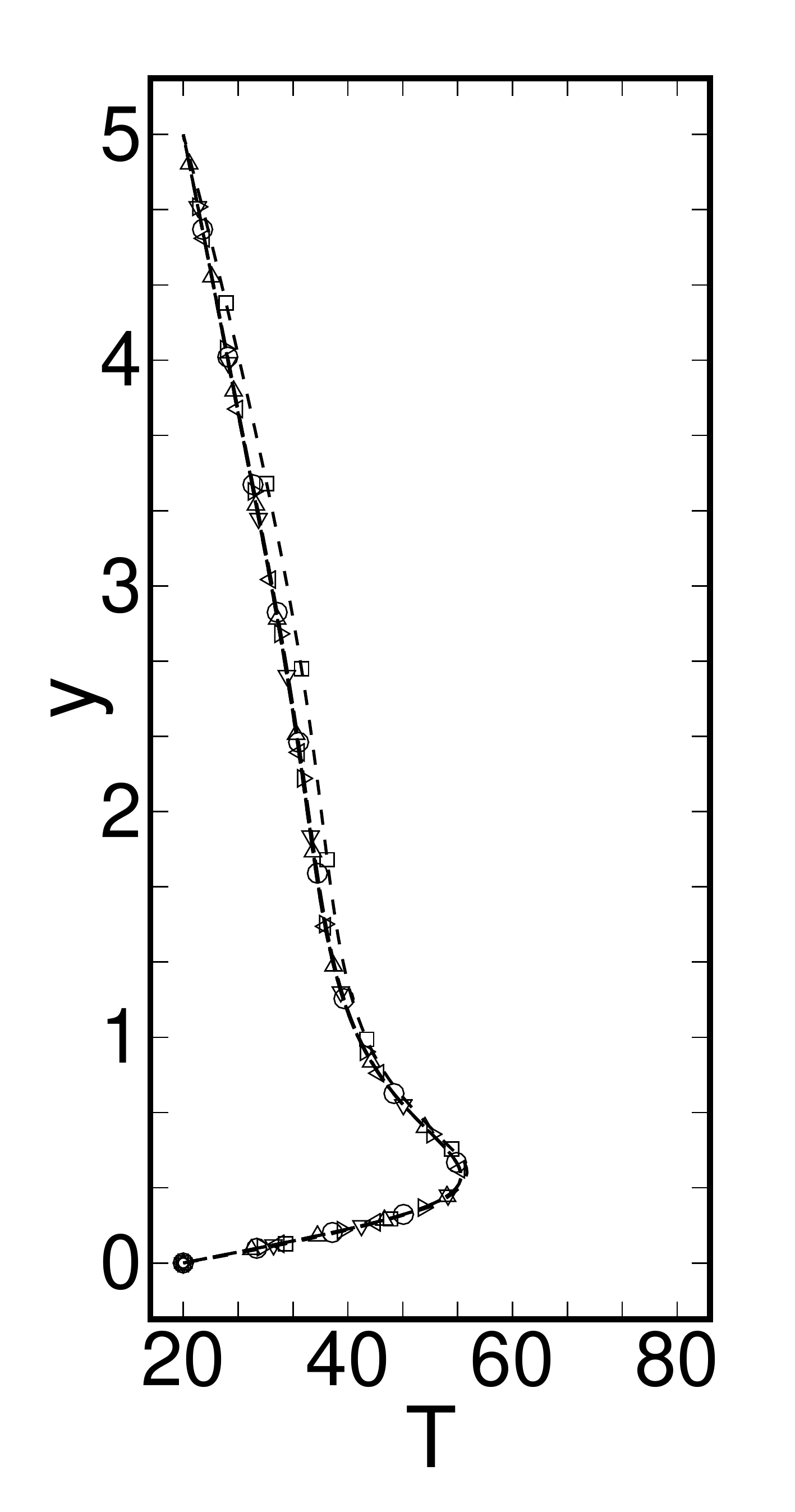}(b)
    \includegraphics[height=3in]{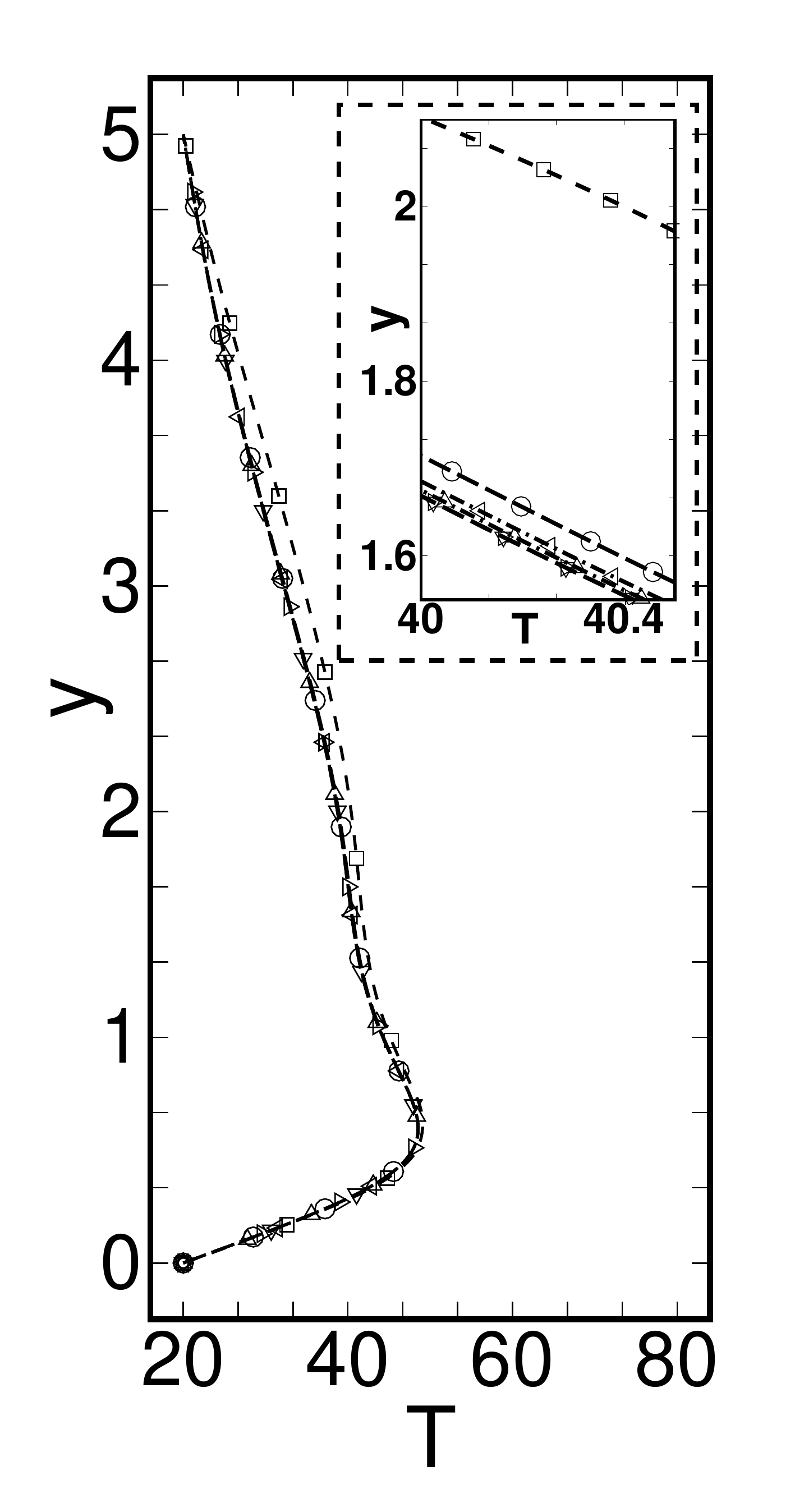}(c)
  }
  \caption{
    Impinging jet ($Re=300$): Comparison of temperature profiles along the vertical direction
    at three horizontal locations, (a) centerline ($x=0$),
    (b) $x=2.0$ and (c) $x=4.0$, computed using element orders
    ranging from $3$ to $8$.
  }
  \label{fig:jet_prof}
\end{figure}

We have computed the time-averaged mean and rms of the temperature
norms based on the history data of $T_{L_2}(t)$ and $T_{H_1}(t)$.
Table \ref{tab:jet_stat} lists the mean ($\overline{T}_{L_2}$
and $\overline{T}_{H_1}$) and rms ($T'_{L_2}$ and $T'_{H_1}$) of
these temperatures  obtained with element orders
ranging from $3$ to $9$. These results for two Reynolds numbers
$Re=300$ and $Re=2000$, and a thermal diffusivity $\alpha=0.01$.
A time step size $\Delta t=2.5e-4$ is employed in these simulations.
Since the flow at $Re=300$ is steady, shown in the table are
the steady-state values and no time-averaging is performed for this
Reynolds number. We observe that for $Re=300$, with element orders
$4$ and above, the computed values for $\overline{T}_{L_2}$
and $\overline{T}_{H_1}$ are essentially the same (with a difference
less than $0.5\%$).
For $Re=2000$, as the element order increases to $6$ and above,
the computed values of $\overline{T}_{L_2}$
and $\overline{T}_{H_1}$ become very close, exhibiting
a sense of convergence.

Figure \ref{fig:jet_prof} shows a comparison of the temperature profiles
along the vertical direction at several horizontal locations
in the domain ($x/d=0$, $2$ and $4$), computed using different element orders
for $Re=300$ and $\alpha=0.01$. It is evident that the temperature
profiles corresponding to element orders $4$ and above all overlap with one
another, again signifying the independence of simulation results
with respect to the mesh resolution.
The majority of simulations reported below are performed
with an element order $6$ for $Re=300$ and element order
$8$ for Reynolds numbers $Re=2000$ and $Re=5000$.


We have also performed simulations at $Re=300$ with several time step
sizes (ranging from $\Delta t=1e-3$ to $\Delta t=2.5e-4$),
and tested the sensitivity of the results with respect to
$\Delta t$. It is observed that the obtained results corresponding to
different $\Delta t$ are basically the same. 
The majority of simulation results reported below are
computed with a time step size $\Delta t=2.5e-4$.

\begin{figure}
  \centerline{
    \includegraphics[width=3.1in]{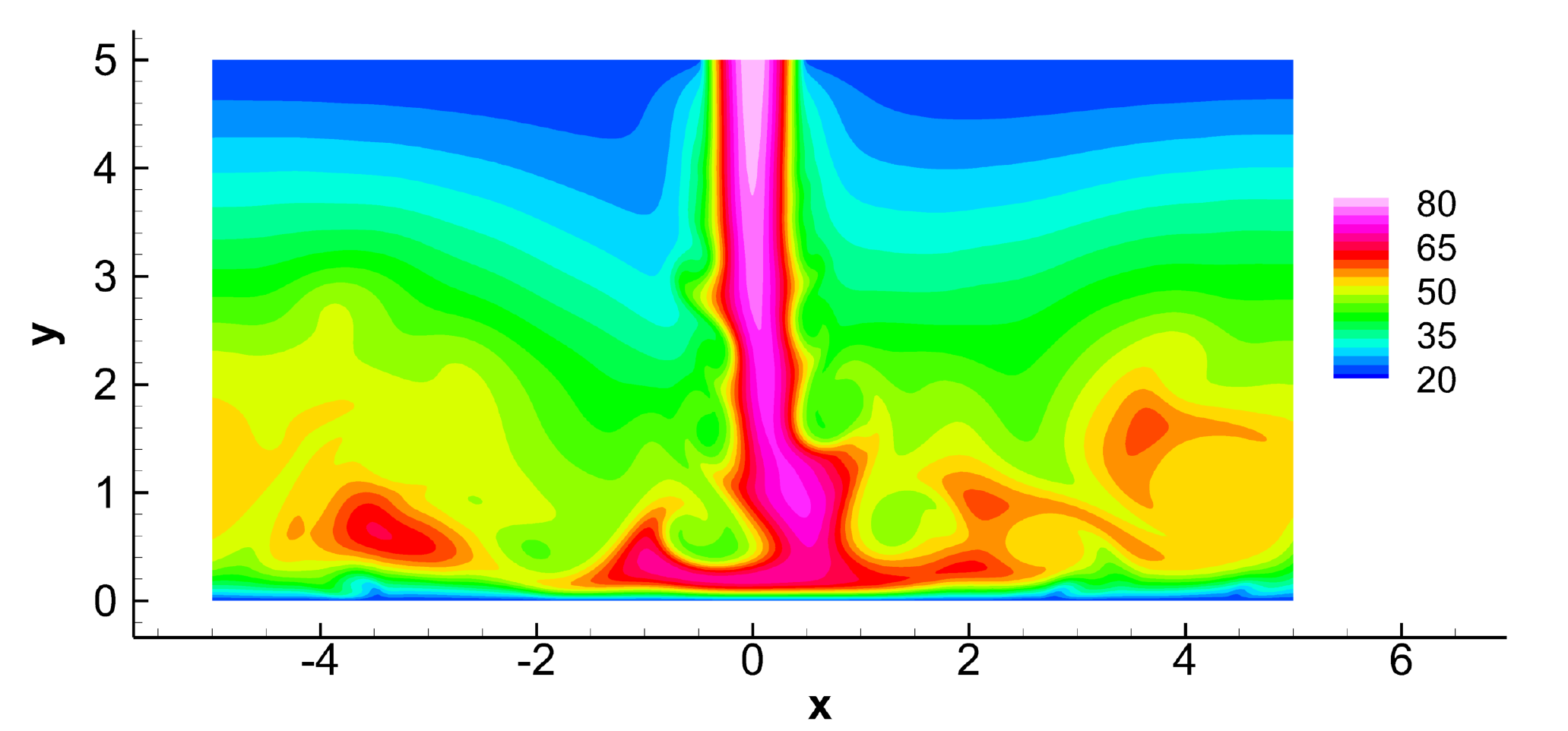}(a)
    \includegraphics[width=3.1in]{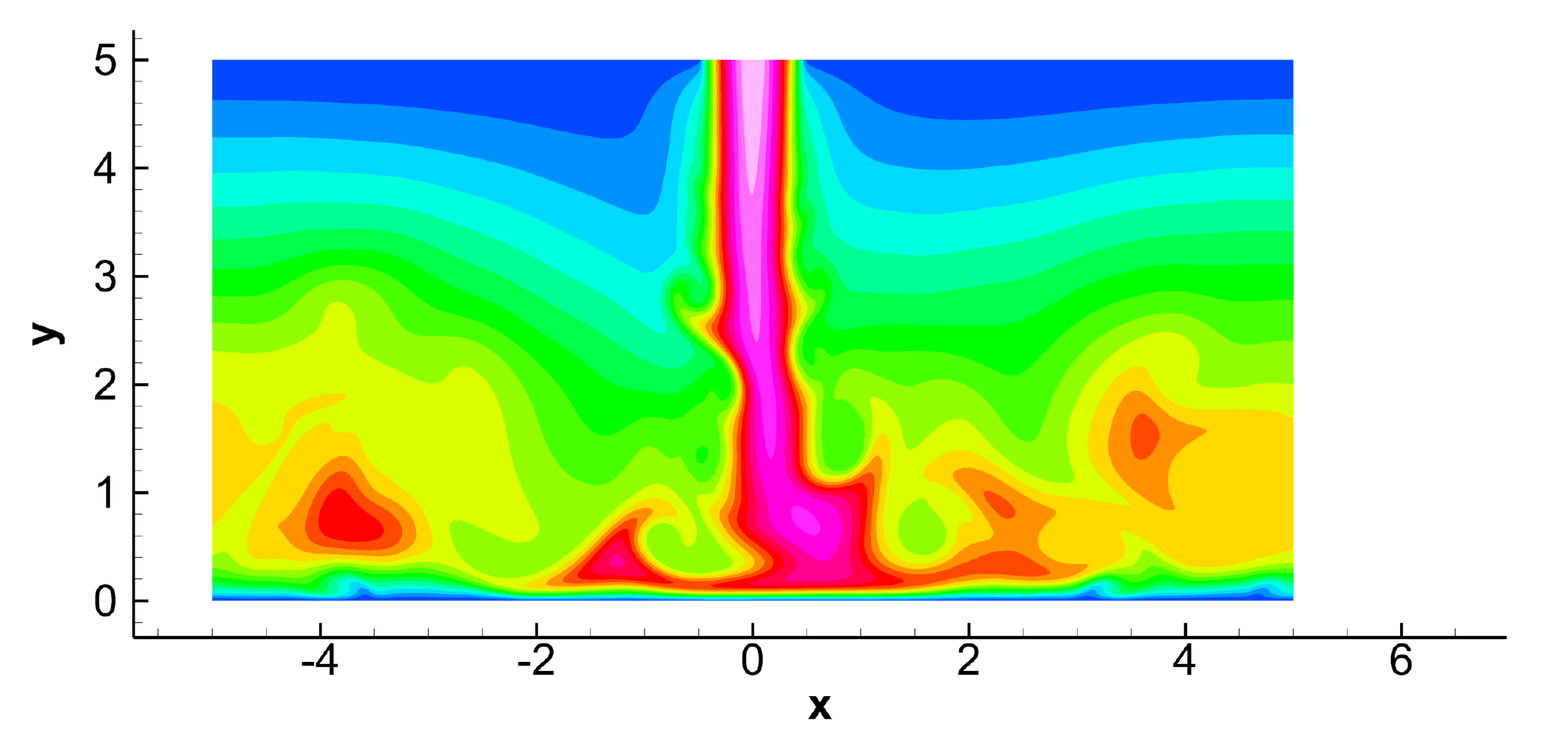}(b)
  }
  \centerline{
    \includegraphics[width=3.1in]{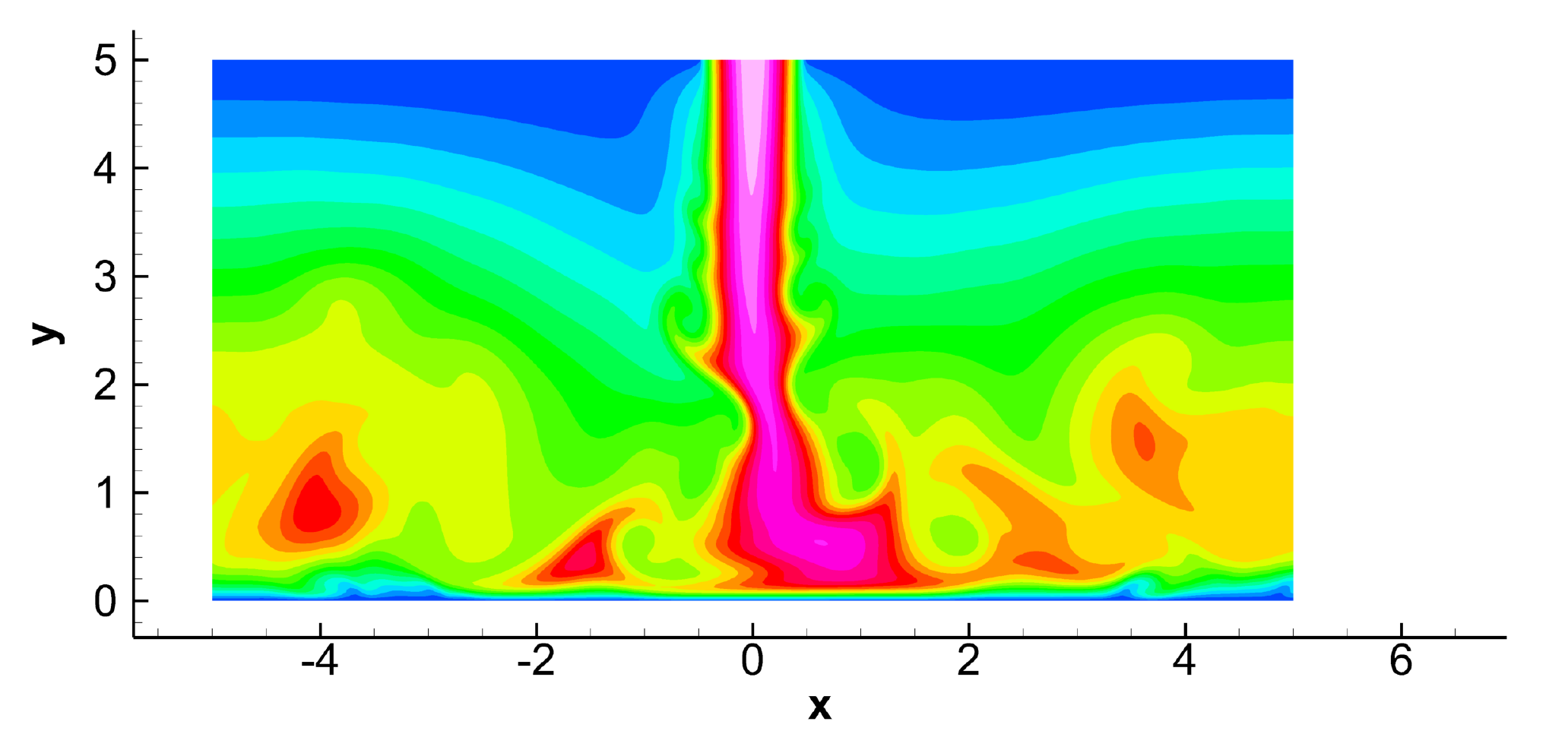}(c)
    \includegraphics[width=3.1in]{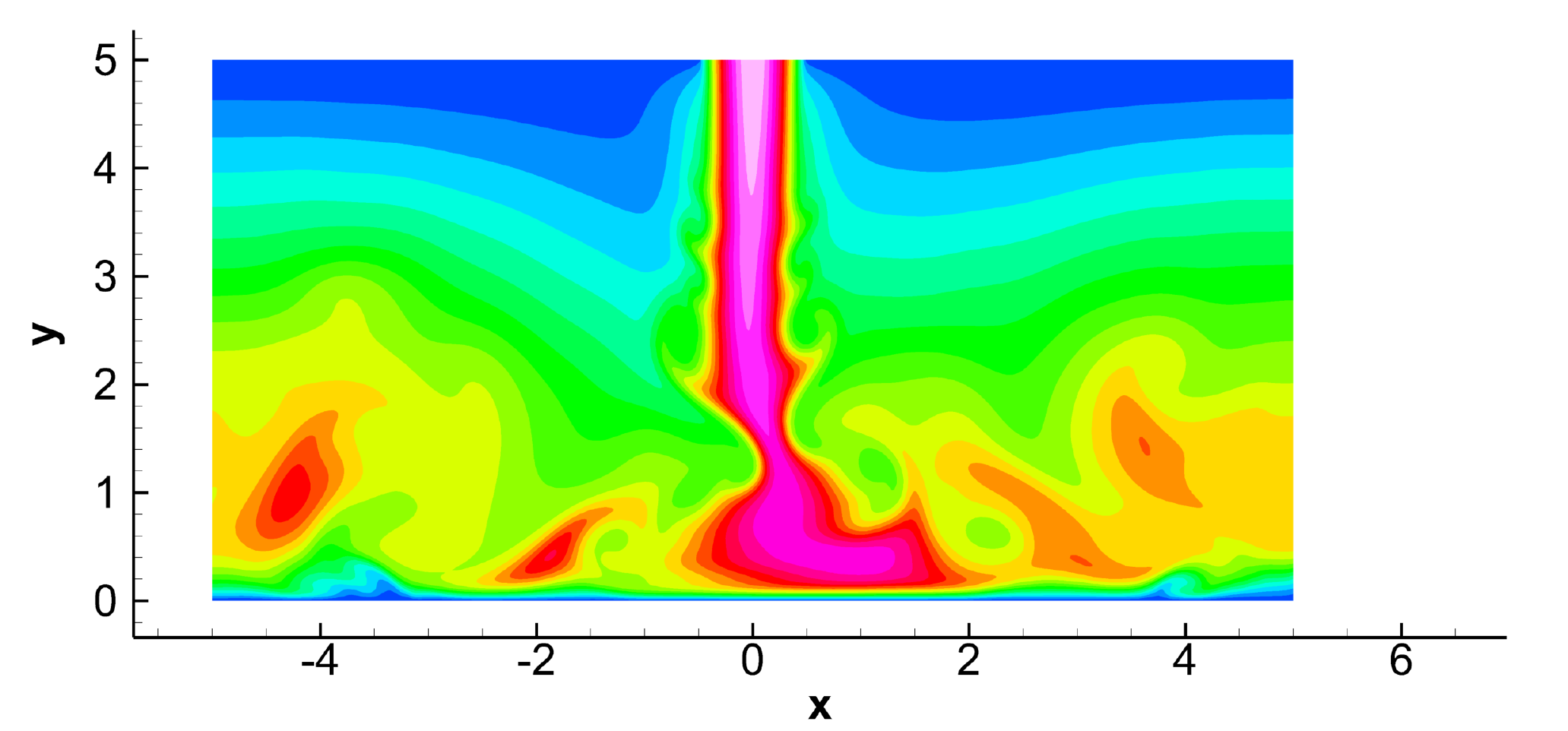}(d)
  }
  \centerline{
    \includegraphics[width=3.1in]{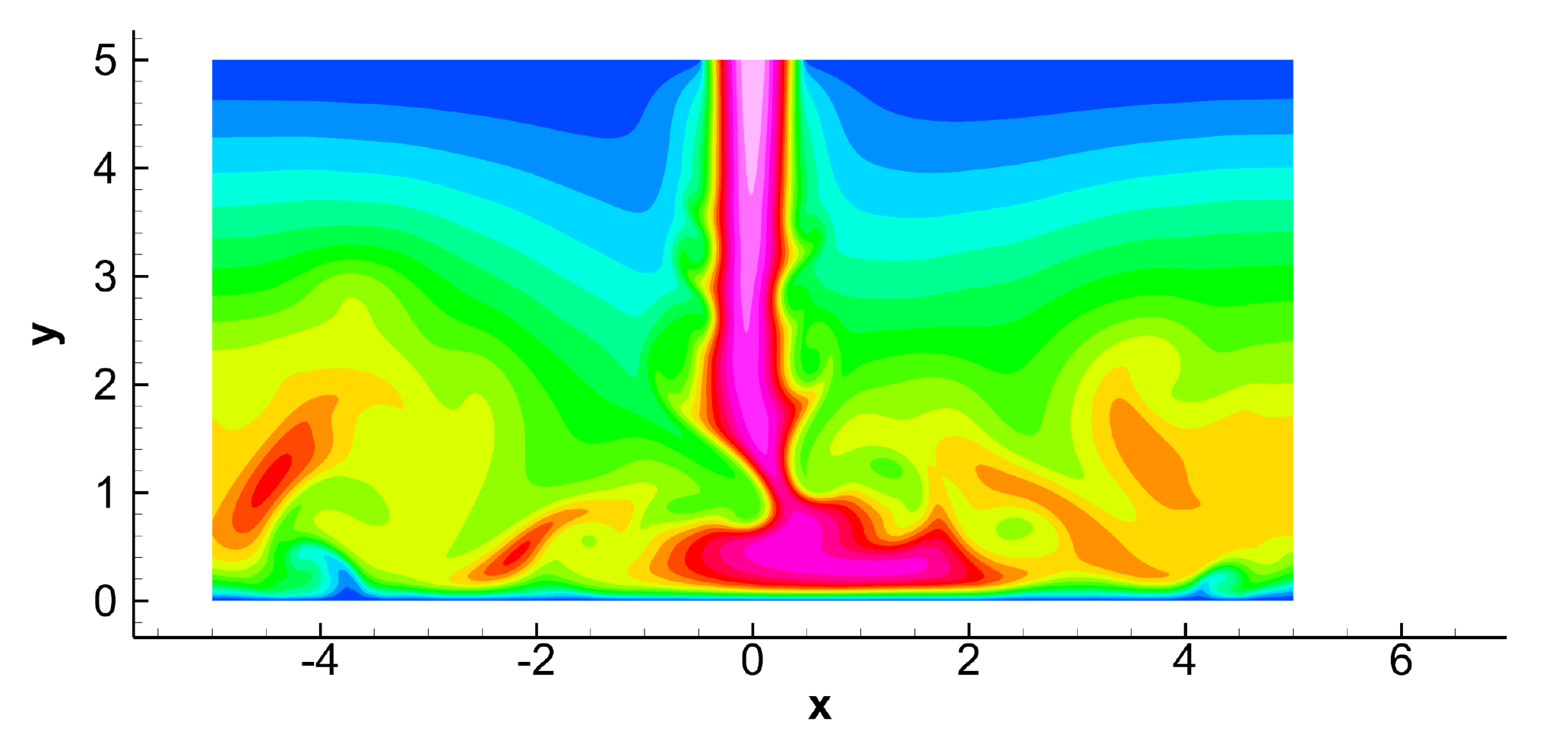}(e)
    \includegraphics[width=3.1in]{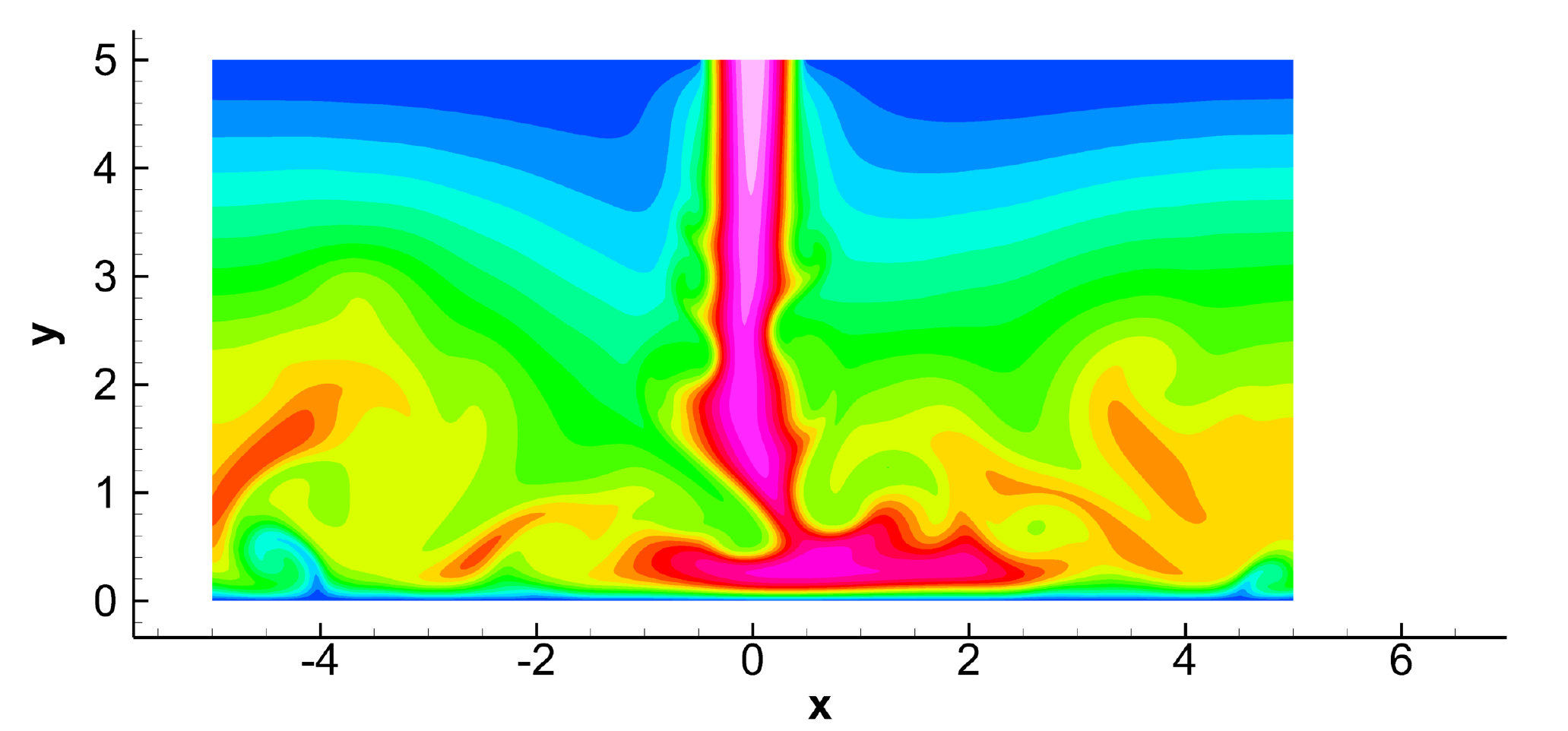}(f)
  }
  \centerline{
    \includegraphics[width=3.1in]{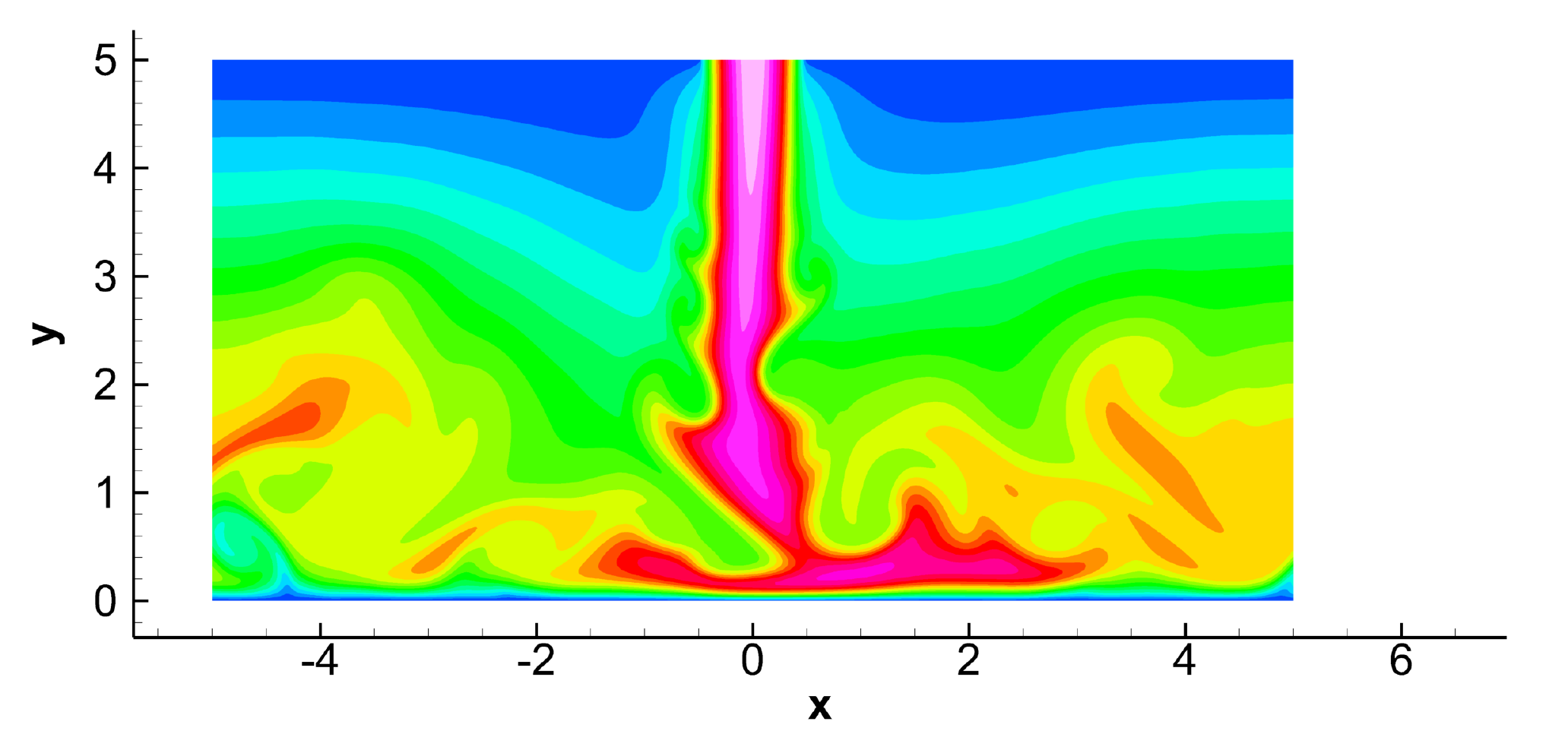}(g)
    \includegraphics[width=3.1in]{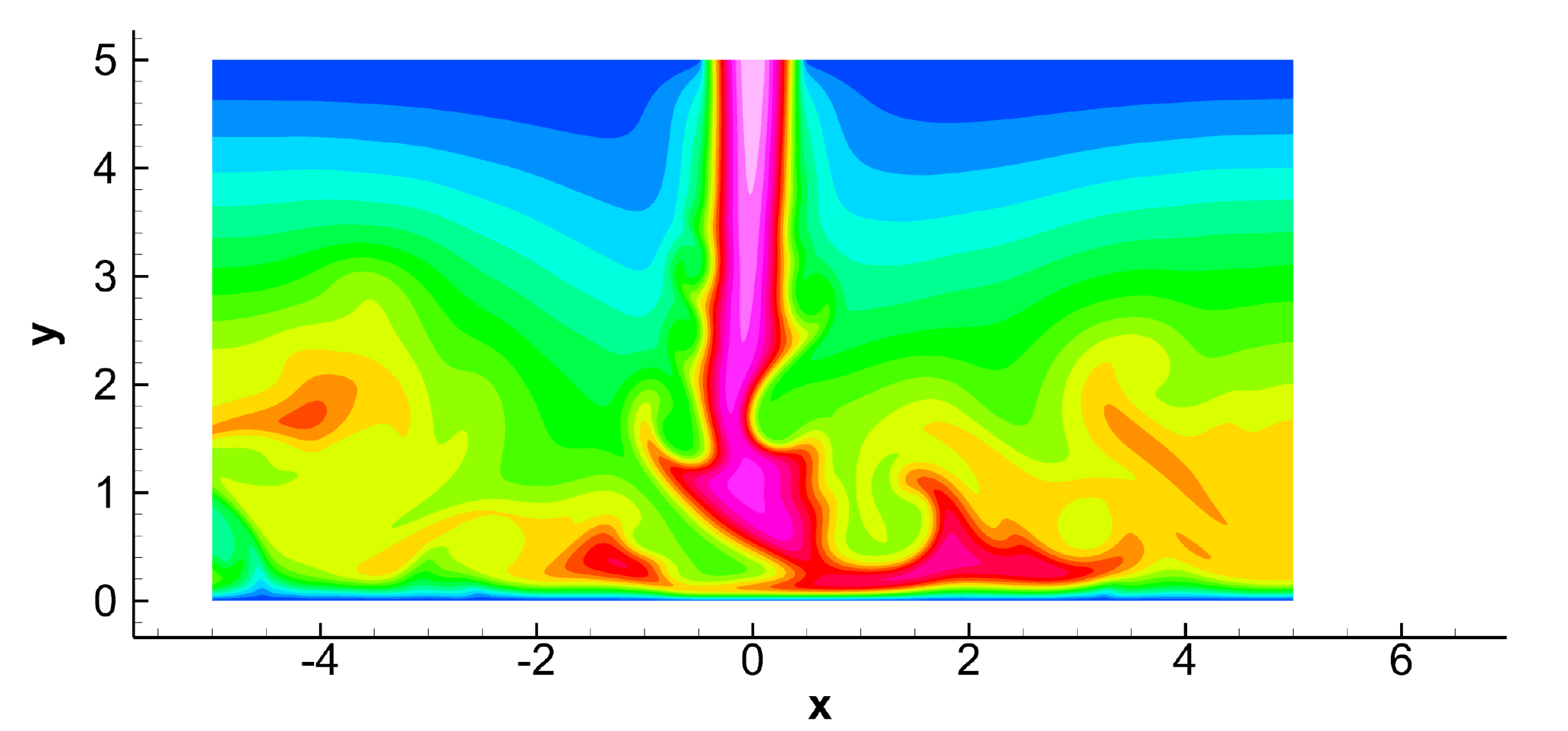}(h)
  }
  \centerline{
    \includegraphics[width=3.1in]{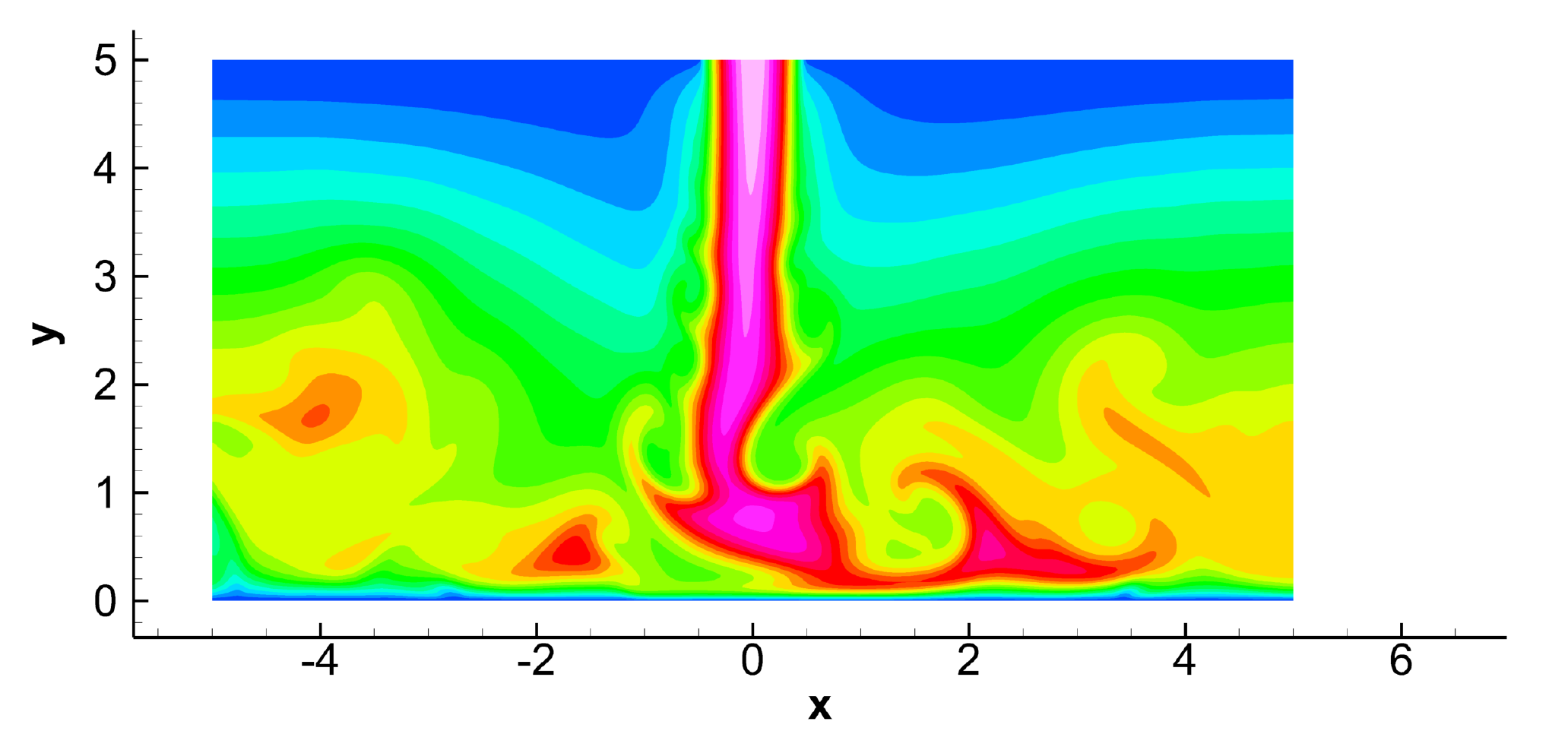}(i)
    \includegraphics[width=3.1in]{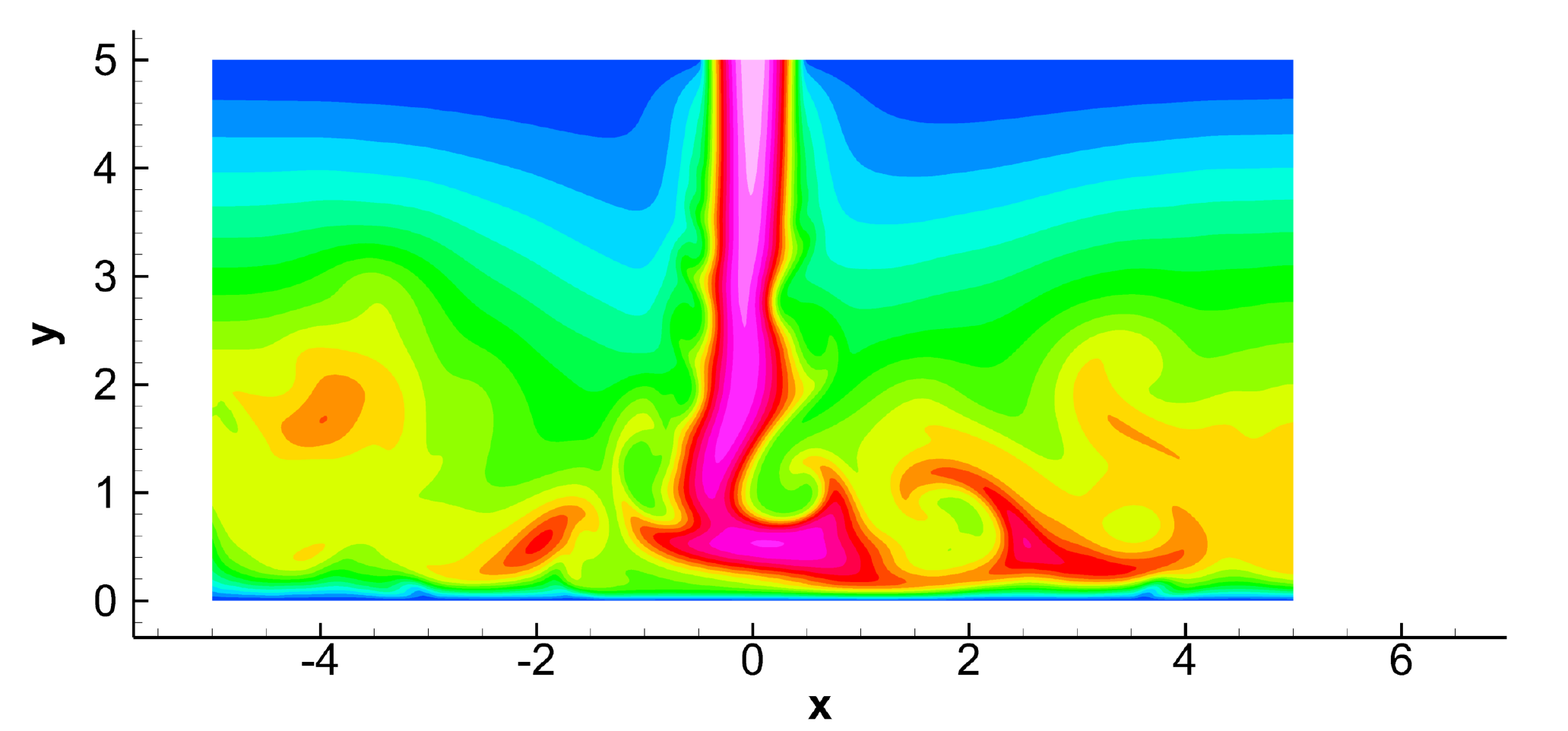}(j)
  }
  \caption{ (color online)
    Impinging jet (Re=5000): temporal sequence of snapshots of
    the temperature distribution
    computed using the current thermal open boundary condition.
    (a) $t=t_0$, (b) $t=t_0+0.5$, (c) $t=t_0+1.0$, (d) $t=t_0+1.5$,
    (e) $t=t_0+2.0$, (f) $t=t_0+2.5$, (g) $t=t_0+3.0$, (h) $t=t_0+3.5$,
    (i) $t=t_0+4.0$, (j) $t=t_0+4.5$.
    Thermal diffusivity is $\alpha=0.005$.
  }
  \label{fig:jet_T}
\end{figure}

\begin{figure}
  \centerline{
    \includegraphics[width=3.1in]{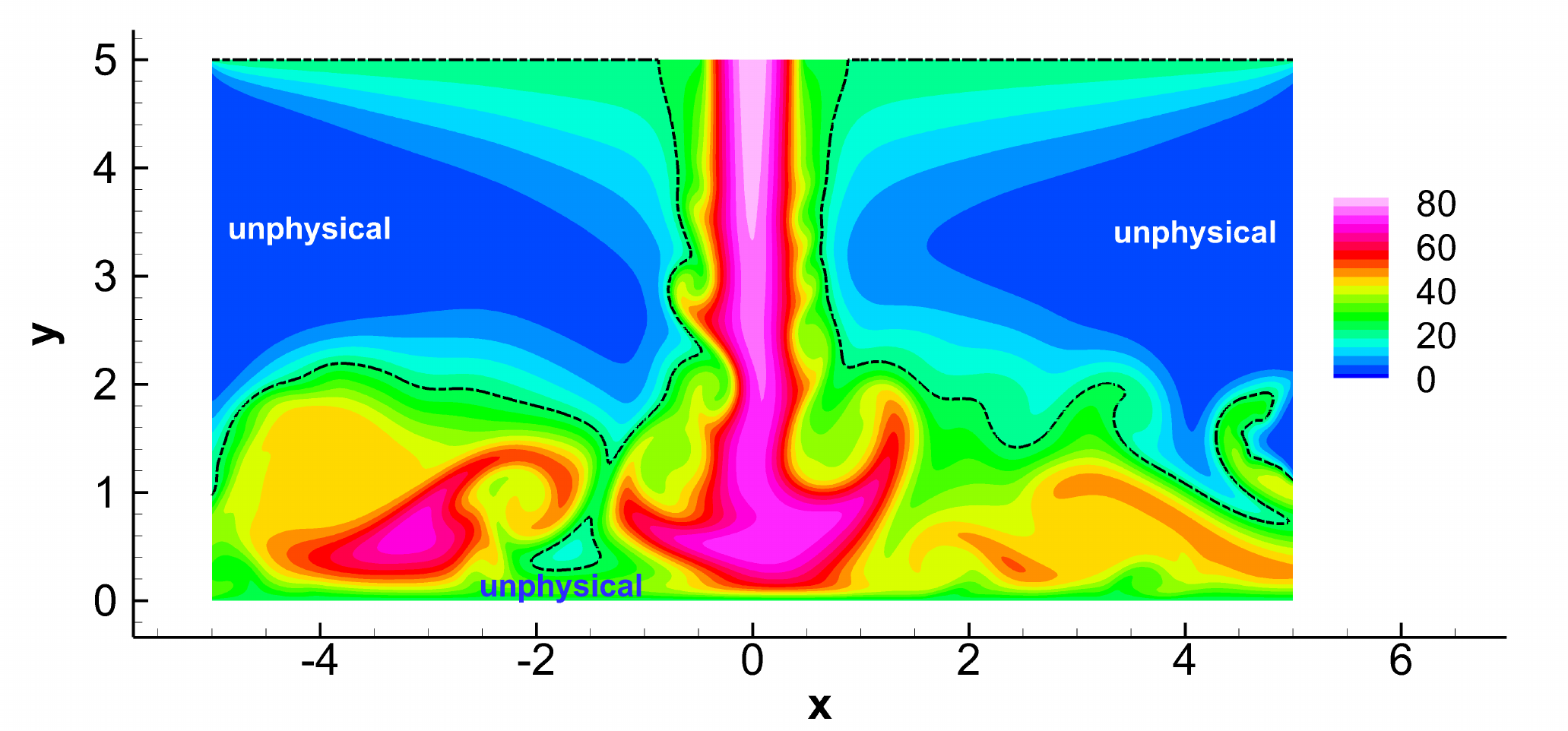}(a)
    \includegraphics[width=3.1in]{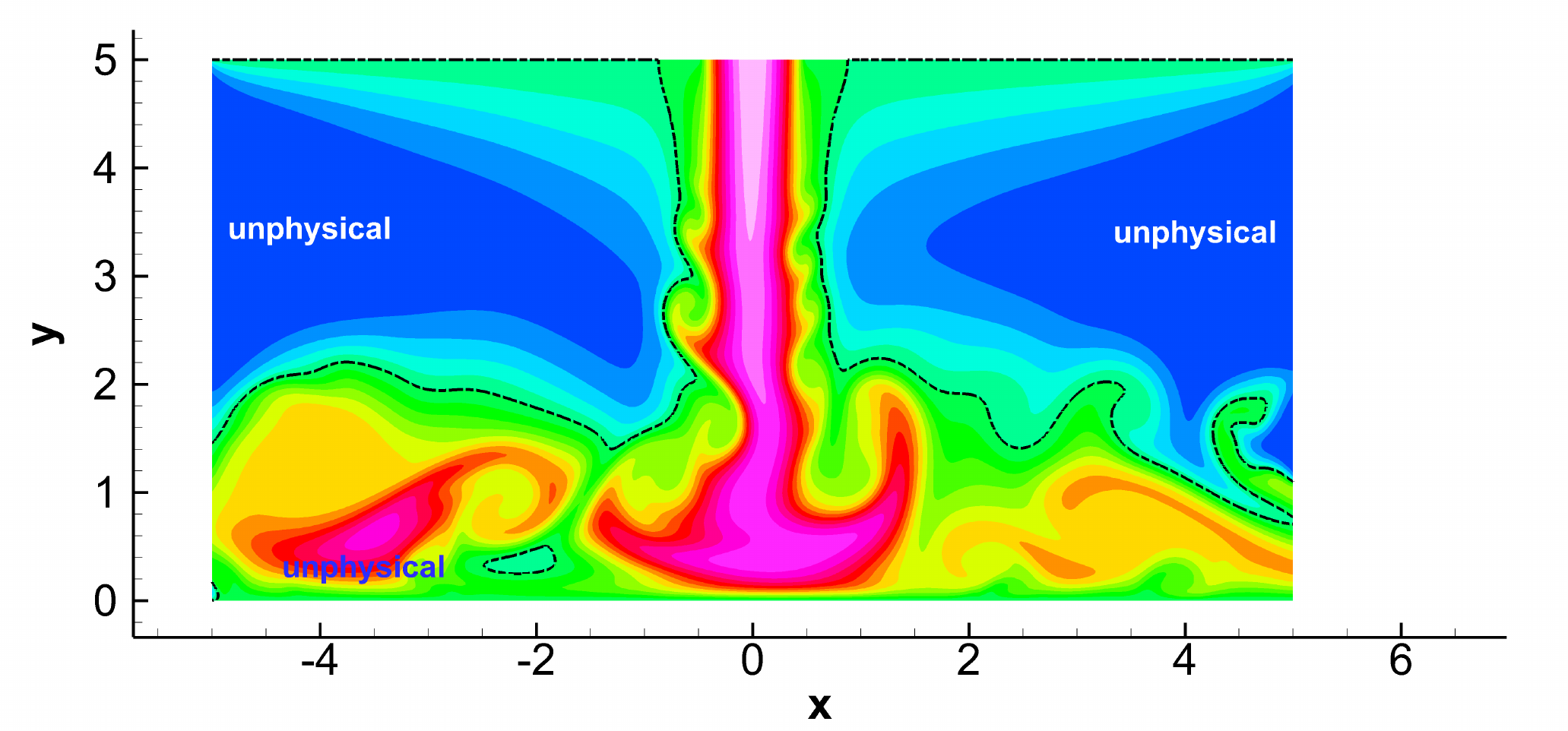}(b)
  }
  \centerline{
    \includegraphics[width=3.1in]{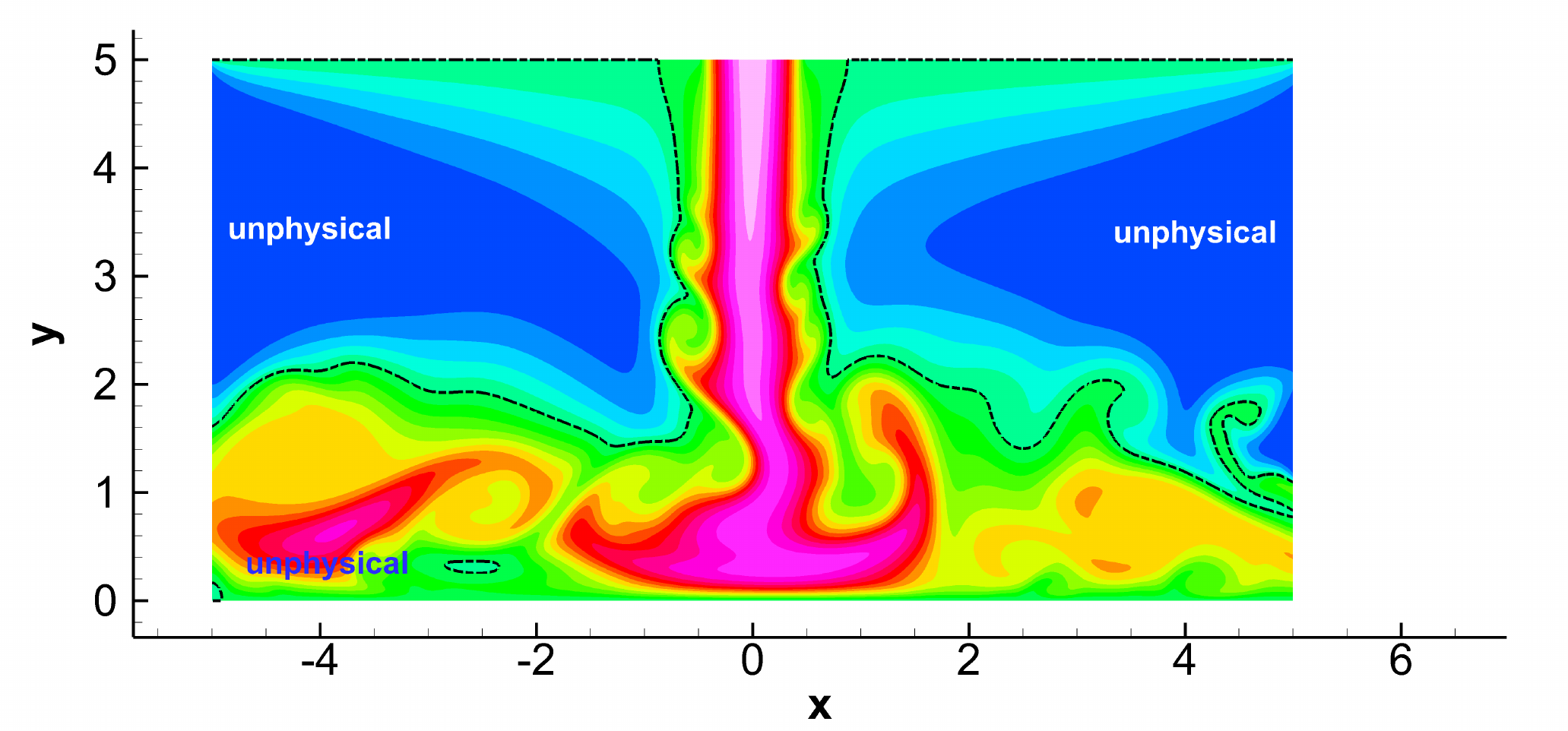}(c)
    \includegraphics[width=3.1in]{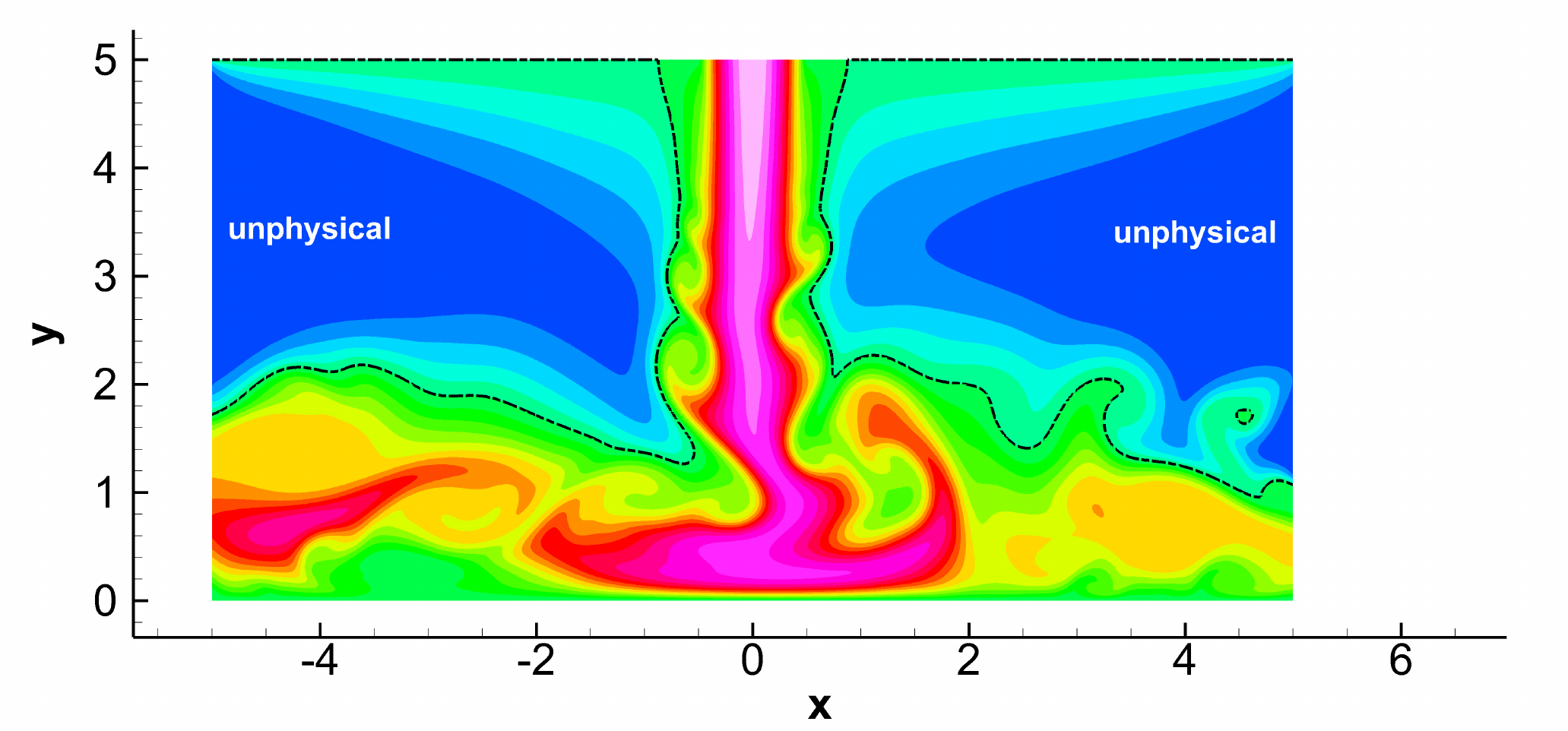}(d)
  }
  \centerline{
    \includegraphics[width=3.1in]{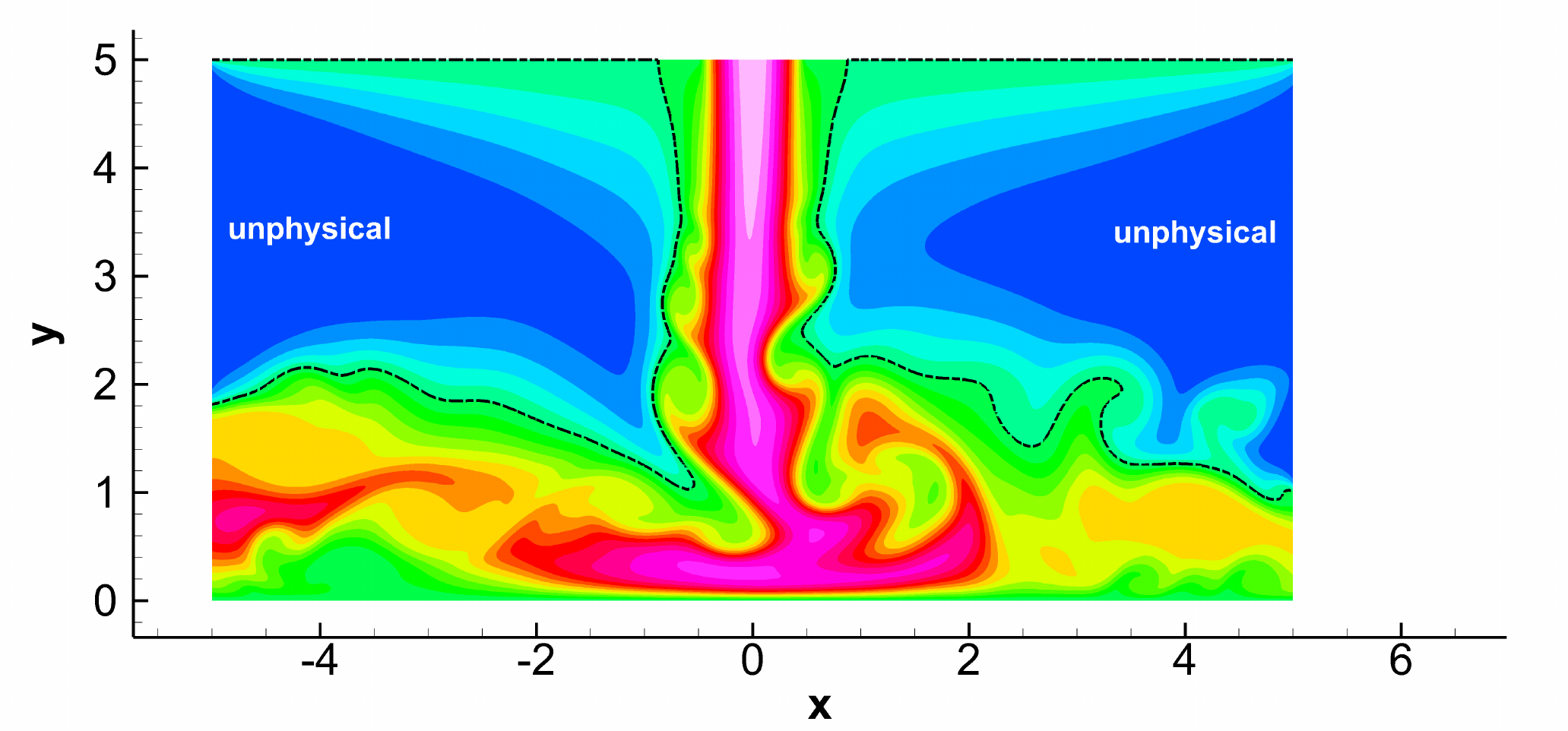}(e)
    \includegraphics[width=3.1in]{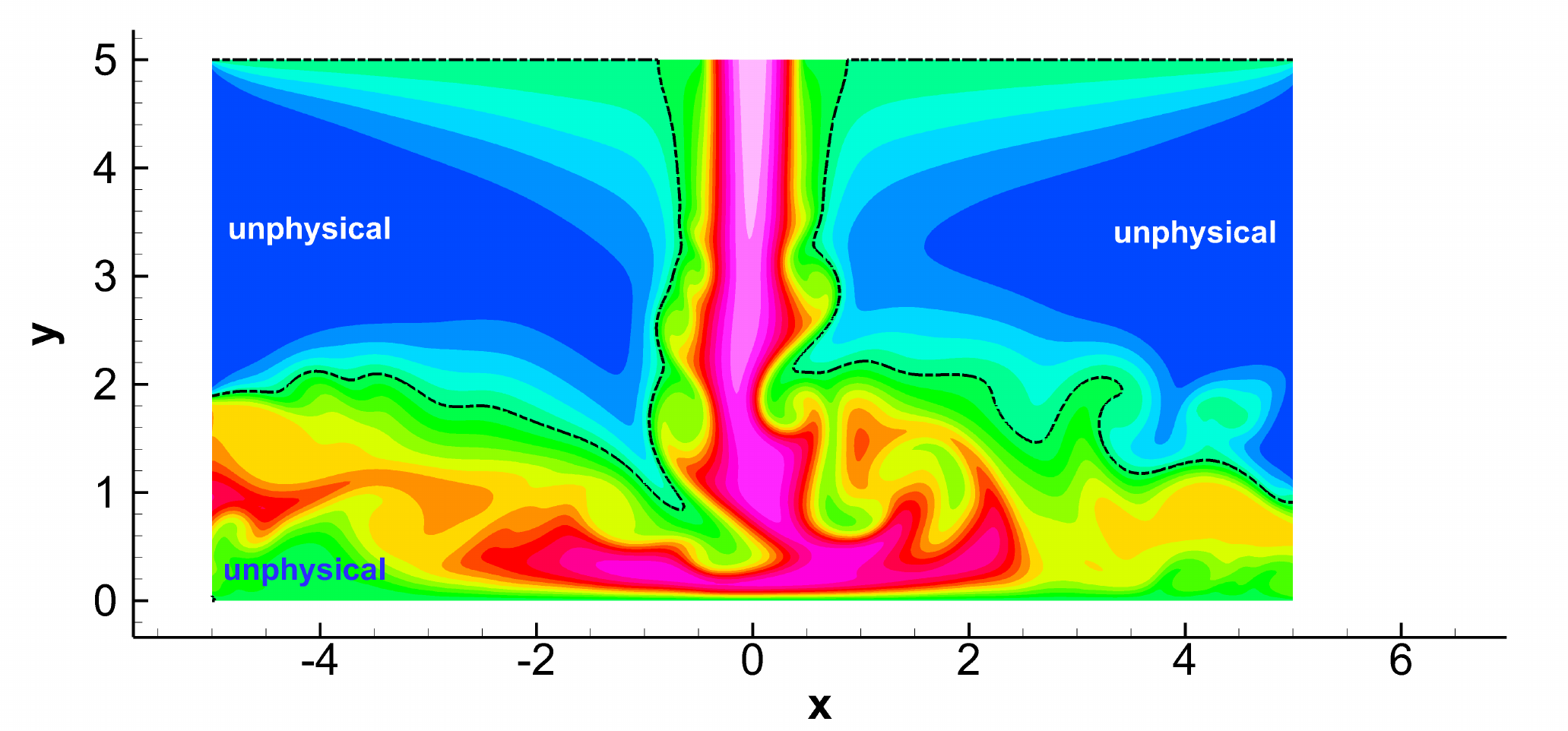}(f)
  }
  \centerline{
    \includegraphics[width=3.1in]{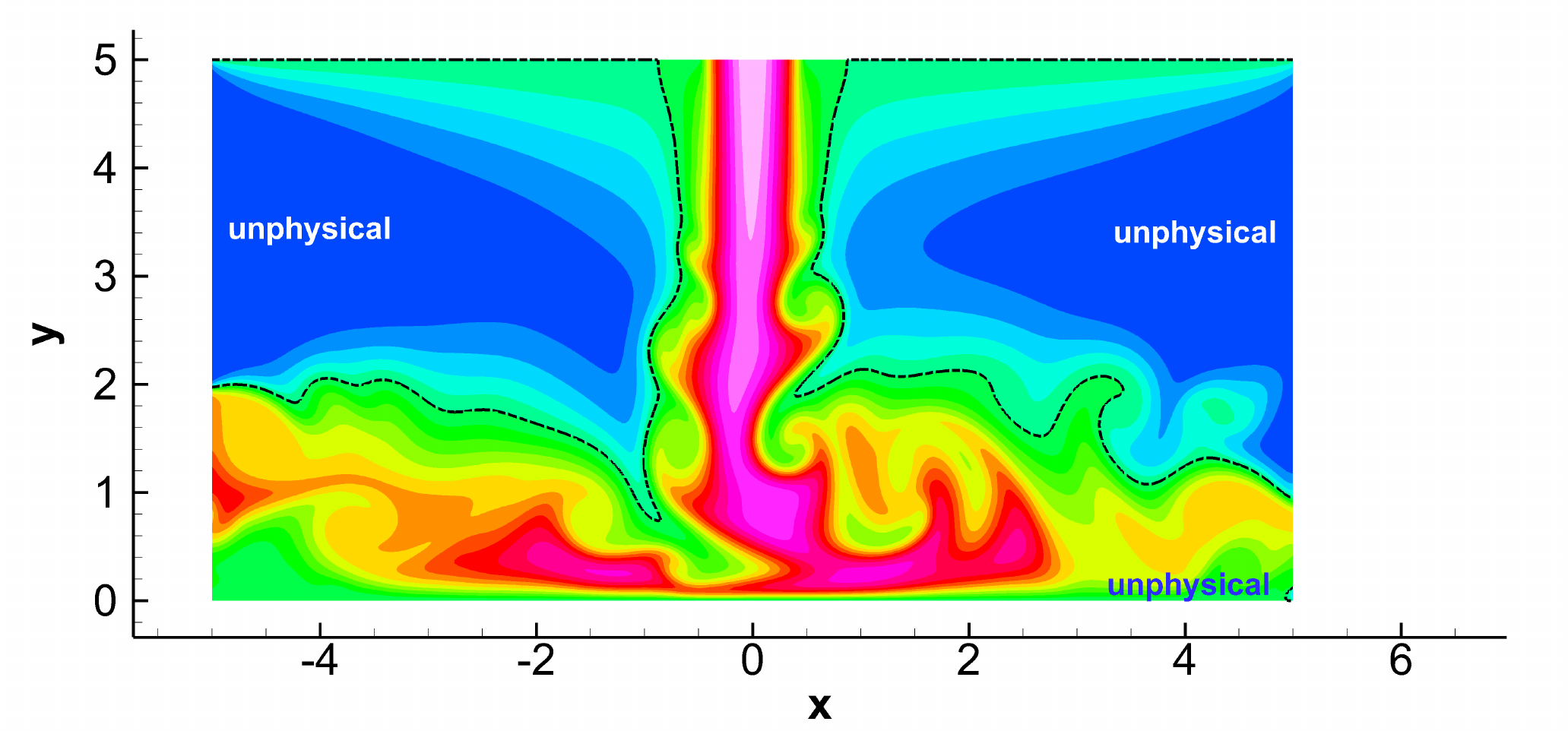}(g)
    \includegraphics[width=3.1in]{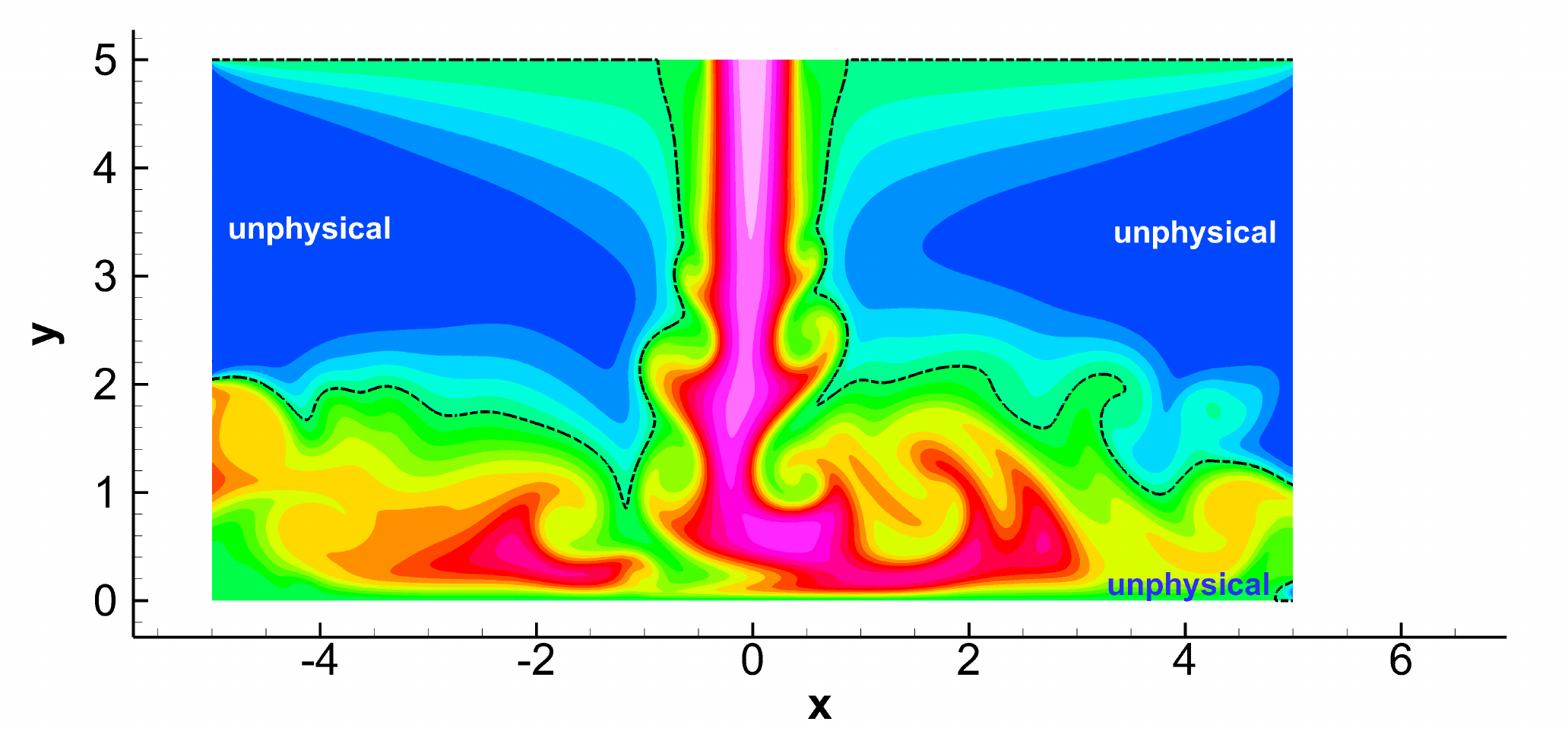}(h)
  }
  \centerline{
    \includegraphics[width=3.1in]{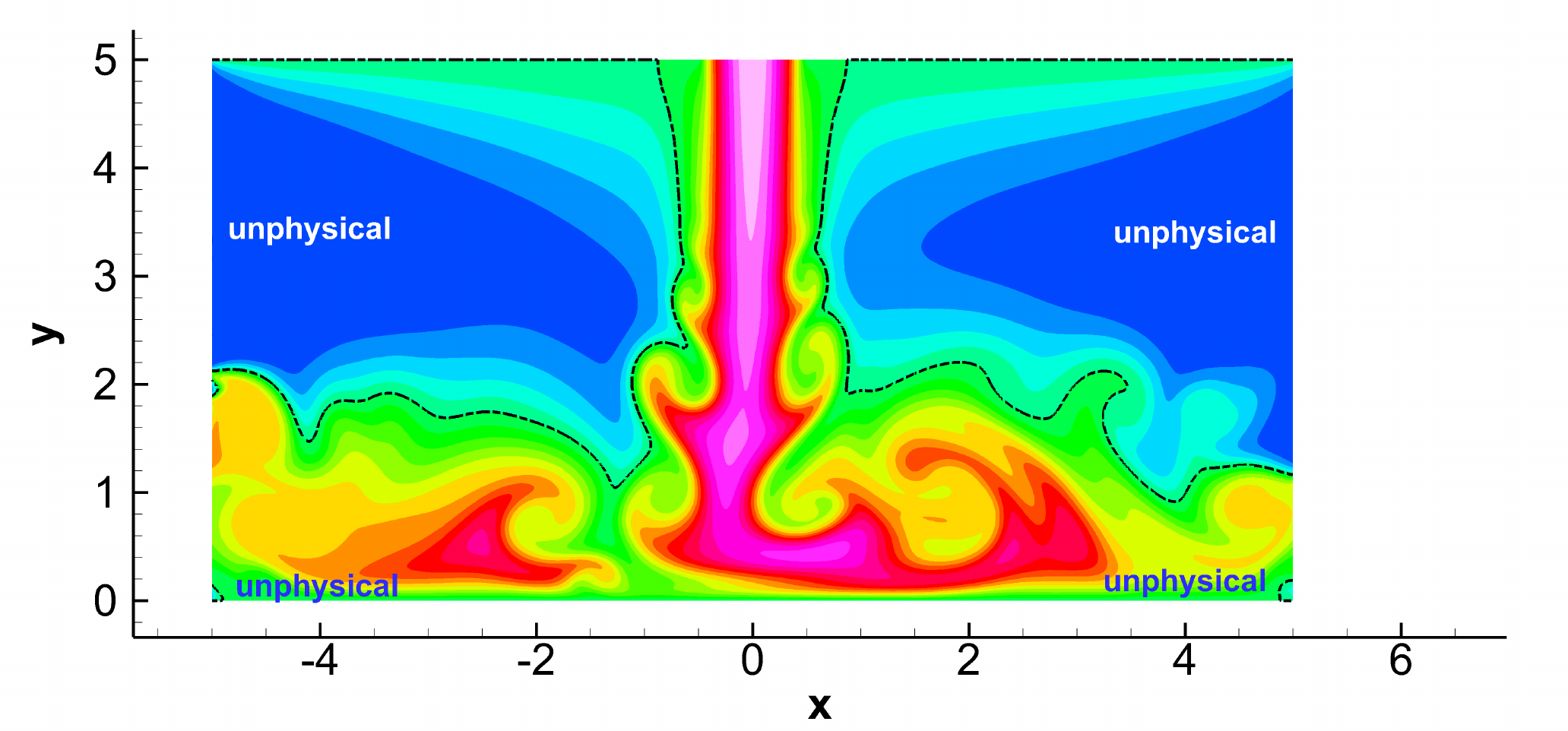}(i)
    \includegraphics[width=3.1in]{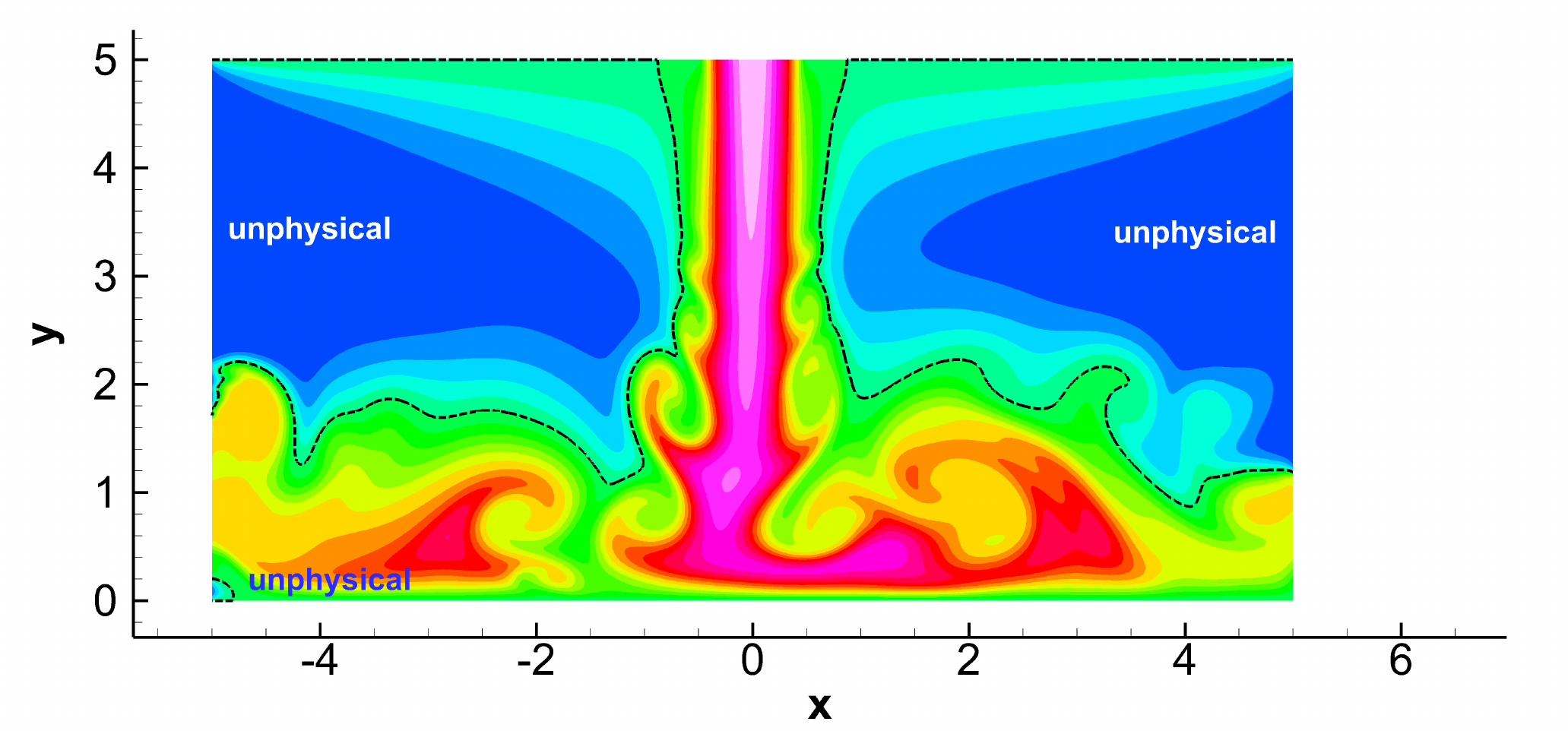}(j)
  }
  \caption{ (color online)
    Impinging jet (Re=5000): temporal sequence of snapshots of
    the temperature distribution
    computed using the Neumann-type zero-flux OBC (equation~\eqref{equ:zero_obc}).
    (a) $t=t_1$, (b) $t=t_1+0.5$, (c) $t=t_1+1.0$, (d) $t=t_1+1.5$,
    (e) $t=t_1+2.0$, (f) $t=t_1+2.5$, (g) $t=t_1+3.0$, (h) $t=t_1+3.5$,
    (i) $t=t_1+4.0$, (j) $t=t_1+4.5$.
    Thermal diffusivity is $\alpha=0.005$.
    Note that the color map represents
    temperature values ranging from $T=0$ to $T=80$.
    The dashed curves mark the contour level $T=20$.
    The Neumann-type OBC leads to unphysical temperature distribution outside
    the jet stream.
  }
  \label{fig:jet_Tfree}
\end{figure}


We now look into the effect of the thermal open boundary condition on
the simulated temperature distributions, and compare the current thermal
OBC (equation~\eqref{equ:obc_A}) and the  Neumann-type zero-flux
OBC (equation~\eqref{equ:zero_obc}) for the impinging jet problem.
We observe that at low Reynolds numbers (such as $Re=300$) these OBCs
produce approximately the same temperature distribution.
However, as the Reynolds number increases to moderate and fairly
large values (about $Re=2000$ and beyond), these OBCs exhibit disparate performance
and produce quite different results for the temperature field.
It is observed that the Neumann-type OBC~\eqref{equ:zero_obc} results in
unphysical temperature distributions in large regions of the domain,
while the current OBC~\eqref{equ:obc_A}
leads to a reasonable temperature distribution in the entire domain. 
These points are demonstrated by  Figures~\ref{fig:jet_T}
to~\ref{fig:jet_Tfree}.
Figure \ref{fig:jet_T} shows the temporal sequence of
snapshots of the temperature fields
computed
using the current thermal OBC~\eqref{equ:obc_A}.
Figure \ref{fig:jet_Tfree} shows the temporal sequence of snapshots
of the temperature fields
computed using the Neumann-type
zero-flux OBC~\eqref{equ:zero_obc}.
The thermal diffusivity is $\alpha=0.005$,
and these results are obtained with an element order $8$ and a time step
size $\Delta t=2.5e-4$.

%

At $Re=5000$, vortices are observed to constantly form
and are convected along the jet streams.
These vortices persist far downstream, and ultimately
discharge from the domain through the left and right
open boundaries. 
It is further observed that, while quite weak,
there is a persistent backflow on the upper portions of
the left and right open boundaries (velocity vectors pointing generally inward),
apparently due to the entrainment
effect of the jet and the flow continuity.
This characteristic is different from that of the cylinder flow
in Section \ref{sec:cyl}, where the backflows on the outflow
boundary are transient and only occur when strong vortices
pass through the open boundary.

We observe that at $Re=5000$ the Neumann-type zero-flux
OBC~\eqref{equ:zero_obc} produces erroneous and
unphysical temperature distributions in vast regions of the domain,
apparently due to
the strong vortices and persistent backflows on the 
open boundary. In Figure \ref{fig:jet_Tfree}, the dashed curves
in each plot mark the temperature contour level $T=20$ degrees
Celsius. The computed temperature in the regions outside the jet streams
have values below $20$ degrees, and in a large region
the temperature is essentially zero.
These results are clearly unphysical.
Similar unphysical temperature distribution has been observed at
$Re=2000$ with this boundary condition.
These results indicate that, while it seems to work well at low Reynolds numbers,
the Neumann-type zero-flux OBC~\eqref{equ:zero_obc}
is inadequate for moderate and high Reynolds numbers, when
strong vortices or backflows are present on the open boundary.

In contrast, the current thermal OBC~\eqref{equ:obc_A}
performs quite well.  It
produces a reasonable temperature distribution
for all Reynolds numbers tested here;
see Figure \ref{fig:jet_T}
for $Re=5000$ and Figures \ref{fig:jet_char}(b,d) for lower Reynolds numbers.
The current method can handle thermal open boundaries well at
moderate and high Reynolds numbers, even when strong vortices or
backflows are prevalent there and pose a significant issue to
other methods.



\section{Concluding Remarks}
\label{sec:summary}


We have presented a simple and effective thermal open boundary condition
 for simulating convective heat transfer
problems on domains involving outflows or open boundaries.
This boundary condition is energy-stable, and ensures that the contribution
of the open boundary will not cause an ``energy-like''
function of the temperature to increase over time,
regardless of the flow situation that occurs on
the outflow/open boundary. 
This open boundary condition can be implemented in
a straightforward way into semi-implicit type schemes
for the heat transfer equation.
Ample numerical experiments show that the presented open boundary condition
has a clear advantage over related methods such as the
Neumann-type zero-flux condition
for high (and moderate) Reynolds numbers,
where strong vortices or backflows might occur on the open boundary.
In the presence of strong vortices or backflows
at the open boundary, the current method  produces reasonable
temperature distributions, while the Neumann-type zero-flux condition
leads to unphysical and erroneous temperature fields.
We anticipate that the presented method can be a powerful tool
and be instrumental in
heat transfer simulations involving inflows/outflows at
large Reynolds numbers.

\section*{Acknowledgement}
This work was partially supported by
NSF (DMS-1522537) and a scholarship from the China Scholarship Council
(CSC-201806080040). 

\section*{Appendix A. Numerical Algorithm and Open Boundary Condition for
  Incompressible Navier-Stokes Equations}

The numerical algorithm and
the open boundary conditions for
the incompressible Navier-Stokes equations
\eqref{equ:nse}--\eqref{equ:div}
employed in the current work stem from
our previous work~\cite{Dong2015clesobc}. 
This Appendix provides a summary of these aspects.
We use the same notation here as in the main text.

For the Navier-Stokes equations \eqref{equ:nse}--\eqref{equ:div},
on the boundary $\partial\Omega_d$
($\partial\Omega_d=\partial\Omega_{dd}\cup\partial\Omega_{dn}$) we
impose the Dirichlet condition
\begin{equation}\label{equ:dbc_v}
\mbs u = \mbs w(\mbs x,t), \quad\text{on}\ \partial\Omega_d,
\end{equation}
where $\mbs w(\mbs x,t)$ is the boundary velocity.
On the open boundary we impose the following condition
(see~\cite{Dong2015clesobc}), 
\begin{equation}
  \nu D_0\frac{\partial\mbs u}{\partial t}
  - p\mbs n + \nu\mbs n\cdot\nabla\mbs u
  - \mbs E(\mbs n, \mbs u) = \mbs f_b(\mbs x,t),
  \quad \text{on} \ \partial\Omega_o,
  \label{equ:obc_v}
\end{equation}
where $D_0\geqslant 0$ is the same constant parameter
as in the temperature condition \eqref{equ:obc_A}, and
$\mbs E(\mbs n,\mbs u)$ is given by
\begin{equation}\label{equ:def_E}
  \mbs E(\mbs n,\mbs u) = \frac12\left[
    (\mbs u\cdot\mbs u)\mbs n + (\mbs n\cdot\mbs u)\mbs u
    \right]\Theta_0(\mbs n,\mbs u).
\end{equation}
Note that $\Theta_0(\mbs n,\mbs u)$ is defined in \eqref{equ:def_Theta0}.
A more general form for $\mbs E(\mbs n,\mbs u)$ is given
by~\cite{Dong2015clesobc},
\begin{equation}\label{equ:def_Egobc}
  \mbs E(\mbs n,\mbs u) = \frac12\left[
    (\beta_0+\beta_2)(\mbs u\cdot\mbs u)\mbs n +
    (1-\beta_0+\beta_1)(\mbs n\cdot\mbs u)\mbs u
    \right]\Theta_0(\mbs n,\mbs u),
\end{equation}
where $\beta_0$, $\beta_1$ and $\beta_2$ are constant
parameters satisfying $0\leqslant\beta_0\leqslant 1$,
$\beta_1\geqslant 0$ and $\beta_2\geqslant 0$.
The open boundary condition \eqref{equ:obc_v}, with $\mbs f_b=0$,
is an energy-stable boundary condition for the
incompressible Navier-Stokes
equations \eqref{equ:nse}--\eqref{equ:div}
as $\delta\rightarrow 0$~\cite{Dong2015clesobc}.
In addition, we impose the following initial condition
for the velocity,
\begin{equation}\label{equ:ic_v}
  \mbs u(\mbs x,0) = \mbs u_{in}(\mbs x)
\end{equation}
where $\mbs u_{in}$ denotes the initial velocity distribution.

Given ($\mbs u^n$,$p^n$) we compute $p^{n+1}$ and $\mbs u^{n+1}$
successively in a de-coupled fashion as follows: \\
\underline{For $p^{n+1}$:}
\begin{subequations}
\begin{align}
&
  \frac{\gamma_0\tilde{\mathbf{u}}^{n+1}-\hat{\mathbf{u}}}{\Delta t}
+ \mathbf{u}^{*,n+1}\cdot\nabla\mathbf{u}^{*,n+1}
+ \nabla p^{n+1}
+ \nu \nabla\times\nabla\times\mathbf{u}^{*,n+1}
= \mathbf{f}^{n+1};
\label{equ:pressure_1}
\\
&
\nabla\cdot\tilde{\mathbf{u}}^{n+1} = 0;
\label{equ:pressure_2}
\\
&
\mathbf{n}\cdot\tilde{\mathbf{u}}^{n+1} 
 = \mathbf{n} \cdot \mathbf{w}^{n+1},
\quad \text{on} \ \partial\Omega_d;
\label{equ:pressure_3}
\\ 
&
\nu D_0\frac{\gamma_0\tilde{\mathbf{u}}^{n+1}-\hat{\mathbf{u}}}{\Delta t}\cdot\mathbf{n}
- p^{n+1}
+ \nu\mathbf{n}\cdot\nabla\mathbf{u}^{*,n+1}\cdot\mathbf{n}
- \mathbf{n}\cdot\mathbf{E}(\mathbf{n},\mathbf{u}^{*,n+1})
= \mathbf{f}_b^{n+1}\cdot\mathbf{n},
\quad \text{on} \ \partial\Omega_o.
\label{equ:pressure_4}
\end{align}
\end{subequations}
\underline{For $\mathbf{u}^{n+1}$:}
\begin{subequations}
\begin{align}
    &
\frac{\gamma_0\mathbf{u}^{n+1}-\gamma_0\tilde{\mathbf{u}}^{n+1}}{\Delta t}
- \nu\nabla^2\mathbf{u}^{n+1} 
= \nu \nabla\times\nabla\times\mathbf{u}^{*,n+1};
\label{equ:velocity_1}
\\
&
\mathbf{u}^{n+1} = \mathbf{w}^{n+1},
\quad \text{on} \ \partial\Omega_d;
\label{equ:velocity_2}
\\ 
&
\nu D_0\frac{\gamma_0\mathbf{u}^{n+1}-\hat{\mathbf{u}}}{\Delta t}
- p^{n+1}\mathbf{n} 
+ \nu\mathbf{n}\cdot\nabla\mathbf{u}^{n+1}
- \mathbf{E}(\mathbf{n},\mathbf{u}^{*,n+1})
+ \nu\left(\nabla\cdot\mathbf{u}^{*,n+1}  \right)\mathbf{n}
= \mathbf{f}_b^{n+1},
\quad \text{on} \ \partial\Omega_o.
\label{equ:velocity_3}
\end{align}
\end{subequations}
%
In the above equations $\tilde{\mbs u}^{n+1}$ is an auxiliary
variable approximating $\mbs u^{n+1}$. We again use
$J$ ($J=1$ or $2$) to denote the temporal order of accuracy.
$\gamma_0$ is defined in \eqref{equ:def_hat}.
$\hat{\mbs u}$ and $\mbs u^{*,n+1}$ are defined by
\begin{equation}
  \hat {\mbs u} = \left\{
  \begin{array}{ll}
    \mbs u^n, & J=1, \\
    2\mbs u^n - \frac12 \mbs u^{n-1}, & J=2;
  \end{array}
  \right.
  \qquad
  \mbs u^{*,n+1} = \left\{
  \begin{array}{ll}
    \mbs u^n, & J=1, \\
    2\mbs u^n-\mbs u^{n-1}, & J=2.
  \end{array}
  \right.
\end{equation}

The weak form for the pressure $p^{n+1}$ can be derived from
equations \eqref{equ:pressure_1}--\eqref{equ:pressure_4}, and
it is given by~\cite{Dong2015clesobc},
\begin{equation}
\begin{split}
\int_{\Omega} \nabla p^{n+1}\cdot \nabla q
&+ \frac{1}{\nu D_0} \int_{\partial\Omega_o}
     p^{n+1} q
=  \int_{\Omega} \mathbf{G}^{n+1}\cdot\nabla q
- \nu\int_{\partial\Omega_d\cup\partial\Omega_o} 
     \mathbf{n}\times \bm{\omega}^{*,n+1}\cdot\nabla q \\
&
+ \int_{\partial\Omega_o} \left\{
  -\frac{1}{\Delta t}\mathbf{n}\cdot\hat{\mathbf{u}}
  + \frac{1}{\nu D_0}\left[
    \nu\mathbf{n}\cdot\nabla\mathbf{u}^{*,n+1}\cdot\mathbf{n}
     - \mathbf{n}\cdot\mathbf{E}(\mathbf{n},\mathbf{u}^{*,n+1})
     - \mathbf{f}_b^{n+1}\cdot\mathbf{n}
  \right]
\right\} q \\
&
- \frac{\gamma_0}{\Delta t}\int_{\partial\Omega_d}
     \mathbf{n}\cdot \mathbf{w}^{n+1} q,
\qquad \forall q \in H^1(\Omega),
\end{split}
\label{equ:p_weakform}
\end{equation}
where
$
\mbs G^{n+1} = \mathbf{f}^{n+1} + \frac{\hat{\mathbf{u}}}{\Delta t}
- \mathbf{u}^{*,n+1}\cdot\nabla\mathbf{u}^{*,n+1}.
$
The weak form for the velocity is given by,
\begin{equation}
\begin{split}
\frac{\gamma_0}{\nu\Delta t} \int_{\Omega}\mathbf{u}^{n+1}\varphi
&+ \int_{\Omega}\nabla\varphi\cdot\nabla\mathbf{u}^{n+1}
+ \frac{\gamma_0 D_0}{\Delta t} \int_{\partial\Omega_o} 
       \mathbf{u}^{n+1}\varphi
= \frac{1}{\nu}\int_{\Omega}\left(
  \mathbf{G}^{n+1}-\nabla p^{n+1}
\right)\varphi \\
&
+ \int_{\partial\Omega_o}\left\{
  \frac{D_0}{\Delta t}\hat{\mathbf{u}}
  + \frac{1}{\nu}\left[
    p^{n+1}\mathbf{n} + \mathbf{E}(\mathbf{n},\mathbf{u}^{*,n+1})
    + \mathbf{f}_b^{n+1}
    - \nu \left(\nabla\cdot\mathbf{u}^{*,n+1}  \right)\mathbf{n}
  \right]
  \right\} \varphi, \\
  &
 \forall \varphi\in H^{1}(\Omega) \ \text{with}\
\varphi|_{\partial\Omega_d}=0.
\end{split}
\label{equ:u_weakform}
\end{equation}
The weak forms in \eqref{equ:p_weakform} and \eqref{equ:u_weakform}
can be discretized using $C^0$ spectral elements in
the standard fashion~\cite{KarniadakisS2005}.
Within each time step, we first solve equation \eqref{equ:p_weakform}
for $p^{n+1}$. Then we solve equation \eqref{equ:u_weakform},
together with the boundary condition \eqref{equ:velocity_2},
for $\mbs u^{n+1}$.
Note that the auxiliary velocity $\tilde{\mbs u}^{n+1}$ is not
explicitly computed.


\bibliographystyle{plain}
\bibliography{heat,obc,mypub,nse,sem,contact_line,interface}

\end{document}